\documentclass[sigconf,screen,nonacm]{acmart}
\newcommand{\citeFullOrAppendix}[1]{\cref{#1}}
\newcommand{\ifFull}[1]{#1}
\newcommand{\ifConference}[1]{}

\copyrightyear{2024}
\acmYear{2024}
\setcopyright{rightsretained}
\acmConference[LICS '24]{39th Annual ACM/IEEE Symposium on Logic in Computer Science}{July 8--11, 2024}{Tallinn, Estonia}
\acmBooktitle{39th Annual ACM/IEEE Symposium on Logic in Computer Science (LICS '24), July 8--11, 2024, Tallinn, Estonia}\acmDOI{10.1145/3661814.3662133}
\acmISBN{979-8-4007-0660-8/24/07}

\usepackage{amsmath,mathtools,amsthm,thmtools}
\usepackage{enumerate}
\usepackage{url,hyperref}
\usepackage[capitalize]{cleveref}
\crefname{enumi}{item}{items}
\crefname{subsection}{Section}{sections}
\AtEndPreamble{%
  \theoremstyle{acmplain}
  \newtheorem{claim}[theorem]{Claim}
  \theoremstyle{acmdefinition}
  \newtheorem{observation}[theorem]{Observation}
  \newtheorem{remark}[theorem]{Remark}}
\usepackage{tikz}
\usetikzlibrary{shapes.multipart, shapes.geometric, positioning, automata}
\usepackage{pgfplots}
\usepackage{pgfpages}

\usepackage{xcolor}
\usepackage{colortbl}

\usepackage{stmaryrd}
\usepackage{mathtools}
\usepackage{bussproofs}

\usepackage{caption}
\usepackage{subcaption}
\usepackage{multirow}
\usepackage{array}
\usepackage{csquotes}
\usepackage{microtype}

\usepackage{float}

\usepackage{listings}
\usepackage{algpseudocode}

\usepackage[textsize=tiny,disable]{todonotes}
\usepackage{datetime2}
\usepackage{newfile} %
\definecolor{ghost-white}{HTML}{F9F8F8}
\definecolor{platinum}{HTML}{E3E2E2}
\definecolor{ash-gray}{HTML}{B6B5B5}
\definecolor{yankees-blue}{HTML}{22223B}
\definecolor{japanese-violet}{HTML}{592B4D}
\definecolor{mountbatten-pink}{HTML}{9F8095}
\definecolor{mountbatten-pink-light}{HTML}{B7A1AE}
\definecolor{deep-ruby}{HTML}{7A3B69}
\definecolor{pewter-blue}{HTML}{8DA1B9}
\definecolor{ming}{HTML}{3A597A}
\definecolor{japanese-indigo}{HTML}{294159}
\definecolor{roman-silver}{HTML}{808D98}
\definecolor{emerald}{HTML}{50C878}
\definecolor{halfgray}{HTML}{7E7E7E}
\definecolor{darkgray}{HTML}{3E3E5E}
\definecolor{solarized-background}{HTML}{FDF6E3}
\definecolor{solarized-background-dark}{HTML}{EEE8D5}
\definecolor{solarized-default}{HTML}{657B83}
\definecolor{solarized-comment}{HTML}{93A1A1}
\definecolor{solarized-green}{HTML}{859900}
\definecolor{solarized-blue}{HTML}{268BD2}
\definecolor{solarized-cyan}{HTML}{2AA198}

\definecolor{wong-green}{HTML}{009E73}
\definecolor{wong-blue}{HTML}{0072B2}
\definecolor{wong-orange}{HTML}{D55E00}
\definecolor{wong-pink}{HTML}{CC79A7}

\colorlet{highlight}{wong-orange}
\colorlet{highlight-2}{wong-pink!80!black}
\colorlet{highlight-3}{wong-green}

\newcommand{\hlA}[1]{\textcolor{highlight}{#1}}
\newcommand{\hlB}[1]{\textcolor{highlight-2}{#1}}
\newcommand{\hlC}[1]{\textcolor{highlight-3}{#1}}

\makeatletter
\newcommand{\ProcessDigit}[1]
{%
    \ifnum\lst@mode=\lst@Pmode\relax%
    {\ttfamily\color{solarized-cyan} #1}%
    \else
    #1%
    \fi
}

\def\addToLiterate#1{\edef\lst@literate{\unexpanded\expandafter{\lst@literate}\unexpanded{#1}}}
\lst@Key{add to literate}{}{\addToLiterate{#1}}
\makeatother

\newcommand{\symbolstyle}{\color{solarized-blue}}

\lstdefinelanguage{agda}{
    morekeywords        = {data,where,module,open,import,using,hiding},
    morekeywords        = {renaming,private,variable,record,with},
    morekeywords        = {constructor,field},
    morecomment         = [l]{--},
    morecomment         = [s]{\{-}{-\}},
    literate            = {bNat}{{\(\mathbb{N}\)}}1
                        {\\bZ}{{\(\mathbb{Z}\)}}1
                        {0}{{{\ProcessDigit{0}}}}1
                        {1}{{{\ProcessDigit{1}}}}1
                        {2}{{{\ProcessDigit{2}}}}1
                        {3}{{{\ProcessDigit{3}}}}1
                        {4}{{{\ProcessDigit{4}}}}1
                        {5}{{{\ProcessDigit{5}}}}1
                        {6}{{{\ProcessDigit{6}}}}1
                        {7}{{{\ProcessDigit{7}}}}1
                        {8}{{{\ProcessDigit{8}}}}1
                        {9}{{{\ProcessDigit{9}}}}1
                        {\ ->\ }{{{\symbolstyle\(\to\)}}}4
                        {:}{{{\symbolstyle{:}}}}1
                        {forall}{{\(\forall\)}}1
                        {\ \\equiv\ }{{{\symbolstyle\(\equiv\)}}}3
                        {\ \\lub\ }{{{\symbolstyle\(\sqcup\)}}}3
                        {\\lub}{{\(\sqcup\)}}1
                        {\\glb}{{\(\sqcap\)}}1
                        {\\lambda}{{{\symbolstyle\(\lambda\)}}}1
                        {-\\leq}{{{\symbolstyle-\(\leq\)}}}2
                        {\ \\leq\ }{{{\symbolstyle\(\leq\)}}}3
                        {\\leq?}{{{\symbolstyle\(\leq\)?}}}2
                        {\\\^r}{{{\(^r\)}}}1
                        {\\\^l}{{{\(^l\)}}}1
                        {\\\_0}{{\(_0\)}}1
                        {\\\_1}{{\(_1\)}}1
                        {\\\_2}{{\(_2\)}}1
                        {\\\_3}{{\(_3\)}}1
                        {\\lceil}{{\(\lceil\)}}1
                        {\\rceil}{{\(\rceil\)}}1
                        {\\lfloor}{{\(\lfloor\)}}1
                        {\\rfloor}{{\(\rfloor\)}}1
                        {\\llog}{{\(\lceil\)log\(_2\)}}5
                        {\\ndiv2ceil}{{\(\lceil\)n/2\(\rceil\)}}5
                        {\\ndiv2floor}{{\(\lfloor\)n/2\(\rfloor\)}}5
                        {=>}{{\(\Rightarrow\)}}1
                        {s\\leqs}{{s\(\leq\)s}}3
                        {z\\leqn}{{z\(\leq\)n}}3
                        {height-\\equiv}{{{\symbolstyle height-\(\equiv\)}}}8
                        {delay-\\leq}{{{\symbolstyle delay-\(\leq\)}}}7
                        {\\equiv}{{\(\equiv\)}}1
                        {\\leq}{{\(\leq\)}}1
                        {\\geq}{{\(\geq\)}}1
                        {\\langle}{{\(\langle\)}}1
                        {\\rangle}{{\(\rangle\)}}1
                        {\\circ}{{\(\circ\)}}1
                        {\\qed}{{\(\qed\)}}1
                        {\\ominus}{{\(\ominus\)}}1
                        {\\nlt}{{\(\not<\)}}1
}
\graphicspath{ {./Images/} }

\pgfplotsset{compat=1.16}

\usepackage{color}

\hypersetup{%
	colorlinks=true,
	linkcolor=blue,
	urlcolor=blue,
	citecolor=blue
}
\makeatletter
\newcommand{\oset}[3][0ex]{%
	\mathrel{\mathop{#3}\limits^{
			\vbox to#1{\kern-2\ex@
				\hbox{$\scriptstyle#2$}\vss}}}}
\makeatother

\DeclarePairedDelimiter\abs{\lvert}{\rvert}

\AtBeginDocument{\renewcommand{\epsilon}{\varepsilon}}
\AtBeginDocument{\renewcommand{\phi}{\varphi}}

\newcommand{\Parikh}[1]{\Psi(#1)}

\newcommand{\Nat}{\mathbb{N}}
\newcommand{\Z}{\mathbb{Z}}
\newcommand{\N}{\mathbb{N}}

\newcommand{\yield}{\mathsf{yield}}
\newcommand{\gap}[2]{\mathsf{G}_{#1,#2}}

\newcommand{\runembed}{\trianglelefteq}
\newcommand{\embedwith}[2][\runembed]{\mathrel{#1_{#2}}}

\newcommand{\Alg}{\mathsf{Alg}}
\newcommand{\Deco}{\mathsf{Deco}}
\DeclarePairedDelimiter{\canonlen}{\lvert}{\rvert_{\mathsf{can}}}
\newcommand{\canonidx}{\mathsf{ix}}
\NewDocumentCommand{\liftmap}{o o m}{%
	\IfNoValueTF{#1}{%
		\IfNoValueTF{#2}{%
			\hat{#3}%
		}{%
			\hat{#3}_{#2}%
		}%
	}{%
		\IfNoValueTF{#2}{%
			\hat{#3}^{#1}%
		}{%
			\hat{#3}^{#1}_{#2}%
		}%
	}%
}
\DeclareMathOperator{\bigdisunion}{\dot{\bigcup}}

\newcommand{\Model}{\mathcal{M}}
\newcommand{\Grammar}{\mathcal{G}}
\newcommand{\Class}[1]{\ensuremath{\mathsf{#1}}}
\newcommand{\Reg}{\Class{Reg}}

\newcommand{\CFL}{\Class{CFL}}
\newcommand{\VASS}{\Class{VASS}}
\newcommand{\PVASS}{\Class{PVASS}}
\newcommand{\VASSnz}{\VASS${}_{\text{nz}}$}
\newcommand{\memo}{\mu}

\newcommand{\complexity}[1]{\textsc{#1}}
\newcommand{\EXPSPACE}{\complexity{ExpSpace}}

\DeclareDocumentCommand{\deriv}{O{}}{\,\Rightarrow_{#1}\,}
\DeclareDocumentCommand{\derivs}{O{}}{\,\oset{*}{\Rightarrow}_{#1}\,}
\DeclareDocumentCommand{\derivv}{O{} m}{\,\Rightarrow_{#1}^{#2}\,}
\DeclareDocumentCommand{\derivvs}{O{} m}{\,\oset{*}{\Rightarrow}_{#1}^{#2}\,}

\DeclareDocumentCommand{\autstep}{O{}}{\oset{#1}{\rightarrow}}
\DeclareDocumentCommand{\autsteps}{O{}}{\oset{#1}{\rightarrow}}

\newcommand{\cA}{\mathcal{A}}
\newcommand{\cC}{\mathcal{C}}%
\newcommand{\cT}{\mathcal{T}}

\AtBeginDocument{\renewcommand{\vec}[1]{\ensuremath{\mathbf{#1}}}}

\newcommand{\eqby}[1]{\mathrel{\raisebox{-.1ex}{\ensuremath{\stackrel{\raisebox{-.25ex}{\scalebox{.5}{\upshape\textrm{#1}}}}{=}}}}}
\newcommand{\eqdef}{\eqby{def}}
\newcommand{\Siep}{\Sigma_{\varepsilon}}%
\newcommand{\Siepp}{\Siep}%

\newcommand{\SC}{\mathsf{SC}}
\newcommand{\Pfour}{\mathsf{P4}}
\newcommand{\Cfour}{\mathsf{C4}}
\newcommand{\M}{\mathbb{M}}
\newcommand{\VA}{\mathsf{VA}}
\newcommand{\PD}{\mathsf{PD}}
\newcommand{\B}{\mathbb{B}}
\newcommand{\can}[1]{\mathsf{can}(#1)}

\newcommand\dc[1]{#1{\downarrow}}

\newcommand\vleq\leq

\newcommand\vstrictlt{\ll}

\newcommand\Trees[1]{{T}(#1)}
\newcommand\leaf[1]{#1[]}
\newcommand\tree[2]{#1[#2]}
\newcommand\subtree[2]{#1/#2}

\let\yieldplain\yield
\renewcommand\yield[1]{\yieldplain(#1)}

\newcommand\treesubst[3]{#1\left[#2 \mapsto #3\right]}

\tikzset{auto,}

\pgfdeclarelayer{back}
\pgfdeclarelayer{front}
\pgfsetlayers{back,main,front}

\makeatletter
\pgfkeys{%
	/tikz/on layer/.code={
		\def\tikz@path@do@at@end{\endpgfonlayer\endgroup\tikz@path@do@at@end}%
		\pgfonlayer{#1}\begingroup%
	}%
}
\makeatother

\usetikzlibrary{arrows,automata,chains,matrix,positioning,scopes,calc,decorations.pathmorphing,shapes.misc,shapes,fit,backgrounds}
\tikzset{mode/.style={font=\scriptsize}}
\tikzset{gadget/.style={->,>=stealth,initial text=,minimum size=7pt,auto,on grid,scale=1,inner sep=1pt,node distance=1cm}}
\tikzset{every state/.style={minimum size=15pt,inner sep=1pt,fill=black!10,draw=black!70,thick}}

\newcommand{\unloopedcycle}[1]{
\begin{tikzpicture}[every circle/.style={}, scale=#1]
\fill (0,0) circle (2pt) node (a) {}    (1,0) circle (2pt) node (b)  {}   (0,-1) circle (2pt) node (c) {}   (1,-1) circle (2pt) node (d) {};
\draw (a.center) -- (b.center) -- (d.center) -- (c.center) -- (a.center);
\end{tikzpicture}
}

\newcommand{\unloopedpathFour}[1]{
\begin{tikzpicture}[every circle/.style={}, scale=#1]
\fill (0,0) circle (2pt) node (a) {}    (1,0) circle (2pt) node (b)  {}   (2,0) circle (2pt) node (c) {}   (3,0) circle (2pt) node (d) {};
\draw (a.center) -- (b.center) -- (c.center) -- (d.center);
\end{tikzpicture}
}

\begin{document}
\title{Verifying Unboundedness via Amalgamation}
\author{Ashwani Anand}
\email{ashwani@mpi-sws.org}
\orcid{0000-0002-9462-8514}
\affiliation{%
  \institution{Max-Planck-Institute for Software Systems (MPI-SWS)}
  \city{Kaiserslautern}
  \country{Germany}}
\author{Sylvain Schmitz}
\email{schmitz@irif.fr}
\orcid{0000-0002-4101-4308}
\affiliation{%
  \institution{Universit\'e Paris Cit\'e, CNRS, IRIF}
  \city{Paris}
  \country{France}}
\author{Lia Sch\"{u}tze}
\email{lschuetze@mpi-sws.org}
\orcid{0000-0003-4002-5491}
\affiliation{%
  \institution{Max-Planck-Institute for Software Systems (MPI-SWS)}
  \city{Kaiserslautern}
  \country{Germany}}
\author{Georg Zetzsche}
\email{georg@mpi-sws.org}
\orcid{0000-0002-6421-4388}
\affiliation{%
  \institution{Max-Planck-Institute for Software Systems (MPI-SWS)}
  \city{Kaiserslautern}
  \country{Germany}}
\begin{abstract}
  Well-structured transition systems (WSTS) are an abstract family of
systems that encompasses a vast landscape of infinite-state
systems. By requiring a well-quasi-ordering (wqo) on the set of
states, a WSTS enables generic algorithms for classic verification
tasks such as coverability and termination.  However, even for systems
that are WSTS like vector addition systems (VAS), the framework is
notoriously ill-equipped to analyse reachability (as opposed to
coverability).  Moreover, some important types of infinite-state
systems fall out of WSTS' scope entirely, such as pushdown systems
(PDS).

Inspired by recent algorithmic techniques on VAS, we propose an abstract notion
of systems where the set of \emph{runs} is equipped with a wqo and supports
\emph{amalgamation} of runs.  We show that it subsumes a large class of
infinite-state systems, including (reachability languages of) VAS and PDS, and
even all systems from the abstract framework of valence systems, except for
those already known to be Turing-complete.

Moreover, this abstract setting enables simple and general
algorithmic solutions to \emph{unboundedness problems}, which have received
much attention in recent years.  We present algorithms for the (i)~simultaneous
unboundedness problem (which implies computability of downward closures and
decidability of separability by piecewise testable languages), (ii)~computing
priority downward closures, (iii)~deciding whether a language is bounded,
meaning included in $w_1^*\cdots w_k^*$ for some words $w_1,\ldots,w_k$, and
(iv)~effective regularity of unary languages.  This leads to either drastically
simpler proofs or new decidability results for a rich variety of systems.

\end{abstract}



\maketitle

\newcommand{\impl}[2]{``$(\labelcref{#1})\Rightarrow(\labelcref{#2})$''}
\newcommand{\equ}[2]{``$(\labelcref{#1})\Leftrightarrow(\labelcref{#2})$''}

\section{Introduction}\label{sec:introduction}
\subsubsection*{Well Structured Transition Systems\nopunct} (WSTS for
short) form an abstract family of systems, for which generic
verification algorithms are available
off-the-shelf~\cite{abdulla00,finkel01}.  They were invented in the
late 1980s~\cite{finkel87} in an attempt to abstract and generalise
from techniques originally designed to reason about vector addition
systems (VASS) and their variants, but they have since been shown to
yield decision procedures outside the field of formal verification,
for various logics, automata models, or proof systems.  

One reason for
their success is the simplicity of the abstract definition: those are
transition systems, where the configurations are equipped with a
\emph{well-quasi-ordering}
(wqo)~\cite{Higman:1952:wqo,kruskal72,SS2012}, which is ``compatible''
with transition steps: if there is a step $c \to c'$ and a
configuration $d\geq c$, then there is a configuration $d'\geq c'$
reached by a step $d\to d'$.  From this simple definition and under
basic effectiveness assumptions, one can decide for instance whether
starting from a source configuration one can reach in finitely many
steps a configuration larger or equal to a target---aka
the \emph{coverability problem}, or restated in terms of formal
languages, the emptiness problem for coverability
languages~\cite{DBLP:journals/acta/GeeraertsRB07}.

Despite their wide applicability and the filiation from vector
addition systems, WSTS have a few essential limitations.  One is that
the framework is of no avail for the \emph{reachability problem},
which is actually undecidable for many classes of
WSTS~\cite{DBLP:conf/icalp/DufourdFS98}---though famously decidable
for vector addition
systems~\cite{mayr81,kosaraju82,lambert92,DBLP:conf/birthday/Leroux12}.
Another is that pushdown systems provide a notable exception to the
applicability of the definition~\cite[end of Sec.~7]{finkel01}---which
might help explain why we still have very limited understanding of vector
addition systems extended by a pushdown
store (PVASS)~\cite{DBLP:conf/icalp/LerouxST15,DBLP:conf/rp/SchmitzZ19}.

\subsubsection*{Towards Unboundedness Problems. }  The case of vector
addition systems is worth further attention: their reachability
problem can be decided through the construction of a
so-called \emph{KLM decomposition}, named
after \citet{kosaraju82}, \citet{lambert92}, and \citet{mayr81}: this
is a structure capturing all the possible runs of the system between
the source and target configurations.  Beyond
reachability, \citet[Sec.~5]{lambert92} already observed that this
decomposition provided considerably more information, allowing in
particular to derive a pumping lemma for VASS (reachability)
languages.  \Citet*{Habermehl:2010:downward-closure-vas} also show that
the \emph{downward closure} of VASS languages can be computed from
the KLM decomposition.

These applications of the KLM decomposition were pushed further
by~\citet*{Czerwinski:2018:unboundedness-problems-vas} to show the
decidability of a family of \emph{unboundedness problems} on VASS
languages---informally, of decision problems for formal languages
where one asks for the existence of infinitely many words of some
shape.
\begin{itemize}
\item One example is provided by \emph{separability problems}~\cite[e.g.,][]{DBLP:conf/icalp/Goubault-Larrecq16,DBLP:journals/dmtcs/CzerwinskiMRZZ17,DBLP:journals/lmcs/CzerwinskiL19,DBLP:conf/lics/CzerwinskiZ20}, where we are
given two languages $K$ and $L$ and we want to decide whether there
exists a language $R$ such that $K\subseteq R$ and $L\cap
R=\emptyset$. Here, $R$ is constrained to belong to a class
$\mathcal{S}$ of \emph{separators}, usually the regular languages or a
subclass thereof.  %
Deciding separability can be seen as an unboundedness
problem: Suppose that for two words $u$, $v$, one defines their distance
by taking the minimal size (of a finite monoid) %
of a language in
$\mathcal{S}$ separating $u$ and $v$.  Then for most classes
$\mathcal{S}$, we have that two languages $K$ and $L$ are inseparable
by $\mathcal{S}$ if and only if this distance between $K$
and $L$ is unbounded.
\item
Another, already mentioned example is computing \emph{downward
closures}.  It is a well known consequence of Higman's
Lemma~\citep{Higman:1952:wqo} that for every language $L$, the set
$\dc{L}$ of all scattered subwords of $L$ is a regular
language~\citep{Haines:1969:downward-closure-regularity}.  However,
effectively computing an automaton for $\dc{L}$ is often a difficult
task~\cite{Courcelle:1991:downward-closures-cfgs,Habermehl:2010:downward-closure-vas}.
In~\cite{DBLP:conf/icalp/Zetzsche15}, it was shown that computing
downward closures often reduces to the \emph{simultaneous
unboundedness problem} (SUP), which asks for a given language
$L\subseteq a_1^*\cdots a_n^*$, whether for every $k\in\N$, there
exists a word $a_1^{x_1}\cdots a_n^{x_n}\in L$ with $x_1,\ldots,x_n\ge
k$. Downward closure computation (also beyond the ordinary subword ordering~\cite{DBLP:conf/lics/Zetzsche18}) based on the SUP have been
studied in several papers in recent
years~\cite{DBLP:conf/popl/HagueKO16,DBLP:conf/fsttcs/Parys17,DBLP:journals/fuin/BarozziniCCP22}.
A refinement of this problem recently investigated in~\cite{anand_et_al_CONCUR}
is to work with a \emph{priority embedding}, allowing to
represent congestion control policies based on priorities
assigned to messages~\cite{haase14}, instead of the scattered
subword ordering. So far, downward closures for the priority
embedding are known to be computable only for context-free
languages~\cite{anand_et_al_CONCUR}.
\item
A third example is deciding whether a given language $L\subseteq\Sigma^*$ is
\emph{(language) bounded}, meaning whether there exist a $k\in\N$ and
$w_1,\ldots,w_k\in\Sigma^*$ such that $L\subseteq w_1^*\cdots w_k^*$.
Deciding language boundednes is motivated by the fact that bounded languages
have pleasant decidability properties \cite[e.g.,][]{DBLP:journals/tcs/ChambartFS16,BaumannDAlessandroGanardiIbarraMcQuillenSchuetzeZetzsche2022a}.
\end{itemize}

\subsubsection*{A Generic Approach. }  The previous applications of KLM
decompositions show the decidability of a whole range of properties
besides reachability in vector addition systems.  Our main motivation
in this paper is to study how to generalise the approach beyond vector
addition systems.  In this, we take our inspiration
from~\citep{Leroux:2015:reachability-vas}, which showed that KLM
decompositions could be recast as computing the downward closure of
the set of runs between source and target configuration with respect
to an embedding relation between runs.  Crucially, this embedding
relation is a well-quasi-ordering, and enjoys an \emph{amalgamation}
property---a notion from model theory, where embeddings of a structure
$A$ into structures~$B$ and~$C$ can be combined into embeddings into a
superstructure~$D$ of $B$ and $C$.  
As the runs of PVASS can be
equipped with an embedding relation enjoying the same
properties~\cite{Leroux:2019:pvass-embedding}, this opens the way for
an abstract framework orthogonal to WSTS, that generalises from vector
addition systems, and where well-quasi-ordered run embeddings and
amalgamation play a role akin to well-quasi-ordered configurations and
compatibility.

\subsection{Contributions} 
We introduce in \cref{sec:amalgamation-systems} a general notion of
\emph{(concatenative) amalgamation systems} that consist of a set of
\emph{runs} equipped with a wqo and such that any two runs with a common subrun
can be \emph{amalgamated}.  We show that our notion of amalgamation systems both
(i)~permits extremely simple decidability arguments for the unboundedness
problems mentioned above, but also (ii)~applies to a wide variety of
infinite-state systems.  In particular, while run amalgamation has been used
for concrete types of systems to prove structural
properties~\cite{Leroux:2019:pvass-embedding,DBLP:conf/fsttcs/FinkelKMMZ23} or
in specialised subroutines of complex
procedures~\cite{DBLP:conf/stacs/ClementeCLP17,DBLP:conf/lics/CzerwinskiZ20},
we identify amalgamation as a powerful algorithmic tool that is often
sufficient on its own for solving prominent problems.  Let us elaborate on (i)
and (ii).

\subsubsection*{Algorithmic Properties. } Regarding the computational
properties of amalgamation systems, we show that under mild
effectiveness assumptions, for concatenative amalgamation systems with
decidable emptiness, downward closures are computable, priority
downward closures are computable, whether the accepted language is a
bounded language is decidable, and all languages over one letter are
effectively regular.  More specifically, we show
in \cref{sec:properties} that if we assume that our class of systems
is effective and closed under rational
transductions~\citep{berstel79}, then all these effectiveness results
hold as soon as emptiness is decidable.

\begin{restatable}{maintheorem}{mainAmalgamation}\label{main-amalgamation}
For every language class that supports concatenative amalgamation and
is effectively closed under rational transductions, the following are
equivalent:
\begin{enumerate}
\item\label{sup-decidable} The simultaneous unboundedness problem is decidable.
\item\label{dc-computable} Downward closures are computable.
\item\label{ptlsep-decidable} Separability by piecewise testable languages is decidable.
\item\label{boundedness-decidable} Language boundedness is decidable.
\item\label{unary-effectively-regular} Unary languages are effectively regular.
\item\label{pdc-computable} Priority downward closures are computable.
\item\label{emptiness-decidable} Emptiness is decidable.
\end{enumerate}
\end{restatable}
Here, by \emph{supports concatenative amalgamation}, we mean that every
language in the class is recognised by some (concatenative) amalgamation system
(see \cref{sec:amalgamation-systems} for the full definition).

problems (\labelcref{sup-decidable})--(\labelcref{pdc-computable}) are
usually considered much more difficult for infinite-state systems than
emptiness~\eqref{emptiness-decidable}.  For instance, emptiness is
decidable for lossy channel systems and lossy counter
machines~\cite{cece96,mayr03}, but e.g., downward closures are not
effective~\cite{cece96} and language boundedness is
undecidable~\cite{DBLP:journals/tcs/ChambartFS16}---incidentally, these
examples show that amalgamation systems are incomparable with~WSTS.
Moreover, when applied to examples of amalgamation
systems, \cref{main-amalgamation} often yields new or drastically
simpler alternative proofs of decidability (see below for
consequences).

\subsubsection*{Amalgamation Systems Everywhere! } Regarding examples of amalgamation systems, we investigate the class of
\emph{valence automata} over graph
monoids~\cite{DBLP:phd/dnb/Zetzsche16,DBLP:journals/iandc/Zetzsche21}.  These form
an abstract model of systems with a finite-state control and a storage
mechanism, which is usually infinite-state.  The storage
mechanism is specified by an undirected graph $\Gamma=(V,E)$ where
self-loops are allowed.  By choosing suitable graphs, one recovers
various concrete infinite-state models from the
literature.  Examples include Turing machines, VASS,
integer VASS, pushdown automata, and combinations, like
pushdown VASS (PVASS).  Valence automata have been studied over the
last
decade~\cite{DBLP:conf/stacs/Zetzsche15,DBLP:phd/dnb/Zetzsche16,DBLP:journals/iandc/Zetzsche21,DBLP:conf/rp/Zetzsche21,DBLP:conf/lics/GanardiMZ22},
and identifying the graphs~$\Gamma$ leading to a decidable emptiness
problem is a challenging open question.

If one rules out graphs that are known to result in
Turing-completeness, then the remaining storage mechanisms %
can be classified into three classes, dubbed $\SC^-$,
$\SC^\pm$, and $\SC^+$ in~\cite[p.~185]{DBLP:phd/dnb/Zetzsche16}, %
obtained by \emph{adding counters} and \emph{building
stacks}. Adding counters means that we take a storage mechanism
and combine it with additional counters: either
``blind'' $\Z$-counters (which can go below zero) or ``partially
blind'' $\N$-counters (which have
to stay non-negative, like in a VASS). Building stacks means that
we take a storage mechanism and define a new one that allows stacks,
where each entry is a configuration of the original storage
mechanism. One can operate on the top-most entry as specified by the
original mechanism; by a push operation, one can create a fresh
empty entry on top, and using a pop, one can remove an empty topmost
entry.  For $\SC^-$ and $\SC^\pm$, emptiness is
decidable~\cite{DBLP:journals/iandc/Zetzsche21}, while for $\SC^+$ it
is open.

We show in \cref{sec:systems} that for every graph $\Gamma$, valence
automata over $\Gamma$ are amalgamation systems, \emph{unless}
reachability is already known to be undecidable.  Note that it is
unavoidable to rule out certain graphs: for instance,
Turing machines cannot be amalgamation systems.  By showing
that valence automata over all the remaining graphs (namely
in~$\SC^+$) lead to amalgamation systems, we obtain a very general
characterisation of decidable unboundedness problems over the entire
class of valence automata.
\begin{restatable}{maintheorem}{mainValence}\label{main-valence}
For every graph $\Gamma$, the following are equivalent for valence automata over $\Gamma$:
\begin{enumerate}
\item The simultaneous unboundedness problem is decidable.
\item Downward closures are computable.
\item Separability by piecewise testable languages is decidable.
\item Language boundedness is decidable.
\item Unary languages are effectively regular.
\item Priority downward closures are computable.
\item Emptiness is decidable.
\end{enumerate}
\end{restatable}

\subsection{Consequences}
There are several examples of infinite-state systems where our
approach either yields new results or new (much simpler) proofs. Let
us mention some of them.

\subsubsection*{Vector Addition Systems.}
A first important insight is that full-fledged KLM decompositions are
not required in order to solve the unboundedness problems
of \cref{main-amalgamation}: our proofs show precisely that a
black-box access to an oracle for reachability along with simple
reasoning on amalgamated runs suffice.  This should be contrasted with
the rather more involved arguments used in the case of VASS to show
the computability of downward
closures~\cite{Habermehl:2010:downward-closure-vas}, decidability of
PTL-separability~\cite{DBLP:journals/dmtcs/CzerwinskiMRZZ17},
decidability of language
boundedness~\cite{Czerwinski:2018:unboundedness-problems-vas}, and
effectively regular unary
languages~\cite{DBLP:journals/acta/HauschildtJ94}.  Also, now that the
proofs are decorrelated from the KLM decomposition, one can use any
algorithm for VASS reachability
(like \citeauthor{DBLP:conf/birthday/Leroux12}'s simple
invariant-based algorithm~\cite{DBLP:conf/birthday/Leroux12}) to
derive the decidability of these problems.  Moreover, the
computability of priority downward closures for VASS languages is a
new result.

\subsubsection*{Models with Decidable Emptiness.}
Beyond VASS, there is a hierarchy of infinite-state systems, within
the framework of valence automata, where decidability of emptiness is
known, namely the classes $\SC^-$ and
$\SC^\pm$~\citep{DBLP:journals/iandc/Zetzsche21}.  The difference
between $\SC^-$ and $\SC^\pm$ is that when we apply ``building
stacks'' and ``adding counters,'' then in $\SC^-$, we can only ever
add $\Z$-counters, while in $\SC^\pm$, we can first add $\N$-counters,
then build stacks, but afterwards only add $\Z$-counters (in
alternation with building stacks).  In particular, by starting with
$\N$-counters and then building stacks, one obtains automata with a
stack, where each entry can contains $\N$-counters; these are
equivalent to the \emph{sequential recursive Petri nets} of
\citet{DBLP:journals/acta/HaddadP07,DBLP:conf/time/HaddadP01} and generalise
both pushdown automata and VASS.

These classes differ in what was already known about decidability.
For $\SC^-$, we knew that emptiness was decidable and that downward
closures were
computable~\cite{DBLP:conf/stacs/Zetzsche15,DBLP:conf/icalp/Zetzsche15},
and effective regularity of unary languages was known for $\SC^-$,
because they have semilinear Parikh
images (this is not the case for
$\SC^\pm$)~\cite{DBLP:conf/mfcs/BuckheisterZ13}.  The decidability of language boundedness was known for
the smaller $\SC^-$-subclass of pushdown automata with
reversal-bounded
counters~\cite{BaumannDAlessandroGanardiIbarraMcQuillenSchuetzeZetzsche2022a}
(which model recursive programs with numeric data
types~\cite{DBLP:conf/cav/HagueL11}).  For $\SC^\pm$, we knew that
emptiness was decidable~\cite{DBLP:journals/iandc/Zetzsche21}, and the
decidability of the SUP could be derived from that proof.  \cref{main-valence} implies 
decidability of all problems (\labelcref{sup-decidable})--(\labelcref{pdc-computable}) for the entire class $\SC^\pm$.

\subsubsection*{PVASS and their Restrictions. }
In $\SC^+$, one can arbitrarily alternate between adding $\N$-counters
and building stacks.  Since $\Z$-counters can always be simulated by
$\N$-counters, $\SC^+$ is more powerful than $\SC^\pm$, but the
decidability of emptiness is open, the simplest open example being
one-dimensional
PVASS~\cite[Prop.~3.6]{DBLP:journals/iandc/Zetzsche21}---also known as
the \emph{Finkel Problem}.  For all these models, our results imply
that decidable emptiness will immediately imply the other properties
of \cref{main-amalgamation}.

Furthermore, since PVASS are amalgamation systems, so are all systems
that have a (language-)equivalent PVASS.  Two examples where emptiness
is known to be decidable come to mind: VASS with \emph{nested zero
tests} (\VASSnz)~\cite{DBLP:conf/fsttcs/AtigG11}, and PVASS where the
stack behaviour is \emph{oscillation
bounded}~\cite{DBLP:journals/corr/GantyV16}.
\begin{itemize}
\item A \VASSnz\ is a VASS that has for each $i$ an operation that tests all counters
$1,\ldots,i$ for zero at the same time.  Reachability is decidable for
\VASSnz~\cite{DBLP:journals/entcs/Reinhardt08,bonnet2013thesis} and
the clover computation of \citet[Thm.~16]{bonnet2013thesis} can be
used to show decidability of the SUP (and thus downward closure
computability and PTL separability);
		the remaining (\labelcref{boundedness-decidable})--(\labelcref{pdc-computable})
in \cref{main-amalgamation} are new results.
\item PVASS with an oscillation
bounded behaviour~\cite{DBLP:journals/corr/GantyV16} are equivalent to
PVASS where the stack behaviour is specified by a finite-index
context-free language, and thus emptiness is
decidable~\cite{DBLP:conf/fsttcs/AtigG11}---while the latter
decidability result relies on \VASSnz, these PVASS appear to be more
expressive in terms of accepted languages.  Thus, all the algorithmic
properties of \cref{main-amalgamation} apply to these as well.
\end{itemize}

\medskip

The paper is structured as follows.  After some preliminaries
in \cref{sec:prelims}, we define the notion of an amalgamation system
in \cref{sec:amalgamation-systems}.  To illustrate the notion, we also
present a few example systems.  In \cref{sec:properties}, we present
simple one-size-fits-all algorithms for the unboundedness properties
in \cref{main-amalgamation} for general amalgamation systems.
Finally, in \cref{sec:systems}, we show that valence automata with
graphs in $\SC^+$ are amalgamation systems and prove \cref{main-valence}.

\section{Well-Quasi-Orders}\label{sec:prelims}
We recall in this section the basic definitions for
well-quasi-orders~\cite{Higman:1952:wqo,kruskal72,SS2012} and
introduce some of the notations used in the paper.

\subsubsection*{Quasi-orders}\label{sub:wqo}
A \emph{quasi-order} (qo) is a pair $(X,{\leq})$ where~$X$ is a set
and~${\leq}\subseteq X\times X$ is a transitive reflexive binary
relation.  We write $x<x'$ if $x\leq x'$ but $x'\not\leq x$%
.  If $Y\subseteq X$, then it defines
an \emph{induced} qo when using the quasi-ordering ${\leq}\cap Y\times
Y$.

\subsubsection*{Well-quasi-orders}A (finite or infinite) sequence $x_0,x_1,\dots$ of elements from~$X$
is \emph{good} if there exist $i<j$ such that $x_i\leq x_j$; the
sequence is otherwise called \emph{bad}.  A qo $(X,{\leq})$ is
a \emph{well-quasi-order} (wqo) if bad sequences are finite.
Well-quasi-orders also enjoy the \emph{finite basis property}: every subset
$Y\subseteq X$ has a finite subset $B\subseteq Y$ such that for every
$y\in Y$ there is $x\in B$ with $x\leq y$.

For instance, if $S$ is a finite set, then $(S,{=})$ is a wqo by the
Pigeonhole Principle.  For another example, $(\Nat,{\leq})$ with the
usual ordering is a wqo because bad sequences are strictly descending
and the ordering is well-founded.  By definition, a qo induced inside
a wqo is also a wqo.

\subsubsection*{Vectors}\label{sub:vec}
By Dickson's Lemma, if $(X,{\leq_X})$ and $(Y,{\leq_Y})$ are two wqos,
then so is their Cartesian product $X\times Y$ with the product (i.e.,
componentwise) ordering defined by $(x,y)\leq(x',y')$ if $x\leq_X x'$
and $y\leq_Y y'$.  In particular, vectors $\vec u\in\N^d$ are
well-quasi-ordered by the product ordering. 

\subsubsection*{Words}\label{sub:words} Let $X$ be a set; we write
$X^\ast$ for the set of finite sequences (or \emph{words}) with
letters taken from~$X$.  We write $\varepsilon$ for the
empty word, and define $X_\varepsilon\eqdef X\cup\{\varepsilon\}$
when we want to treat it as a letter.

If $(X,{\leq})$ is a wqo, then by Higman's Lemma, so is
$(X^\ast,{\leq_\ast})$ where $w\leq_\ast w'$ if there exists a
(scattered word) \emph{embedding}, i.e., a strictly monotone map
$f\colon [1,|w|]\to[1,|w'|]$ such that $w(i)\leq w'(f(i))$ for all
$i\in[1,|w|]$.  For instance, over the wqo $(\{a,b,c,r\},{=})$,
$acab\leq_\ast \underline{a}bra\underline{c}ad\underline{a}\underline{b}ra$,
where we have underlined the positions in the image of the embedding.

\subsubsection*{Trees}\label{sub:trees}
Let $X$ be a set.  The set of (finite, ordered) \emph{$X$-labelled
trees} $\Trees{X}$ is the smallest set such that, if $x\in X$,
$n\in\N$, and $t_1,\dots,t_n\in\Trees{X}$ then the tree with root
labelled by~$x$ and with immediate subtrees $t_1, \ldots, t_n$,
denoted by $\tree{x}{t_1, \ldots, t_n}$, is in~$\Trees X$.
If $t\in \Trees X$ and $p \in \Nat^*$, the \emph{subtree of~$t$
at~$p$}, written $\subtree t{p}$, if it exists, is defined inductively by
$\subtree t\varepsilon\eqdef t$ and $\subtree{\tree x{t_1, \ldots,
t_n}}{(i \cdot p')}\eqdef \subtree{t_i}{p'}$ when $i\in[1,n]$. %

If $(X,{\leq})$ is a wqo, then $(\Trees X,{\leq_T})$ is also a wqo by
Kruskal's Tree Theorem, where $\leq_T$ denotes the \emph{homeomorphic
tree embedding} relation: $\tree x{t_1,\ldots,t_n}\leq_T t$ if there
exists a subtree $\subtree tp=\tree{x'}{t'_1,\ldots,t'_m}$ of~$t$ for
some~$p$%
, such that (i)~$x\leq x'$ and (ii)~there
exist $1\leq j_1<\cdots<j_n\leq m$ such that $t_i\leq_T t'_{j_i}$ for
all $i\in[1,n]$---this second condition corresponds to finding a word
embedding between $t_1\cdots t_n$ and $t'_1\cdots t'_m$ with respect
to~$\leq_T$.%

\section{Amalgamation Systems}\label{sec:amalgamation-systems}

Broadly speaking, an amalgamation system consists of an infinite set
of runs that are ordered by an embedding relation.  Moreover, under
certain circumstances, we can combine multiple runs into a new one.
We define amalgamation systems in the
forthcoming \cref{sub:amalgamation}, before illustrating the concept
with some concrete examples
in \crefrange{par:example-nfa}{par:example-cfgs}---some of the proofs
details that they are indeed amalgamation systems are deferred to
in \cref{sec:systems}.  For these basic examples, \cref{tbl:amal}
presents the known results pertaining to \cref{main-amalgamation}.

\begin{table}
\caption{\label{tbl:amal}Known effectiveness results for regular
(\cref{par:example-nfa}), VASS (\cref{par:example-vass}), and
context-free (\cref{par:example-cfgs}) languages.}
\begin{tabular}{lrrr}
  \toprule
  & \textsf{Reg}  & \textsf{VASS} & \textsf{CFL}\\
  \midrule
  \eqref{sup-decidable} SUP                          &\cite{DBLP:conf/icalp/Zetzsche15} & \cite{DBLP:conf/icalp/Zetzsche15} & \cite{DBLP:conf/icalp/Zetzsche15}\\
  \eqref{dc-computable} $\dc{L}$                     & folklore & \cite{Habermehl:2010:downward-closure-vas} & \cite{vanleeuwen78,Courcelle:1991:downward-closures-cfgs}    \\
  \eqref{ptlsep-decidable} PTL sep.                  & \cite{place13}  & \cite{DBLP:journals/dmtcs/CzerwinskiMRZZ17}  & \cite{DBLP:journals/dmtcs/CzerwinskiMRZZ17}\\
  \eqref{boundedness-decidable} boundedness          & \citep{GinsburgSpanier} & \cite{Czerwinski:2018:unboundedness-problems-vas} & \citep{GinsburgSpanier}     \\
  \eqref{unary-effectively-regular} unary eff.\ reg. & \citep{parikh64} & \cite{DBLP:journals/acta/HauschildtJ94} & \citep{parikh64}\\
  \eqref{pdc-computable}~$L{\downarrow_{\mathsf{P}}}$  & \cite{anand_et_al_CONCUR} & Thm.~\ref{main-amalgamation} & \cite{anand_et_al_CONCUR}\\
  \eqref{emptiness-decidable} emptiness & folklore   & \citep{mayr81,kosaraju82,DBLP:conf/birthday/Leroux12} & folklore \\
  \bottomrule
\end{tabular}
\end{table}

\subsection{Concatenative Amalgamation Systems}\label{sub:amalgamation}

A \emph{(concatenative) amalgamation system} is a tuple
$S=(\Sigma,R,E,\mathsf{can})$, where $\Sigma$ is a finite alphabet
and $R$ is a (usually infinite) set
of \emph{runs}%
. The set $E$ describes how runs from $R$ embed into each other. Each run $\rho\in R$ has an
associated \emph{canonical decomposition} $\can{\rho}=u_1\cdots
u_n\in\Siepp^\ast$ of some length $\canonlen{\rho} = n$, with
every~$u_i\in\Siep$.  The corresponding \emph{accepted word}
$\yield\rho\in\Sigma^\ast$ is obtained by
concatenating the~$u_i$'s; the \emph{language} accepted by the system
is then $L(S)\eqdef\bigcup_{\rho\in R}\yield\rho$.  A
language class \emph{supports amalgamation} if, for every language $L$
in the class, there exists an amalgamation system $S$ such that $L(S)=L$.

\subsubsection{Admissible Embeddings and Gaps. }
Furthermore, for any two runs $\rho,\sigma$, $E(\rho,\sigma)$ is a set
of \emph{admissible embeddings} between their canonical
decompositions.  Here, an embedding over the alphabet~$(\Siep,{=})$ is
defined as in \cref{sub:words}: if $\can{\rho}=u_1\cdots u_n$ and
$\can{\sigma}=v_1\cdots v_m$ are the canonical decompositions of the
two runs, then an embedding of $\rho$ in $\sigma$ is a strictly
monotone map $f\colon[1,n]\to [1,m]$ with $v_{f(i)}=u_i$ for every
$i\in[1,n]$.  For each embedding $f\in E(\rho,\sigma)$ and
$i\in[0,n]$, we define the \emph{gap word} $\gap{i}{f}\in\Sigma^*$,
such that $v_1\cdots v_m=\gap{0}{f}u_1\gap{1}{f}\cdots u_n\gap{n}{f}$.
Formally,%
\begin{align*}
\gap{0}{f}&\eqdef v_1\cdots v_{f(1)-1}\;, \\
\gap{i}{f}&\eqdef v_{f(i)+1}\cdots v_{f(i+1)-1} & &\text{for all $i\in[1,n-1]$},\\
\gap{n}{f}&\eqdef v_{f(n)+1}\cdots v_m\;. 
\end{align*}
If the set $E(\rho,\sigma)$ is non-empty, we write
$\rho\runembed\sigma$; if we wish to refer to a specific $f \in
E(\rho, \sigma)$, we write $\rho \embedwith{f}\sigma$.  Note that
$\rho\runembed\sigma$ implies $\can\rho\leq_*\can\sigma$ but the
converse might not hold: $E(\sigma,\rho)$ need not contain all the
possible embeddings between the canonical decompositions of~$\rho$
and~$\sigma$.

\subsubsection{Conditions. }\label{def:conds} Finally, we require the following:
\begin{description}
\item[composition\label{composition}] if $f\in E(\rho,\sigma)$ and $g\in E(\sigma,\tau)$,
then $f\circ g\in E(\rho,\tau)$,%
\item[wqo\label{wqo}] $(R,\runembed)$ is a well-quasi-order, and
\item[(concatenative) amalgamation\label{amalgamation}] if $\rho_0\embedwith{f}\rho_1$ and $\rho_0\embedwith{g}\rho_2$ with canonical decomposition $\can{\rho_0}=w_1\cdots w_n$, then for every choice of $i\in[0,n]$, there exists a run $\rho_3\in R$ such that $\rho_1 \embedwith{f'} \rho_3$, $\rho_2 \embedwith{g'} \rho_3$ with $f' \circ f = g' \circ g$ (we write $h$ for this composition) and %
\begin{align*}
\gap{j}{h} &\in \{\gap{j}{f}\gap{j}{g},~\gap{j}{g}\gap{j}{f}\} &  &\text{for every $j\in[0,n]$, and} \\
\gap{i}{h} &= \gap{i}{f}\gap{i}{g} & & \text{for the chosen $i$.}
\end{align*}
\end{description}
Thus the embedding $h$ of $\rho_0$ into $\rho_3$ has the property that
each gap word is the concatenation of the two gap words from the
embeddings into $\rho_1$ and $\rho_2$, in some order.  Moreover, in
the particular gap~$i$, we know that the gap word from $\rho_1$ comes
first, and then the gap word from $\rho_2$.  This tell us that, given
a gap $i\in[1,n]$, we can choose a run $\rho_3$ such that the
concatenation in gap~$i$ happens in an order of our choice.

\subsubsection{Gap Languages. }Given an amalgamation system and some
run $\rho\in R$ %
we also define
the \emph{gap language} for a gap~$i\in[0,\canonlen{\rho}]$
\begin{equation*}L_{\rho,i}\eqdef\{\gap{i}{f} \mid \exists \sigma\in R\colon \rho\embedwith f \sigma \}\;.\end{equation*}
In other words, $L_{\rho,i}$ is the set of all words that can be
inserted in the $i$-th gap of $\rho$'s canonical decomposition when
$\rho$ embeds into some larger run.
A language $L\subseteq\Sigma^*$ is a \emph{subsemigroup} if for any two words $u,v\in L$, we have $uv\in L$.
The following is a direct consequence of concatenative amalgamation.
\begin{observation}\label{gap-semigroup}
For each run $\rho$ and $i$, the language $L_{\rho,i}$ is a subsemigroup.
\end{observation}

\subsubsection{Effectiveness}\label{ssub:eff}
In order to derive the algorithmic results of \cref{sec:properties},
we need to make some effectiveness assumptions.  These are always
clear from our constructions and will be used tacitly in the
algorithms.  Specifically, we assume that each run in~$R$ has some
finite representation such that
\begin{enumerate}[(i)]
\item the set $R$ is recursively enumerable,
\item the function $\can{\cdot}$ is computable, and
\item for any two runs $\rho,\sigma$ we can compute the set
  $E(\rho,\sigma)$ of admissible embeddings (and hence decide whether
  $\rho\runembed\sigma$ and compute the various gap words).
\end{enumerate}

\subsection{Example: Regular Languages}\label{par:example-nfa}

Let us start with a very simple example, namely a non-deterministic
finite automaton (NFA).  Suppose we have an NFA
$\cA=(Q,\Sigma,\Delta,I,F)$ with finite state set~$Q$, input
alphabet~$\Sigma$, transition set $\Delta\subseteq Q\times\Siepp\times
Q$, initial states $I\subseteq Q$, and final states $F\subseteq Q$.
Then we can view it as an amalgamation system by taking
$R\subseteq \Delta^*$ to be the set of all finite sequences
$(q_0,a_1,q_1)(q_1,a_2,q_2)\cdots (q_{n-1},a_n,q_n) \in \Delta^\ast$
starting in some $q_0\in I$ and ending in some $q_n\in F$, and where
the end state of each individual element of the sequence is the same
as first state of the next element.  Such a transition sequence can
therefore actually be executed by the automaton.  %
The canonical decomposition of such a run is simply $a_1\cdots a_n$,
and $L(\cA)=\bigcup_{\rho \in R}\yield{\rho}$ as desired.

For two runs $\rho=r_1\cdots r_n\in R$ and $\sigma=s_1\cdots s_m\in
R$, we then let $E(\rho,\sigma)$ be the set of strictly monotone maps
$f\colon [1,n]\to[1,m]$ such that $s_{f(i)}=r_i$ for every
$i\in[1,n]$.  Thus $(R,{\runembed})$ is the qo induced by
$R\subseteq\Delta^\ast$ inside the word embedding
qo~$(\Delta^\ast,{\leq_\ast})$ over the alphabet~$(\Delta,{=})$, and
the \nameref{composition} and \nameref{wqo} conditions follow.

Concatenative amalgamation also holds.  Let $\rho_0,\rho_1,\rho_2$ be
runs with $\rho_0 \embedwith f \rho_1$ and $\rho_0 \embedwith
g \rho_2$, where the canonical decomposition of $\can{\rho_0}$ has
$n$ transitions. Because the individual transitions of $\rho_0$ must be
compatible with each other, we know that the $i$-th transition ends in
the same state $q_i$ that the $(i+1)$-st transition begins in.
However, that means that the gap words $\gap{i}{f}$ and $\gap{i}{g}$
must be read on loops from $q_i$ to $q_i$.  Therefore, we can
concatenate them in any order we wish to obtain a new run $\rho_3$ as
required.

\begin{figure}[bth]
	\begin{center}
		\begin{tikzpicture}[node distance=2em]
			\node [state, initial by arrow, initial text={}] at (0,0) (q0) {$\hlA{q_0}$};
			\node [state] at (2,0) (q1) {$\hlB{q_1}$};
			\node [state] at (1,-1) (q2) {$\hlC{q_2}$};
			\node [state] at (3,-1) (q3) {$\hlC{q_3}$};
			\node [state,accepting] at (4,0) (qf) {$\hlA{q_f}$};
			
			\draw [-{latex}] (q0) edge node {$a$} (q1);
			\draw [-{latex}] (q1) edge [loop above] node {$b$} ();
			\draw [-{latex}] (q1) edge node {$a$} (qf);
			\draw [-{latex}] (q0) edge node {$a$} (q2);
			\draw [-{latex}] (q2) edge node {$b$} (q3);
			\draw [-{latex}] (q3) edge node {$a$} (qf);
		\end{tikzpicture}
	\end{center}
	\caption{A finite automaton for \cref{ex:nfa}.}
	\label{fig:example-nfa}
\end{figure}
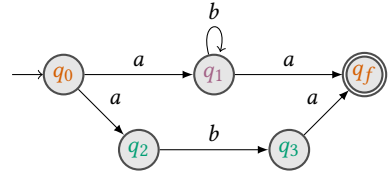
\begin{example}\label{ex:nfa}
As a concrete example, consider the automaton
in \cref{fig:example-nfa}.  A possible run is $\rho_0\eqdef (\hlA{q_0},
a, \hlB{q_1})(\hlB{q_1}, a, \hlA{q_f})$. Another run is $\rho\eqdef
(\hlA{q_0}, a, \hlC{q_2})(\hlC{q_2}, b, \hlC{q_3})(\hlC{q_3},
a, \hlA{q_f})$.  Although $\can{\rho_0}=aa\leq_\ast
aba=\can{\rho}$, there is no a run embedding between the two: the 
transition $(\hlA{q_0}, a, \hlB{q_1})$ of $\rho_0$ cannot be mapped to
a corresponding transition in $\rho$.

If we consider the run $\rho_1\eqdef (\hlA{q_0},
a, \hlB{q_1})(\hlB{q_1}, b, \hlB{q_1})(\hlB{q_1}, a, \hlA{q_f})$, then
we can embed $\rho_0$ into $\rho_1$ via the map $\{ 1 \mapsto 1,
2 \mapsto 3 \}$. Indeed, we get a decomposition of the yield of
$\rho_1$ into $\gap{0}{f} a \gap{1}{f} a \gap{2}{f}$ with $\gap0f
= \gap2f = \varepsilon$ and $\gap1f = b$.  And clearly we can take the
$b$-labelled loop several times, yielding for instance the run
$\rho_2\eqdef (\hlA{q_0}, a, \hlB{q_1})(\hlB{q_1},
b, \hlB{q_1})(\hlB{q_1}, b, \hlB{q_1})(\hlB{q_1}, a, \hlA{q_f})$.
\end{example}

\subsection{Example: Vector Addition Systems}\label{par:example-vass}

A $d$-dimensional (labelled) \emph{vector addition system with states
(VASS)} is a finite automaton $\mathcal V=(Q,\Sigma,\Delta,I,F)$ with
transition labels in $\Siep\times\Z^d$: $\Delta$ is now a finite
subset of~$Q\times\Siepp\times\Z^d\times Q$.  We revisit here the
results of~\citep{Leroux:2015:reachability-vas} to show that VASS are
amalgamation systems.

\subsubsection{Configurations and Semantics. }Let
$\mathit{Confs}\eqdef Q \times \Nat^d$.  A \emph{configuration} is a
pair $q(\vec u) \in \mathit{Confs}$ of a state and the values of
the $d$~counters.  Configurations are ordered through the product
ordering: $q(\vec u)\leq q'(\vec u')$ if $q=q'$ and $\vec u\leq\vec
u'$.  If $c = q(\vec u)$ is a configuration and $\vec d\in\Nat^d$, we
write $c + \vec d$ for the configuration $q(\vec u + \vec d)$; note
that $c\leq c'$ if and only if there exists $\vec d\in\Nat^d$ such
that $c'=c+\vec d$.

The counters can be incremented and decremented along
transitions, but not tested for zero: for configurations $c,
c'\in \mathit{Confs}$ and a transition $t\in\Delta$, we write
$c \autstep[t] c'$ if $t = (q, x, \vec v, q')$, $c = q(\vec c)$, and
$c' = q'(\vec c + \vec v)$, and extend this notation to sequences of
transitions: $c_0\xrightarrow{t_1\cdots t_n}c_n$ if there exist
$c_1,\dots,c_{n-1}$ such that $c_{i-1}\autstep[t_i]c_i$ for all
$i\in[1,n]$.  VASS transitions are \emph{monotonic}, in that if
$c\autstep[t]c'$ and $\vec d\in\Nat^d$, then $c+\vec
d\autstep[t]c'+\vec d$.

\subsubsection{Runs and Admissible Embeddings.}
A \emph{run}
is a sequence $\rho=(c_0, t_1, c_1)\allowbreak\cdots(c_{n-1},t_n,c_n)$ such that
$c_0=q_0(\vec 0)$ for some $q_0\in I$, $c_n=q_f(\vec 0)$ for some
$q_f\in F$, and $c_{i-1} \autstep[t_{i}] c_{i}$ at each step.  Let
$t_i$ be $(q_{i-1},a_i,\vec v_i,q_i)$ for each~$i$ in this run; the
associated canonical decomposition is $\can\rho\eqdef a_1\cdots a_n$
and we recover the usual notion of a VASS (reachability) language:
$L(\mathcal V)=\bigcup_{\rho\in R}\yield\rho$.%

Let $(R,{\runembed})$ be the qo induced by $R\subseteq
(\mathit{Confs}\times\Delta\times\mathit{Confs})^\ast$ inside the word
embedding qo
$((\mathit{Confs}\times\Delta\times\mathit{Confs})^\ast,{\leq_\ast})$
over the product alphabet
$(\mathit{Confs},{\leq})\times(\Delta,{=})\times(\mathit{Confs},{\leq})$.
Thus for two runs $\rho_0= (c_0,t_1,c_1)\cdots(c_{n-1},t_n,c_n)$ and
$\rho_1 = (c'_0,t'_1,c'_1)\cdots(c'_{m-1},t'_m,c'_m)$, we have
$\rho_0\runembed \rho_1$ if there exists a strictly monotone map $f
: [1,n] \to [1,m]$ such that, for all $i\in[1,n]$, $c_{i-1}\leq
c'_{f(i)-1}$, $t_i=t'_{f(i)}$, and $c_i\leq c'_{f(i)}$.  By
definition, the \nameref{composition} and \nameref{wqo} conditions follow.  Also,
because~$\rho_1$ is a run, there exist $\vec d_1,\dots,\vec
d_n\in\Nat^d$ such that for all $i\in[1,n]$, 
\begin{equation}
c'_{f(i)-1}=c_{i-1}+\vec d_i\xrightarrow{t_{i}}c_i+\vec
d_i=c'_{f(i)}\;.\label{vas-step-i}
\end{equation}  Furthermore, letting $\vec d_0\eqdef\vec 0$, $\vec
d_{n+1}\eqdef\vec 0$, $f(0)\eqdef 0$, and $f(n+1)\eqdef m+1$, then
each $\gap{i}{f}$ for $i\in[0,n]$ is the yield of the transition sequence
\begin{equation}
c_i+\vec
d_i=c'_{f(i)}\xrightarrow{t'_{f(i)+1}\cdots t'_{f(i+1)-1}}c'_{f(i+1)-1}=c_i+\vec
d_{i+1}\label{vas-gap-i}\;.
\end{equation}

\subsubsection{Concatenative Amalgamation. }\label{ssub:vass-amalg} Assume
$\rho_0\embedwith{f}\rho_1$ as above, and
$\rho_0\embedwith{g}\rho_2=(c''_0,t''_1,c''_1)\cdots(c''_{p-1},t''_p,c''_p)$.
Then there exist $\vec d_1,\dots,\vec
d_n$ (resp.\ $\vec d'_1,\dots,\vec d'_n$) such that we can
decompose~$\rho_1$ (resp.\ ~$\rho_2$) as
in \eqref{vas-step-i} and \eqref{vas-gap-i}.  We can amalgamate to a
third run~$\rho_3$ by observing that, by monotonicity, $c_i+\vec
d_i+\vec d'_i=c'_{f(i)}+\vec d'_i\xrightarrow{t'_{f(i)+1}\cdots t'_{f(i+1)-1}}c'_{f(i+1)-1}+\vec d'_i=c_i+\vec
d_{i+1}+\vec d'_i=c''_{g(i)}+\vec d_{i+1}\xrightarrow{t''_{g(i)+1}\cdots t''_{g(i+1)-1}}c''_{g(i+1)-1}+\vec d_{i+1}=c_i+\vec
d_{i+1}+\vec d'_{i+1}$ yields $\gap{i}{f}\gap{i}{g}$ for all $i\in[0,n]$.

\begin{figure}[htbp]
	\begin{center}
		\begin{tikzpicture}
			\node [state,initial, initial text={}] at (0,0) (q0) {$\hlA{q_0}$};
			\node [state] at (2,0) (q1) {$\hlB{q_1}$};
			\node [state,accepting] at (4,0) (qf) {$\hlA{q_f}$};
			
			\draw [-{latex}] (q0) edge node {$t_3: a / +\tikz{\draw [fill=highlight] circle(2.5pt);}$} (q1);
			\draw [-{latex}] (q1) edge node {$t_5: a / -\tikz{\draw [fill=highlight] circle(2.5pt);}$} (qf);
			
			\draw [-{latex}] (q0) edge [loop below] node {$t_2:b / +\tikz{\draw [fill=highlight-2] rectangle(4pt,4pt);}$} ();
			\draw [-{latex}] (q0) edge [loop above] node {$t_1:c / +\tikz{\draw [fill=highlight-3,rotate=45] rectangle(3.5pt,3.5pt);}$} ();
			\draw [-{latex}] (q1) edge [loop below] node {$t_4: c / -\tikz{\draw [fill=highlight-3,rotate=45] rectangle(3.5pt,3.5pt);} $} ();
			\draw [-{latex}] (qf) edge [loop below] node {$t_6: b / -\tikz{\draw [fill=highlight-2] rectangle(4pt,4pt);}$} ();
		\end{tikzpicture}
	\end{center}
	\caption{A VASS for \cref{ex:vass}.}
	\label{fig:example-vass}
\end{figure}
\begin{example}\label{ex:vass}
As a concrete example, consider the VASS in \cref{fig:example-vass}. The three runs we will consider are 
{\footnotesize\begin{align*}
\rho_0&\eqdef(\hlA{q_0}(), t_3, \hlB{q_1}(\tikz{\draw[fill=highlight]circle(2pt);}))\:
(\hlB{q_1}(\tikz{\draw [fill=highlight]circle(2pt);}),t_5,\hlA{q_f}())\\
\rho_1&\eqdef(\hlA{q_0}(),t_2,\hlA{q_0}(\tikz{\draw [fill=highlight-2]rectangle(3.5pt,3.5pt);}))\:
(\hlA{q_0}(\tikz{\draw [fill=highlight-2]rectangle(3.5pt,3.5pt);}),t_3,\hlB{q_1}(\tikz{\draw [fill=highlight]circle(2pt);}\,\tikz{\draw [fill=highlight-2] rectangle(3.5pt,3.5pt);}))\:
(\hlB{q_1}(\tikz{\draw [fill=highlight] circle(2pt);}\,\tikz{\draw[fill=highlight-2]rectangle(3.5pt,3.5pt);}),t_5,\hlA{q_f}(\tikz{\draw[fill=highlight-2] rectangle(3.5pt,3.5pt);}))\:
(\hlA{q_f}(\tikz{\draw [fill=highlight-2]rectangle(3.5pt,3.5pt);}),t_6,\hlA{q_f}())\\
\rho_2&\eqdef(\hlA{q_0}(),t_1,\hlA{q_0}(\tikz{\draw [fill=highlight-3,rotate=45] rectangle(3pt,3pt);}))\:
(\hlA{q_0}(\tikz{\draw [fill=highlight-3,rotate=45] rectangle(3pt,3pt);}),t_3,\hlB{q_1}(\tikz{\draw [fill=highlight] circle(2pt);}\,\tikz{\draw [fill=highlight-3,rotate=45] rectangle(3pt,3pt);}))\:
(\hlB{q_1}(\tikz{\draw [fill=highlight] circle(2pt);}\,\tikz{\draw [fill=highlight-3,rotate=45] rectangle(3pt,3pt);}),t_4,\hlB{q_1}(\tikz{\draw [fill=highlight] circle(2pt);}))\:
(\hlB{q_1}(\tikz{\draw [fill=highlight] circle(2pt);}),t_5,\hlA{q_f}())
\end{align*}}

We have $\rho_0 \embedwith{f} \rho_1$ and $\gap{0}{f} = b, \gap{1}{f}
= \epsilon$, and $\gap{2}{f} = b$.  Similarly, we have $\rho_0 \embedwith{g} \rho_2$ and $\gap{0}{g} = c, \gap{1}{g} = c$, and $\gap{2}{g} = \epsilon$.
We can amalgamate to obtain {\footnotesize$\rho_3 =
(\hlA{q_0}(),t_2,\hlA{q_0}(\tikz{\draw [fill=highlight-2]rectangle(4pt,4pt);}))\:\allowbreak
(\hlA{q_0}(\tikz{\draw[fill=highlight-2]rectangle(3.5pt,3.5pt);}),t_1,\hlA{q_0}(\tikz{\draw[fill=highlight-2] rectangle(3.5pt,3.5pt);}\hspace{.3pt}\tikz{\draw[fill=highlight-3,rotate=45] rectangle(3pt,3pt);}))\allowbreak
(\hlA{q_0}(\tikz{\draw[fill=highlight-2]rectangle(3.5pt,3.5pt);}\hspace{.3pt}\tikz{\draw[fill=highlight-3,rotate=45]rectangle(3pt,3pt);}),t_3,\hlB{q_1}(\tikz{\draw[fill=highlight]circle(2pt);}\hspace{.3pt}\tikz{\draw[fill=highlight-2]rectangle(3.5pt,3.5pt);}\hspace{.3pt}\tikz{\draw[fill=highlight-3,rotate=45]rectangle(3pt,3pt);}))\:\allowbreak
(\hlB{q_1}(\tikz{\draw[fill=highlight]circle(2pt);}\hspace{.3pt}\tikz{\draw[fill=highlight-2]rectangle(3.5pt,3.5pt);}\hspace{.3pt}\tikz{\draw[fill=highlight-3,rotate=45]rectangle(3pt,3pt);}),t_4,\hlB{q_1}(\tikz{\draw[fill=highlight]circle(2pt);}\hspace{.3pt}\tikz{\draw[fill=highlight-2]rectangle(3.5pt,3.5pt);}))\:\allowbreak
(\hlB{q_1}(\tikz{\draw[fill=highlight]circle(2pt);}\hspace{.3pt}\tikz{\draw[fill=highlight-2]rectangle(3.5pt,3.5pt);}),t_5,\hlA{q_f}(\tikz{\draw[fill=highlight-2]rectangle(3.5pt,3.5pt);}))\:
(\hlA{q_f}(\tikz{\draw[fill=highlight-2]rectangle(3.5pt,3.5pt);}),t_6,\hlA{q_f}())$}.
One can check that $\rho_0 \embedwith{h} \rho_3$ and $\gap{i}{h}
= \gap{i}{f}\gap{i}{g}$ for all $i\in[0,2]$.
\end{example}

VASS can be construed as ``adding counters'' to finite automata;
see \cref{sec:counter-extension} for a general construction showing
that this can be achieved more generally in amalgamation systems.

\subsection{Example: Context-Free Languages}
\label{par:example-cfgs}
A \emph{context-free grammar} (CFG) is a tuple $\mathcal G=(N, \Sigma,
S, \Delta)$ with finite non-terminal alphabet $N$, finite terminal
alphabet $\Sigma$, initial non-terminal $S \in N$, and finite set of
productions $\Delta \subseteq N \times (N \cup \Sigma)^*$; each such
production is written $A\to\alpha$ with $A\in N$ and $\alpha\in
(N \cup \Sigma)^*$.

We mostly consider grammars from the perspective of their derivation
trees.  In a slight divergence from the usual definition (but
consistent with e.g., \citep[App.~A]{Leroux:2019:pvass-embedding}),
we label nodes in derivation trees not with symbols from
$N \cup \Siep$, but with productions from~$\Delta$.  Indeed, the usual
homeomorphic tree embedding (c.f.\ \cref{sub:trees}) over trees
labelled by $(N\cup\Siep,{=})$ is a well-quasi-ordering, but this
labelling is not sufficient for amalgamation and we shall rather rely
on the homeomorphic tree embedding over trees labelled by
$(\Delta,{=})$.

\begin{example}\label{ex:cfg}
Consider the grammar with $N\eqdef\{A, B\}$, $\Sigma\eqdef\{a, b\}$,
and $\Delta\eqdef \{A \to a, A \to B, B \to a, B \to b, B \to AB \}$
and the two trees in \cref{fig:homeomorphic-but-not-pumpable}.  The
left tree homeomorphically embeds into the right one, but no larger
trees can be derived from this ordering, making it unsuitable for
amalgamation.
\end{example}

Let us assume wlog.\ that the productions in~$\Delta$ are either
non-terminal productions of the form $A \to B_1\cdots B_k$ with $k>0$
and $B_1,\dots,B_k\in N$, or terminal productions of the form $A \to
a$ with $a \in \Siep$.  We call a tree in $\Trees\Delta$
\emph{$A$-rooted} if its root label is a production $A\to\alpha$
for some $\alpha$.  Using the notations from \cref{sub:trees},
a \emph{derivation tree} is either a leaf $\tree{(A \to a)}{}$
labelled by a terminal production $(A \to a)\in\Delta$, or a tree
$\tree{(A \to B_1\cdots B_k)}{t_1,\ldots,t_k}$ where $(A\to B_1\cdots
B_k)\in\Delta$ is a non-terminal production and for all $i\in[1,k]$,
$t_i$ is a $B_i$-rooted derivation tree.

\begin{figure}[thbp]
	\begin{subfigure}[t]{.43\columnwidth}
		\centering	
		\begin{tikzpicture}[node distance=1.2em,inner sep=2pt]
			\node [draw,circle] (a) {$A$};
			\node [below=of a] (a') {$a$};
			\draw [-] (a) -- (a');
			
			\node [draw,circle,right=of a] (b) {$A$};
			\node [draw,circle,below=of b] (b') {$B$};
			\node [below=of b'] (b'') {$a$};
			\draw [-] (b) -- (b') -- (b'');
			
			\draw [-{latex},dashed,color=highlight] (a) -- (b);
			\draw [-{latex},dashed,color=highlight] (a')
			-- (b'');
		\end{tikzpicture}
		\caption{The left tree embeds
			into the right tree, but amalgamation is not possible.}
		\label{fig:homeomorphic-but-not-pumpable}
	\end{subfigure}%
	\quad%
	\begin{subfigure}[t]{.53\columnwidth}
		\centering
		\begin{tikzpicture}[node distance=2em,inner sep=2pt]			
			\node [draw,rectangle%
                        ] (b'') {$A \to a$};
			\node
                        [draw=highlight-3,thick,dashed,rectangle,right=4em
                        of b''%
                                ] (c'') {$A \to B$};
			\node [draw=highlight-3,thick,dashed,rectangle,below=of c''] (c'''') {$B \to AB$};
			\node [draw,rectangle,below=of c''''] (c'5) {$A \to a$};
			\node [draw=highlight-3,thick,dashed,rectangle,right=.6em of c'5] (c'6) {$B \to b$};
			
			\draw [-] %
                          (c'') -- (c'''') -- (c'5);
			\draw [-] (c'''') -- (c'6);

			\draw [-{latex},dashed,color=highlight] (b'') -- (c'5);
		\end{tikzpicture}
		\caption{The embedding between the left and
			the right trees can be used to create
			successively larger trees. %
		}
		\label{fig:pumpable-cfg-embedding}
	\end{subfigure}\caption{Derivation tree embeddings
		for \cref{ex:cfg}.}
\end{figure}
Observe that, if two $A$-rooted derivation trees $t$ and~$t'$
homeomorphically embed into each other, i.e., $t\leq_T t'$, then $t$
has a production $A\to\alpha$ as root label, and there exists a
subtree $t'/p=\tree{(A\to
\alpha)}{t'_1,\dots,t'_k}$ for some $k$.  Then, putting a hole at position~$p$
in~$t'$ gives rise to an \emph{$A$-context}, i.e., a context where
plugging any $A$-rooted derivation tree will produce an $A$-rooted
derivation tree; this context can thus be iterated at will.
Furthermore, if $A\to\alpha$ is a non-terminal production, then
$\alpha=B_1\cdots B_k$, $t=\tree{(A\to B_1\cdots
B_k)}{t_1,\dots,t_k}$, and $t'/p=\tree{(A\to B_1\cdots
B_k)}{t'_1,\dots,t'_k}$, and inductively for each $i\in[1,k]$ $t_i$
and $t'_i$ are two $B_i$-rooted derivation trees with $t_i\leq_Tt'_i$,
allowing to repeat the same reasoning. 
\renewcommand{\thmcontinues}[1]{continued}
\begin{example}[continues=ex:cfg]
In \cref{fig:pumpable-cfg-embedding} the nodes in dashed
boxes in the right tree form an $A$-context corresponding to the
derivation $A \deriv B \deriv AB \deriv Ab$.  This $A$-context can be
iterated (and thus the derivation as well) to derive larger and larger
trees before plugging the node $A\to a$ from the image of the left tree.
\end{example}

Finally, the canonical decomposition of a derivation tree is defined
inductively by $\can{\tree{(A \to a)}{}}\eqdef \epsilon a\epsilon$ for
$a\in\Siep$ and
$\can{\tree{(A \to \alpha)}{t_1,\ldots,t_n}}\eqdef\epsilon\can{t_1}\cdots\can{t_n}\epsilon$
otherwise.  A \emph{run} is then an $S$-rooted derivation tree, and
this matches the usual definition of the language of a context-free
grammar: $L(\mathcal G)=\bigcup_{\rho\in R}\yield\rho$.  A comment is
in order for those explicit $\epsilon$'s.  This can be seen
as a transformation of the grammar into its associated parenthetical
grammar (followed by an erasure of the parentheses), so that the
extra~$\varepsilon$ reflect the tree structure.  In turn, this serves
to break up gap words that would otherwise span several levels of the
derivation trees, and forces them to reflect how the trees embed.  We
can then interleave these smaller gap words as may be required for the
concatenative amalgamation of trees.  We will generalise this whole
construction in \cref{sec:alg-closure}, where we consider algebraic
extensions of amalgamation systems.

\section{Algorithms for Amalgamation Systems}\label{sec:properties}

In this section, we prove that amalgamation systems have all the
algorithmic properties in \cref{main-amalgamation}.  We work with
amalgamation systems satisfying the implicit effectiveness
assumptions of \cref{ssub:eff}, and language classes that
are effectively closed under rational transductions (and thus under
morphisms and intersection with regular languages)---also known
as \emph{full trios}~\cite[see, e.g.,][]{berstel79}.

\subsection{The Simultaneous Unboundedness Problem}
\label{sec:sup}
We now prove the first main result of this work, about
the \emph{simultaneous unboundedness problem} (SUP) for formal
languages.
\begin{description}
	\item[Given] An alphabet $\{a_1,\ldots,a_n\}$ and a language
		$L\subseteq a_1^*\cdots a_n^*$.
	\item[Question] Is it true that for every $k\in\N$, there exist
		$x_1,\ldots,x_n\in\N$ such that $x_1,\ldots,x_n\ge k$ and
		$a_1^{x_1}\cdots a_n^{x_n}\in L$?
\end{description}
There has been some interest in this problem because
in~\cite[Thm.~1]{DBLP:conf/icalp/Zetzsche15}, it was shown that for any full
trio $\mathcal{C}$, downward closures are computable if and only if
the SUP is decidable for $\mathcal{C}$, thus under the hypotheses
of \cref{main-amalgamation}, \equ{sup-decidable}{dc-computable}.
Moreover, in~\cite[Thm.~2.6]{DBLP:journals/dmtcs/CzerwinskiMRZZ17}, it was shown
that for any full trio $\mathcal{C}$, separability by piecewise
testable languages is decidable if and only if the SUP is decidable,
thus \equ{sup-decidable}{ptlsep-decidable}. In fact, analogous results hold also for some orderings beyond the subword ordering~\cite{DBLP:conf/lics/Zetzsche18}. Given this motivation, the
SUP is known to be decidable for
VASS~\cite{Habermehl:2010:downward-closure-vas,Czerwinski:2018:unboundedness-problems-vas},
higher-order pushdown automata~\cite{DBLP:conf/popl/HagueKO16}, and
even higher-order recursion schemes~\cite{DBLP:conf/lics/ClementePSW16,DBLP:conf/fsttcs/Parys17,DBLP:conf/icalp/BarozziniPW22}.

\begin{proof}[Proof of \impl{emptiness-decidable}{sup-decidable}]%
Let us first define the \emph{Parikh image}
$\Parikh w$ of a word $w\in\{a_1,\ldots,a_n\}^\ast$ as the vector in
$\N^n$ where $\Parikh w(i)$ is the number of occurrences of $a_i$
inside~$w$. 
We also write $\Parikh\rho$ as shorthand for $\Parikh{\yield\rho}$. For two vectors $\vec u,\vec v\in\N^n$, we write $\vec
u\vstrictlt \vec v$ if, for all $i\in[1,n]$, $\vec u(i) < \vec v(i)$.

Our algorithm consists of two semi-decision procedures.  The first
enumerates $k\in\N$ and then checks whether there exists a word with at least $k$ repetitions of each letter. This can be decided in a full trio with an oracle for the
emptiness problem by checking whether $L \cap a_1^{\geq k} \cdots a_n^{\geq k}=\emptyset$.  The other one enumerates pairs of runs
$\rho,\sigma$ and checks whether (i)~$\rho\runembed\sigma$ and
(ii)~$\Parikh{\rho}\vstrictlt\Parikh{\sigma}$.
Clearly, if the first semi-decision procedure terminates, then our
system cannot be a positive instance of the SUP, because in any word
in~$L$, there are at most~$k$ occurrences of $a_i$. Moreover, if the
second semi-decision procedure finds~$\rho$ and~$\sigma$ as above,
then by amalgamation, we obtain runs $\sigma_1,\sigma_2,\ldots$ such
that $\Parikh{\sigma_k}(i)\ge k$, meaning we have a positive
instance.

It remains to argue that one of the two procedures will terminate.
This is trivial if our system is a negative instance.  Conversely, if
our system is a positive instance, then there is an infinite sequence
of runs $\rho_1,\rho_2,\ldots$ such that $\Parikh{\rho_k}(i)\ge k$
for every $i\in[1,n]$.  This sequence has an infinite subsequence
$\rho'_1,\rho'_2,\ldots$ such that
$\Parikh{\rho'_{k+1}}(i)>\Parikh{\rho'_k}(i)$ for every
$i\in[1,n]$.  Since $(R,{\runembed})$ is a \nameref{wqo}, we can find
$j<\ell$ such that %
$\rho'_j\runembed\rho'_\ell$.  By definition of the $\rho'_k$, this pair satisfies
$\Parikh{\rho'_j}\vstrictlt\Parikh{\rho'_\ell}$.
\end{proof}

\begin{proof}[Proof of \impl{sup-decidable}{emptiness-decidable}]%
Assuming decidable SUP, for any 
$L\subseteq\Sigma^*$, take the rational transduction
$T\eqdef\Sigma^*\times\{a_1\}^*$ and observe that $TL\subseteq a_1^*$ is a positive
instance of the SUP if and only if $L\ne\emptyset$.
\end{proof}%

\subsection{Language Boundedness}
\newcommand{\prefixes}{\mathsf{prefixes}}
Our next main result is about deciding language boundedness.
\begin{description}
\item[Given] A language $L\subseteq\Sigma^*$.
\item[Question] Does there exist $k\in\N$ and words
$w_1,\ldots,w_k\in\Sigma^*$ such that $L\subseteq w_1^*\cdots w_k^*$?
\end{description}

The decidability of language boundedness was known for pushdown
automata since the 1960's \cite[e.g.,][]{GinsburgSpanier}---and is
even in polynomial time~\cite{DBLP:journals/ijfcs/GawrychowskiKRS10}.
The question was open for many years in the case of reversal-bounded
counter machines
(RBCM)~\cite{DBLP:journals/ijfcs/CadilhacFM12,DBLP:conf/dlt/EremondiIM15}
before it was settled for VASS
in~\cite{Czerwinski:2018:unboundedness-problems-vas}. For RBCM with a
pushdown, it was settled even more
recently~\cite{BaumannDAlessandroGanardiIbarraMcQuillenSchuetzeZetzsche2022a}.
The proof here is substantially simpler.  Our proof also easily yields
that all amalgamation systems enjoy a \emph{growth dichotomy}: their
languages either have polynomial growth (if they're bounded) or
otherwise exponential growth
(see~\cite{DBLP:journals/ijfcs/GawrychowskiKRS10} for a precise
definition). After being open for RBCM for a long
time~\cite{DBLP:conf/stacs/IbarraR86}, this was shown
in~\cite{BaumannDAlessandroGanardiIbarraMcQuillenSchuetzeZetzsche2022a}
for RBCM and for pushdown RBCM.

We begin with a simple
characterisation for subsemigroups.

\begin{lemma}\label{boundedness-semigroup}
  Let $L\subseteq\Sigma^*$ be a subsemigroup. Then exactly one of the
  following holds: (i)~$L$ is bounded or (ii)~$L$ contains two
  prefix-incomparable words.
\end{lemma}
\begin{proof}
  First, suppose that (i) and (ii) both hold. Then for some words $w_1,\ldots,w_k$, we have $L\subseteq
  w_1^*\cdots w_k^*$, which implies that for every $n\in\N$, $L$
  contains at most $n^{k-1}$ words of length $n$.  On the other hand,
  if $u,v\in L$ are prefix-incomparable, meaning that neither is a
  prefix of the other, then the two words $uv$ and $vu$ have equal
  length, but are distinct.  Moreover, as $L$ is a subsemigroup, we
  have $\{uv,vu\}^n\subseteq L$ for every $n$.  But the set $\{uv,vu\}^n$
  contains $2^{n}$ distinct words of length $|uv|\cdot n$.  Yet
  $2^n\le (|uv|\cdot n)^{k-1}$ cannot hold for every $n\in\N$.  Hence,
  (i) and (ii) are mutually exclusive.
  Now suppose (ii) does not hold.  If $L\subseteq\{\varepsilon\}$, then $L$ is bounded.  Otherwise, $L$ contains some $w\ne\varepsilon$. As~$L$ is a subsemigroup, $w^n\in L$ for all
  $n>0$, and since any two words in $L$ are prefix-comparable, we have
  $L\subseteq\prefixes(w^*)$.  Since $w^*$ is bounded,
  $\prefixes(w^*)$ and thus $L$ are bounded as well.
\end{proof}

\begin{proof}[Proof of \impl{emptiness-decidable}{boundedness-decidable}]
We again provide two semi-decision procedures. The procedure for
positive instances simply enumerates expressions $w_1^*\cdots w_k^*$
and checks whether $L\subseteq w_1^*\cdots w_k^*$, which is decidable
for a full trio with an oracle for the emptiness problem.  The more
interesting case is the procedure for negative instances, which looks
for the following \emph{non-boundedness witness}: three runs $\rho_0$,
$\rho_1$, and $\rho_2$ with $\rho_0 \embedwith f \rho_1$ and
$\rho_0\embedwith g\rho_2$ such that for some $i$, the words
$\gap{i}{f}$ and $\gap{i}{g}$ are prefix-incomparable.  Let us show
that non-boundedness witnesses characterise negative instances.

First, suppose there is a non-boundedness witness.  Then the gap
language $L_{\rho_0,i}$ contains two prefix-incomparable words.
Moreover, by \cref{gap-semigroup}, $L_{\rho_0,i}$ is a subsemigroup,
and thus by \cref{boundedness-semigroup}, the language $L_{\rho_0,i}$
is not bounded.  However, every word in $L_{\rho_0,i}$ appears as a
factor in a word of $L$.  Then $L$ is not bounded, as otherwise the
set of factors of $L$ would be bounded and thus also $L_{\rho_0,i}$.

Conversely, suppose there is no non-boundedness witness.  As
$(R,{\runembed})$ is a \nameref{wqo}, $R$ itself has a finite basis
$\{\rho_1,\ldots,\rho_m\}$. Let $\can{\rho_i}=a_{i,1}\cdots
a_{i,n_i}$ be the canonical decomposition of $\rho_i$.  Then we have
\begin{equation}
L\subseteq\bigcup_{i=1}^m L_{\rho_i,0} a_{i,1} L_{\rho_i,1} \cdots
 a_{i,n_i} L_{\rho_i,n_i}\;.
\end{equation}
Since there is no non-boundedness witness, each language
$L_{\rho_i,j}$ is linearly ordered by the prefix ordering.  Moreover,
by \cref{gap-semigroup} each language $L_{\rho_i,j}$ is a subsemigroup
and thus \cref{boundedness-semigroup} implies that $L_{\rho_i,j}$ is
bounded. As boundedness is preserved by finite products, finite
unions, and taking subsets, $L$ must be bounded.
\end{proof}

\begin{proof}[Proof
of \impl{boundedness-decidable}{emptiness-decidable}] Given a language
$L\subseteq\Sigma^*$, consider the rational transduction
$T\eqdef\Sigma^*\times\{a,b\}^*$.  Then $TL$ is a bounded language
(actually, the empty set) if and only if $L=\emptyset$ (as otherwise,
$TL=\{a,b\}^*$ is not bounded).
\end{proof}

\subsection{Unary Languages}
We now come to results on languages over single-letter alphabets
$\Sigma = \{ a \}$.  To simplify the exposition, we slightly abuse
notation and identify each word $a^k$ with the number $k\in\N$ and
thus assume that each $\yield{\rho}$ for a run $\rho$ is a natural
number.

First, we provide in \cref{sub:leroux19} a very simple proof due
to \citet{Leroux2019} that shows that all VASS languages over a single
letter are regular.  
We then show how to make the proof effective in \cref{sub:unary-eff};
the resulting proof is still markedly simpler than the one of
effective regularity in VASS
by \citet{DBLP:journals/acta/HauschildtJ94}, which relies
on \citeauthor{DBLP:phd/dnb/Hauschildt90}'s
dissertation~\cite{DBLP:phd/dnb/Hauschildt90}.

\subsubsection{Regularity. }\label{sub:leroux19}
There is a proof due to \citeauthor{Leroux2019} which shows, only
using amalgamation, that all unary VASS languages are
regular~\cite{Leroux2019}, and happens to apply to our notion of
amalgamation systems.  It relies on a folklore result from number
theory~(see, e.g.\ \cite{wilf1978circle}).
\begin{lemma}[folklore]\label{semigroups-ultimately-periodic}
If $S\subseteq\N$ is a subsemigroup of~$\N$, then $S$ is ultimately
periodic.  Moreover, $S$ is ultimately identical to $\N\cdot\gcd(S)$, that is, there exists some threshold $k$ such that for all $n \geq k$ we have $n \in S \Leftrightarrow n \in \N \cdot \gcd(S)$.
\end{lemma}

Since \citeauthor{Leroux2019}'s proof has not been published, we
reproduce it here, in a general form for amalgamation systems.
\begin{lemma}[\citep{Leroux2019}]\label{lem:leroux19}
Every unary language accepted by an amalgamation system is regular.
\end{lemma}
\begin{proof}
Let $L\subseteq\N$ be the language of an amalgamation system
$(\{a\},R,E,\mathsf{can})$.  We want to show that $L$ is ultimately periodic.
Since $(R,\runembed)$ is a \nameref{wqo}, the set $R$ has a finite
basis, say $\{\rho_1,\ldots,\rho_m\}$. For each $\rho_i$, consider the
set
\begin{equation}
  L_i\eqdef\{\yield{\sigma}-\yield{\rho_i} \mid \exists\sigma\in R\colon \rho_i\runembed \sigma \}\;.
  \label{unary-above-basis}
\end{equation}
Since every run of $R$ embeds one of the runs $\rho_1,\ldots,\rho_m$, we have
$L=(\yield{\rho_1}+L_1)~\cup~\cdots~\cup~(\yield{\rho_m}+L_m)$.  By
concatenative amalgamation, each
$L_i$ is a subsemigroup of $\N$.  By \cref{semigroups-ultimately-periodic},
this implies that $L_i$ is ultimately periodic, and thus so is $L$.
\end{proof}

\subsubsection{Effectiveness. }\label{sub:unary-eff}
Unfortunately, \citeauthor{Leroux2019}'s proof in \cref{lem:leroux19}
is not effective: even for VASS, one cannot compute a basis of the set
of runs (see \citeFullOrAppendix{appendix:minimal-runs}).  Therefore, we prove an
effective version, which works by enumeration.  It enumerates certain
combinations of runs that yield an ultimately periodic subset of
numbers.  Moreover, we will show that for every amalgamation system,
there exists such a combination of runs that yields exactly its entire
language.

\begin{proof}[Proof of \impl{emptiness-decidable}{unary-effectively-regular}]
Define a \emph{unary witness} as a pair $(F,T)$, where $F\subseteq R$
is a finite set of runs and $T\subseteq R\times R\times R$ is a finite
set of triples $(\rho,\sigma,\tau)$ of runs such that
$\rho\runembed\sigma$ and $\rho\runembed\tau$.  The
set \emph{represented by $(F,T)$} is defined as
\begin{align*}
S(F,T)&\eqdef\{\yield{\rho} \mid \rho\in F\} ~\cup~ \bigcup_{(\rho,\sigma,\tau)\in T} S(\rho,\sigma,\tau)\;,
\shortintertext{where, for runs $\rho,\sigma,\tau$ with $\rho\runembed\sigma$ and $\rho\runembed\tau$,}
S(\rho,\sigma,\tau)&\begin{multlined}[t]\eqdef \yield{\rho}~+~\N\cdot (\yield{\sigma}-\yield{\rho}) \\
~+~\N\cdot(\yield{\tau}-\yield{\rho})\;.\end{multlined}
\end{align*}

Our algorithm works as follows.  It enumerates unary witnesses $(F,T)$
and for each of them, checks whether $L\subseteq S(F,T)$.  The latter
is decidable because $S(F,T)$ is an effectively regular language
and we can check if the set $L\cap (\N\setminus S(F,T))$ is empty in a
full trio with an oracle for emptiness.  Since we always have
$S(F,T)\subseteq L$ by construction, this algorithm is correct: if it
finds $(F,T)$ with $L\subseteq S(F,T)$, then we know $L=S(F,T)$.
It remains to show termination.
	\vspace{-0.5cm}
\begin{claim}\label{cl-unary}
There is a unary witness $(F,T)$ with $L=S(F,T)$.
\end{claim}
To prove \cref{cl-unary}, let $\{\rho_1,\ldots,\rho_m\}$ be a finite
basis of the \nameref{wqo} $(R,{\runembed})$ and define the sets $L_i$
as in \eqref{unary-above-basis}.  Since each $L_i$ is a semigroup,
\cref{semigroups-ultimately-periodic} tells us that $L_i$ ultimately agrees with
$\N\cdot \gcd(L_i)$.  In particular, there are $k,\ell\in L_i$ with
$k-\ell=\gcd(L_i)$.  This means that there are runs $\sigma_i$ and
$\tau_i$ with $\rho_i\runembed\sigma_i$ and $\rho_i\runembed\tau_i$
with $\yield{\sigma_i}=\yield{\rho_i}+k$ and
$\yield{\tau_i}=\yield{\rho_i}+\ell$.  We choose
$T\eqdef\{(\rho_i,\sigma_i,\tau_i) \mid i\in[1,m]\}$.  We now claim
that the set $L\setminus S(\emptyset,T)$ is finite. Note that if this
is true, we are done, because we can choose $F$ by picking a run for
each number in $L\setminus S(\emptyset,T)$.  For finiteness of $L\setminus
S(\emptyset,T)$, it suffices to show finiteness of $L_i\setminus G_i$ for each $i$, where
\[ G_i \eqdef \N\cdot(\yield{\sigma_i}-\yield{\rho_i})+\N\cdot(\yield{\tau_i}-\yield{\rho_i}). \]
To show that $L_i\setminus G_i$ is finite, we claim that
$\gcd(G_i)$ divides $\gcd(L_i)$. This will imply finiteness of $L_i\setminus G_i$ because $G_i$ is a subsemigroup of $\N$ and thus
ultimately agrees with $\N\cdot\gcd(G_i)$. 

Since $\yield{\sigma_i}-\yield{\rho_i}$ and $\yield{\tau_i}-\yield{\rho_i}$
both belong to $G_i$, we know that $\gcd(G_i)$ divides both numbers, and
therefore $\gcd(G_i)$ also divides their difference, which is
$\yield{\sigma_i}-\yield{\tau_i}=\gcd(L_i)$ by the choice of $\sigma_i$ and
$\tau_i$. The claim is established.
\end{proof}

\begin{proof}[Proof
of \impl{unary-effectively-regular}{emptiness-decidable}]
Given $L\subseteq\Sigma^*$, consider the rational transduction
$T\eqdef\{(w,a^{|w|}) \mid w\in\Sigma^*\}$.  Then $TL\subseteq\{a\}^*$ is a unary
language.  Moreover, $TL=\emptyset$ if and only if $L=\emptyset$. Thus, we can
decide emptiness of $L$ using an NFA for $TL$ constructed by the
oracle for effectively regular unary languages of amalgamation
systems.
\end{proof}

\subsection{Computing Priority Downward Closures}

\newcommand{\colred}{\color{red!60}}
\newcommand{\priority}{\ensuremath{\mathcal{P}}}
\newcommand{\ld}{\lessdot}
\newcommand{\lowerletters}[1]{\ensuremath{\Sigma_{\leq #1}}}
\newcommand{\equalletters}[1]{\ensuremath{\Sigma_{= #1}}}
\NewDocumentCommand\bo{o}{
	\IfNoValueTF{#1}{\preccurlyeq_{\mathsf{B}}}{\preccurlyeq_{\mathsf{B}}^{#1}}
}
\NewDocumentCommand\so{o}{
\IfNoValueTF{#1}{\preccurlyeq_{\mathsf{S}}}{\preccurlyeq_{\mathsf{S}}^{#1}}
}
\NewDocumentCommand\sbo{o}{
\IfNoValueTF{#1}{\preccurlyeq_{\mathsf{S}}}{\preccurlyeq_{\mathsf{S}}^{#1}}
}

\NewDocumentCommand\po{o}{
        \IfNoValueTF{#1}{\preccurlyeq_{\mathsf{P}}}{\preccurlyeq_{\mathsf{P}}^{#1}}
}
\NewDocumentCommand\sd{o}{
        \IfNoValueTF{#1}{\mathord{\downarrow_{\mathsf{S}}}}{\mathord{\downarrow_{\mathsf{S}}^{#1}}}
}
\NewDocumentCommand\bd{o}{
        \IfNoValueTF{#1}{\mathord{\downarrow_{\mathsf{B}}}}{\mathord{\downarrow_{\mathsf{B}}^{#1}}}
}
\NewDocumentCommand\pd{o}{
        \IfNoValueTF{#1}{\mathord{\downarrow_{\mathsf{P}}}}{\mathord{\downarrow_{\mathsf{P}}^{#1}}}
}

\newcommand{\s}{\ensuremath{\Sigma}}
\newcommand{\expression}{\mathfrak{R}}
\newcommand{\new}{\mathcal{B}}

Motivated by the verification of systems that communicate via channels
with congestion control, recent work~\cite{anand_et_al_CONCUR} considered the
problem of computing downward closures with respect to
the \emph{priority ordering}, which was introduced
in~\cite{haase14}.  In that setting, one has an alphabet $\Sigma$ with
associated priorities in $[0,d]$, specified by a priority map
$p\colon\Sigma\to[0,d]$.  Then $u\po v$ holds if $u=u_1\cdots u_n$,
$u_1,\ldots,u_n\in\Sigma$, and $v=v_1u_1\cdots v_nu_n$,
$v_1,\ldots,v_n\in\Sigma^*$, such that for each $i$, the letters in
$v_i$ have priority at most $p(u_i)$.  In other words, letters can
only be dropped from $v$ if they are followed by some (undropped)
letter of higher or equal priority.  In a channel with congestion
control, sending message sequences from a set $L\subseteq\Sigma^*$
will result in received messages in the \emph{priority downward
closure}
\begin{equation}L\pd=\{u\in\Sigma^* \mid \exists v\in L\colon u\po v \}\;. \end{equation}
As with the ordinary word embedding, because $\po$ well-quasi-orders
the set of words with priorities~\cite[Lem.~3.2]{haase14}, the language $L\pd$ is regular for any
language $L\subseteq\Sigma^*$, hence the problem of \emph{computing priority downward
closures}, i.e., computing an NFA for $L\pd$ for an input language~$L$.

The proof of \impl{pdc-computable}{emptiness-decidable} consists in
observing that $L=\emptyset$ if and only if $L\pd=\emptyset$, the
latter being straightforward with an NFA recognising $L\pd$.  Here, we
only want to sketch the proof
of \impl{emptiness-decidable}{pdc-computable}, and point
to \citeFullOrAppendix{appendix-priority-order} for the actual proof.

To compute the priority downward closure of a language
$L\subseteq\Sigma^*$, the algorithm uses a strategy from
\cite{DBLP:conf/icalp/Zetzsche15}.  It essentially enumerates $\po$-downward
closed sets $D$, decompose them into finitely many ideals as
$D=I_1\cup\cdots\cup I_n$, and then decides (i)~whether $L$ is
included in $D$ and (ii)~whether each $I_i$ is included in $L\pd$.
Here, in a wqo $(X,{\le})$, an \emph{ideal} is a non-empty subset
$I\subseteq X$ that is downwards closed and \emph{upwards directed},
meaning that for any $x,y\in I$, there is a $z\in I$ with $x\le z$ and
$y\le z$.  It is a general property of wqos that every downwards closed
set decomposes into finitely many
ideals~\cite[e.g.,][]{Leroux:2015:reachability-vas,DBLP:conf/icalp/Goubault-Larrecq16}.
In this algorithm, deciding
(i) is easy, because $D$ is already a regular language, and since we
assume decidable emptiness and closure under rational transductions,
we can decide the emptiness of $L\cap (\Sigma^*\setminus D)$.

The challenging part is deciding~(ii). For ordinary downward closures, this
problem reduces to the SUP~\cite{DBLP:conf/icalp/Zetzsche15}. For priority
downward closures, this also leads to an unboundedness problem, but a more
intricate one. Instead of some measures (in the SUP: the number of occurrences
of each letter) to be unbounded simultaneously, we here also need to decide
\emph{nested unboundedness}. However, using amalgamation, it will be possible
to detect such nested unboundedness properties using certain ``run
constellations.''

\subsubsection{Nested Unboundedness}
Let us illustrate this with an example. A particular unboundedness
property that is required for ideal inclusion is whether in a language
$L\subseteq\{0,1\}^*$, for every $k\ge 0$, there is a word $w\in L$
with $\ge k$ factors, each containing $\ge k$ contiguous~$0$'s, and
they are separated by~$1$'s. In other words, we need to find
arbitrarily many arbitrarily long blocks of~$0$'s.  Let us call this
property \emph{nested unboundedness}.  Intuitively, this is more
complicated that the SUP%
. 

However, using run amalgamation, this amounts to checking a simple kind of witness.
First, we need to slightly transform our language. Consider the
language 
	\begin{multline*}
		K=\{u_1\cdots u_n \mid n\in\N, ~v_0u_1v_1\cdots u_nv_n \in L,~\text{$\forall i\in[0,n]$:}\\ \text{either (a)~$v_i\in 0^*$, (b)~$v_i\in (10^*)^*$ and $u_{i+1}\in (10^*)^+$, or}\\ \text{(c)~$i=n$ and $v_i\in(10^*)^+$} \}
	\end{multline*}
Hence, words in $K$ are obtained from words in~$L$ by removing (a)~factors in
$0^*$ or (b)~factors from $(10^*)^*$, but the latter only if that
	factor was followed by a non-removed~$1$, or is a suffix in $(10^*)^+$. This means that we can either
make individual blocks of~$0$'s smaller, or remove maximal factors
of~$0$'s, including exactly one neighbouring~$1$.  One
can see that~$K$ can be obtained from~$L$ using a
rational transduction and thus we can construct an
amalgamation system for~$K$.  Moreover, $K$ has the nested unboundedness property if and only if~$L$ does.
	However, the advantage of working with~$K$ is that if nested unboundedness holds, then~$K$ contains every word in~$(10^*)^*$.

\subsubsection{Witnesses for Nested Unboundedness} 
In an amalgamation system for $K$, we have a simple kind of witness
for our property: runs $\rho_0\runembed\rho_1\runembed \rho_2$ such
that (i)~some gap of $\rho_0\runembed\rho_1$ between some positions
$i<j$ of $\rho_1$ contains a~$1$ and (ii)~some gap of
$\rho_1\runembed\rho_2$, which is between~$i$ and~$j$ in~$\rho_1$,
belongs to~$0^+$. Then, by amalgamating~$\rho_2$ with itself
above~$\rho_1$ again and again, we can create arbitrarily long blocks
of contiguous~$0$'s. The resulting run~$\rho_2'$ still embeds~$\rho_1$
and thus~$\rho_0$, such that one gap of~$\rho'_2$ in~$\rho_0$ contains
both our long block of~$0$'s and also a~$1$. Thus,
amalgamating~$\rho'_2$ again and again above~$\rho_0$, we obtain
arbitrarily many long blocks of~$0$'s.

\subsubsection{Existence of Witnesses}
Of course, we need to show that if nested unboundedness is satisfied, then the
	witness exists.  Let $S$ be an amalgamation system for $K$ and
	suppose $K$ has nested unboundedness. Let $M$ be the maximal
	length of the canonical decompositions of runs in a finite
	basis of $S$.
	Consider the sequence $w_1,w_2,\ldots$ with $w_i=(0^i1)^{2M}$ for $i\ge
	1$.  Then $w_1,w_2,\ldots\in K$, so there must be runs $\rho_1$ and
	$\rho_2$ with $\rho_1\runembed \rho_2$ such that $\yield{\rho_1}=w_i$ and
	$\yield{\rho_2}=w_j$.  Then every non-empty gap of $\rho_2$ in
	$\rho_1$ belongs to $0^+$.  Moreover, any embedding of a minimal run
	$\rho_0$ of $S$ into $\rho_1$ will have some gap containing two $1$'s,
	and thus have $10^i1$ as a factor. Thus the runs $\rho_0,\rho_1,\rho_2$
	constitute a witness.  

\subsubsection*{Other Concepts}
The full proof in \citeFullOrAppendix{appendix-priority-order} involves several
steps.  First, to simplify the exposition, we work with a slightly
different ordering, called the \emph{simple block ordering} and
denoted $\so$, such that downward closure computability of $L\sd$
with respect to $\so$ implies that of $L\pd$ with respect to $\po$.
We then provide a syntactic description of the ideals of $\so$.  To
avoid some technicalities, we introduce the related notion
of \emph{pseudo-ideals} and devise our algorithm to work with those.
We define a notion of $I$-witness for each pseudo-ideal $I$,
which is a particular constellation of run embeddings in a system
for $L\sd$ such that we have $I\subseteq L\sd$ if and only if an
amalgamation system for $L\sd$ possesses an $I$-witness.  There, we
use that $L\sd$ is obtained using a rational transduction from $L$.

\subsection{Factor Universality}
Finally, we want to mention that while amalgamation allows us to
perform many algorithmic tasks (and applies to a significant class of
systems---see the next section), it is not quite enough to cover the
axiomatically defined class of \emph{unboundedness predicates} of
formal languages, as introduced
by~\citet{Czerwinski:2018:unboundedness-problems-vas}.  That paper
shows that for each unboundedness predicate, if it can be decided for
regular languages, then it can even be decided for VASS.

Consider the \emph{factor universality problem}: given a language
$L\subseteq\Sigma^*$, does every word from $\Sigma^*$ appear as a factor in
$L$?  This is an unboundedness predicate in the sense of
\cite{Czerwinski:2018:unboundedness-problems-vas}, but it is not decidable for
all amalgamation systems.  Indeed, as observed in
\cite{Czerwinski:2018:unboundedness-problems-vas}, factor universality is
undecidable for one-counter automata, which, as a subclass of context-free
languages, support amalgamation.

\section{Examples of Amalgamation Systems}
\label{sec:systems}

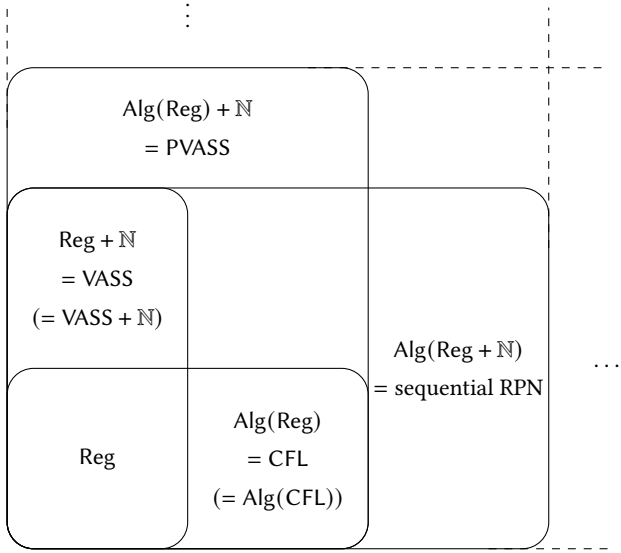
\begin{figure}[tbp]
	\begin{center}
		\begin{tikzpicture}[scale=0.8]
			\node at (1.5,1.5) {$\Reg$};
			
			\draw [rounded corners=10] (0,0) rectangle (3,6);
			\node at (1.5,4.5) {$\begin{array}{c}
					\Reg + \Nat \\
					= \VASS \\
					(= \VASS + \Nat)
				\end{array}$};
			
			\draw [rounded corners=10] (0,0) rectangle (6,3);
			\node at (4.5,1.5) {$\begin{array}{c}
					\Alg(\Reg) \\
		 			= \CFL \\
		 			(= \Alg(\CFL))
			\end{array}$};
		
			\draw [rounded corners=10] (0,0) rectangle (6,8);
			\node at (3,7) {$\begin{array}{c}
					\Alg(\Reg) + \Nat \\
					= \PVASS
			\end{array}$};
			
			\draw [rounded corners=10] (0,0) rectangle (9,6);
			\node at (7.5,3) {$\begin{array}{c}
					\Alg(\Reg + \Nat) \\
					= \text{sequential RPN}
			\end{array}$};
		
			\node at (10,3) {$\cdots$};
			\node at (3,9) {$\vdots$};
			
			\draw [dashed] (8,0) -- (10,0);
			\draw [dashed] (5,8) -- (10,8);
			\draw [dashed] (0,7) -- (0,9);
			\draw [dashed] (9,5) -- (9,9);
		
		\end{tikzpicture}
	\end{center}
	\caption{The hierarchy of classes obtained by the operations $\cdot + \Nat$ and $\Alg(\cdot)$.}
	\label{fig:hierarchy}
\end{figure}

In this section, we show that amalgamation systems are not limited to
the few examples from \cref{sec:amalgamation-systems}.  On the
contrary, we present a whole hierarchy of language classes that
support amalgamation (and satisfy the implicit effectiveness
assumptions of \cref{ssub:eff}), starting from regular languages as
already discussed in \cref{par:example-nfa} (see \cref{sub:reg}), by
repeatedly applying the operations of ``adding counters'' (see
\cref{sec:counter-extension}) and ``building stacks'' (see
\cref{sec:alg-closure})---those operations are denoted by $\cdot +
\Nat$ and $\Alg(\cdot)$, respectively, in \cref{fig:hierarchy}.
This hierarchy includes some well-known classes---for example VASS
languages and context-free languages as already mentioned in
\cref{par:example-vass,par:example-cfgs}---, along with perhaps
lesser-known classes---like PVASS languages or the languages of
sequential recursive Petri nets
(RPN)~\citep{DBLP:journals/acta/HaddadP07,DBLP:conf/time/HaddadP01}.

As explained in the introduction, this entire hierarchy can also be
envisioned from the perspective of \emph{valence automata} by choosing
appropriate graph
monoids~\cite{DBLP:phd/dnb/Zetzsche16,DBLP:conf/rp/Zetzsche21}.  In
that framework, barring the classes that are already known to have an
undecidable emptiness problem, one can show that our hierarchy
actually exhausts all the remaining cases, yielding a proof of
\cref{main-valence}; see \cref{sub:va} for a sketch and
\citeFullOrAppendix{appendix-valence-automata} for more details.

\subsubsection*{Well-quasi-ordered Decorations}
Our constructions in this section actually capture more than the
hierarchy of \cref{fig:hierarchy}: we show in
\cref{sec:counter-extension,sec:alg-closure} how to ``add counters''
and ``build stacks'' to any language class $\cC$ that supports
amalgamation---provided an additional technical requirement is met by
that class.  To show that the classes $\cC + \Nat$ and $\Alg(\cC)$
support amalgamation, we need the property that $\cC$ supports
``wqo decorations,'' as we define now.

Given a run $\rho$ and a set $X$, an $X$-\emph{decoration} of $\rho$
is a pair $(\rho, w)$ where $w$ is a word in~$X^*$ of length $\abs{w}
= |\yield{\rho}|$.  %
Intuitively, one might think of a decoration as adding additional
information from $X$ to every letter in $\yield{\rho}$.  For a set of
runs $R$, we write $\Deco^X\!(R)$ for the set of all $X$-decorations of
runs from $R$.  If $(X,{\leq})$ is a qo, we define the set of
admissible embeddings between $X$-decorated runs by $E^X((\rho,
u),(\sigma,v))\eqdef\{f\in E(\rho,\sigma)\mid u_i \leq
v_{f(i)}\text{ for all }i\in[1,|\yield{\rho}|]\}$; this defines a
quasi-ordering $\runembed^X$ on $\Deco^X\!(R)$ such that $(\rho, u)
\runembed^X (\sigma, v)$ if and only if there is $f$ such that
$\rho \runembed_f \sigma$ and $u_i \leq v_{f(i)}$ for all
$i\in[1,|\yield{\rho}|]$.

\begin{definition}[wqo decorations]
\label{def:wqo-deco}	
An amalgamation system with set of runs $(R,{\runembed})$
supports \emph{wqo decorations} if, for \emph{every} wqo
$(X,{\leq})$, the set of decorated runs $(\Deco^X\!(R),{\runembed^X})$
is also a wqo.
\end{definition}
By extension, we say a class of languages (that supports amalgamation)
\emph{supports wqo decorations} if for every language $L$ in the class, there
exists an amalgamation system that supports wqo decorations and whose
language is $L$.
In practice, supporting wqo decorations is not a major
restriction: as part of our proofs, we will show that regular
languages support wqo decorations, and furthermore that this property
is maintained by the operations $\cdot + \Nat$ and $\Alg(\cdot)$.

\begin{remark}
  Not every amalgamation system supports wqo decorations.  Here is an
  example: let $\Sigma\eqdef\{a\}$ and define $R\eqdef\Sigma^\ast$
  with $\can{w}\eqdef w$ for all $w\in \Sigma^\ast$.  If $n\leq m$, we
  define the identity function $f\colon i\mapsto i$ for every $i\in
  [1,n]$ as the sole admissible embedding between $a^n$ and $a^m$; all
  the gap words are $\varepsilon$ except possibly $G_{n,f} = a^{m-n}$.
  Thus $\runembed$ is the prefix ordering, but over a unary alphabet
  it gives rise to a wqo $(R,{\runembed})$ isomorphic to
  $(\mathbb{N},{\leq})$.  Also, those embeddings compose, and if
  $a^n\runembed a^r$ and $a^n\runembed a^s$ (thus with $n\leq r$ and
  $n\leq s$), then letting $m \eqdef r + s - n$ allows to amalgamate
  into~$a^m$.
  
  Consider now the wqo $X\eqdef (\{A,B\},{=})$, i.e., the finite set
  $\{A,B\}$ with the equality relation.  Then
  $(\Deco^X\!(R),{\runembed^X})$ is isomorphic with the prefix ordering
  over $\{A,B\}^\ast$, which is not a wqo.  Indeed, among the
  decorated runs in $\Deco^X\!(R)$, one finds the decorated pairs $(a^n,
  B^{n-1}A)$ for all $n\geq 2$, and those decorated pairs form an
  infinite antichain: whenever $n < m$, when attempting to compare
  $(a^n, B^{n-1}A)$ with $(a^m, B^{m-1}A)$, the embedding $f\in
  E(a^n,a^m)$ maps~$n$ to itself, but the $n$th letter in the
  decoration of the first pair is~$A$ and the $n$th letter in the
  decoration of the second pair is~$B$, thus $(a^n, B^{n-1}A)
  \not\runembed^X (a^m, B^{m-1}A)$.
\end{remark}

\subsection{Regular Languages}\label{sub:reg}

We already discussed the case of regular languages in
\cref{par:example-nfa}.  Let us provide here a more formal statement.
\begin{restatable}{theorem}{regamalgamation}
	\label{thm:regular-amalgamation}
  The class of regular languages supports
  amalgamation and wqo decorations.
\end{restatable}
\begin{proof}
  The class of regular languages is produced exactly by finite-state
  automata, which we can assume wlog.\ to be $\varepsilon$-free.  The
  definitions of runs, their canonical decompositions, admissible
  embeddings, and how to amalgamate them were already given in
  \cref{par:example-nfa}.  It remains to be shown that $\varepsilon$-free
  finite-state automata support wqo decorations.  Let $(X,\leq)$ be a
  wqo, and $(\rho,w)$ an $X$-decoration of a
  run~$\rho=(q_0,a_1,q_1)\cdots(q_{n-1},a_n,q_n)\in R$. Since the
  automaton is $\varepsilon$-free, $w$ is of length $|w|=n$ and can be
  written as $w=x_1\cdots x_n$.  Then the map $r\colon \Deco^X\!(R)\to
  (\Delta\times X)^\ast$ defined by
  $r\colon(\rho,w)\mapsto((q_0,a_1,q_1),x_1)\cdots((q_{n-1},a_n,q_n),x_n)$
  is an \emph{order reflection}, in that if $r(\rho,w)\leq_\ast
  r(\rho',w')$, then $(\rho,w)\runembed^X(\rho',w')$.  By Dickson's
  and Higman's lemmata, $((\Delta\times X)^\ast,{\leq_\ast})$ is a
  wqo, and this order reflection shows that $(\Deco^X\!(R),
  {\runembed^X})$ is also a wqo: $r$ pointwise maps bad
  sequences $(\rho_0,w_0), \allowbreak(\rho_1,w_1),\ldots$ over
  the ordering $(\Deco^X\!(R), \runembed^X)$ to bad sequences
  $r(\rho_0,w_0),\allowbreak r(\rho_1,w_1),\ldots$ of the same length
  over the ordering $((\Delta\times X)^\ast,{\leq_\ast})$. Therefore, bad sequences over
  $(\Deco^X\!(R), \runembed^X)$ must be finite.
\end{proof}
\subsection{Counter Extension}
\label{sec:counter-extension}

Vector addition systems were presented in \cref{par:example-vass} as
finite-state automata that additionally modify a set of~$d$ counters.
A natural question is whether we can generalise this operation of
``adding counters'' to arbitrary amalgamation systems.  To that end,
we first define in \cref{sub:lang-counters} a generic operator that
takes a language class~$\cC$ and forms a language class $\cC+\Nat$ of
languages in~$\cC$ extended with $d>0$ counters, such that for
instance $\Reg + \Nat = \VASS$.  We then show in
\cref{sub-amalg-counters} that, if $\cC$ supports amalgamation and
wqo decorations, then so does $\cC+\Nat$.

\subsubsection{Extending Languages.}\label{sub:lang-counters}
Fix $d>0$ the number of counters we wish to add.  We use a finite
alphabet $\mathbb U_d\subseteq\Z^d$ of unit updates, defined by
$\mathbb U_d\eqdef\{\vec 0\}\cup\{\vec e_i,-\vec e_i\mid i\in[1,d]\}$
where each $\vec e_i$ is the unit vector such that $\vec e_i(i)=1$ and
$\vec e_i(j)=0$ for all $j\neq i$.  Then the morphism
$\delta\colon\mathbb U_d^\ast\to\Z^d$ maps words over~$\mathbb U_d$ to
their effect, and is defined by $\delta(\vec u_1\cdots\vec
u_n)\eqdef\sum_{j=1}^n\vec u_j$.  Finally, let $N_d\subseteq\mathbb
U_d^\ast$ be the language of all $\N$-counter-like words over~$\mathbb
U_d$; formally, 
\begin{equation*}
  N_d\eqdef\{ w \in \mathbb U_d^\ast\mid \delta(v) \geq \vec 0 \text{
    for all prefixes } v \text{ of } w \text{ and } \delta(w) = \vec 0
  \}\;.\end{equation*}
Put differently, $N_d$ is the language of the VASS with a single state
$q$ and a transition $(q,\vec u,\vec u,q)$ for each $\vec u\in\mathbb
U_d$.

Let $\Delta$ be a finite alphabet; a morphism $\eta\colon\Delta^*
\to\mathbb U_d^*$ is \emph{tame} if, for all $a\in\Delta$ and all
$i\in[1,d]$, all the occurrences of $\vec e_i$ in $\eta(a)$ occur
before all the occurrences of~$-\vec e_i$.  Let $\Sigma$ also be a
finite alphabet; a morphism $\alpha\colon\Delta^\ast\to\Sigma^\ast$ is
\emph{alphabetic} if $\alpha(\Delta)\subseteq\Siep$.

\begin{definition}[Counter extension]\label{def:counters}
  Let $L$ be a language over a finite alphabet $\Delta$.  For a tame
  morphism $\eta\colon\Delta^* \to\mathbb U_d^*$ and an alphabetic
  morphism $\alpha\colon\Delta^\ast\to\Sigma^\ast$ into a finite
  alphabet~$\Sigma$, let $L_{\eta,\alpha}$ be the language
  $\alpha(\eta^{-1}(N_d) \cap L)$.
  For a class of languages~$\cC$, $\cC + \Nat$ is the class of
  languages $L_{\eta,\alpha}$ when $L$ ranges over~$\cC$. 
\end{definition}\noindent
Very informally, for $\cC=\Reg$, $L\subseteq\Delta^\ast$ describes
transition sequences, $\eta$ the effect of each transition encoded as
a word over $\mathbb U_d$, and $\alpha$ its label in
$\Sigma_\varepsilon$, resulting in $L_{\eta,\alpha}$ being a VASS
language.

\subsubsection{Extending Amalgamation Systems.}\label{sub-amalg-counters}
We are going to show that this construction is well behaved in the
following sense.
\begin{restatable}{theorem}{countersamalgamation}
\label{thm-counters-amalgamation}
If $\cC$ is a class of languages that supports concatenative
amalgamation and wqo decorations, then so does $\cC + \Nat$.
\end{restatable}

To this end, fix $d>0$ and let $S=(\Delta,R,E,\mathsf{can})$ be an
amalgamation system supporting wqo decorations and accepting a
language $L \in \cC$, and let $\eta$ and $\alpha$ be morphisms as in
\cref{def:counters}.  Our goal is to define an amalgamation system
$S_{\eta,\alpha}$ that supports wqo decorations such that
$L(S_{\eta,\alpha})=L_{\eta,\alpha}$.

We decorate runs $\rho\in R$ with pairs of counter valuations from
$P\eqdef\Nat^{d}\times\Nat^d$.  Consider a $P$-decorated run
$(\rho,w)$ and let $a_1\cdots a_n=\yield\rho$ be the word accepted
by~$\rho$.  Then $w=(\vec u_1,\vec v_1)\cdots(\vec u_n,\vec v_n)$.  We
say that $(\rho, w)$ is \emph{coherent} if $\vec v_i = \vec u_i +
\delta(\eta(a_i))$ for all $i\in[1,n]$ and $\vec v_i=\vec u_{i+1}$ for
all $i\in[1,n-1]$.  We say that $(\rho, w)$ is \emph{accepting} if it
is coherent and additionally the initial counters are $\vec u_1 = \vec
0$ and the final counters are $\vec v_n = \vec 0$. 

Let $R_{\eta}$ be the set of accepting decorated runs in
$\Deco^{P}(R)$.  If the canonical decomposition of a run $\rho \in
R$ is $a_1 \cdots a_m$ with each $a_i \in \Delta_\epsilon$, then the
canonical decomposition of a decorated run $(\rho,w)\in
R_{\eta}$ is defined as $\alpha(a_1) \cdots \alpha(a_m)$, with
each $\alpha(a_i) \in \Siep$ by definition of~$\alpha$.
Let
$S_{\eta,\alpha}\eqdef(\Sigma,R_{\eta},E^{P},\alpha\circ\mathsf{can})$.
The following claim  shows that
this yields the intended language (see \citeFullOrAppendix{appendix:counter-ext}).

\begin{restatable}{claim}{counterextlangsysequiv}
  If $S$ is an amalgamation system and $L$ is its language, then
  $L(S_{\eta,\alpha}) =
  L_{\eta,\alpha}$.
\end{restatable}

To complete the proof of \cref{thm-counters-amalgamation}, it
remains to show that $S_{\eta,\alpha}$ satisfies the conditions of
\cref{def:conds} (see \cref{cl:counters-amalg}) and supports wqo
decorations in the sense of \cref{def:wqo-deco} (see
\cref{cl:counters-wqodec}).

\begin{claim}\label{cl:counters-amalg}
  If $S$ is an amalgamation system that supports wqo decorations, then
  $S_{\eta,\alpha}$ is an amalgamation system. 	
\end{claim}
\begin{proof}We show that $S_{\eta,\alpha}$ satisfies the conditions of
\cref{def:conds}.
\textbf{Composition. } As we have defined above, the admissible
embeddings in $E^{P}((\rho,u), (\sigma,v))$ of two runs
in $R_{\eta}$ are those of $E(\rho,\sigma)$ that respect the
ordering on decorations in~$P$.  %
These embeddings compose, %
because the admissible
embeddings of $S$ compose and because additionally $\vleq$ on
$P\eqdef\Nat^d\times\Nat^d$ is transitive%
.

\noindent\textbf{Well-quasi-order. } Because we assume $S$ supports
wqo decorations and $(P,{\vleq})$ is a wqo,
$(\Deco^{P}(R), \runembed^{P})$
is a wqo, and the induced
$(R_{\eta},\runembed^{P})$ as well.

\noindent\textbf{Amalgamation. }  The construction mirrors the one
presented in \cref{ssub:vass-amalg}.  Let $(\rho_0,w_0)\in R_{\eta}$
be a run with $a_1\cdots a_n$ the accepted word of~$\rho_0$ and
$w_0=(\vec u_1,\vec v_1)\allowbreak\cdots\allowbreak(\vec u_n,\vec
v_n)$.  Assume $(\rho_0,w_0)\runembed_{f}^P(\rho_1,w_1)$ and
$(\rho_0,w_0)\runembed_{g}^P(\rho_2,w_2)$ for $(\rho_1,w_1),
(\rho_2,w_2)\in R_\eta$.  By definition of our decorated embedding, we
have $\rho_0\embedwith{f}\rho_1$ and $\rho_0\embedwith{g}\rho_2$, and
for each $i\in[1,n]$ there exists $\vec c_i,\vec d_i\in\N^d$ such that
the $f(i)$th pair in~$w_1$ is $(\vec u_i+\vec c_i,\vec v_i+\vec c_i)$
and the $g(i)$th pair in~$w_2$ is $(\vec u_i+\vec d_i,\vec v_i+\vec
d_i)$.  Because~$C$ is an amalgamation system, there exists a run
$\rho_3\in R$ with $\rho_1 \embedwith{f'} \rho_3$,
$\rho_2\embedwith{g'}\rho_3$, and $\rho_0\embedwith{h}\rho_3$ with
$h=f\circ f'=g\circ g'$ that satisfies the concatenative amalgamation
condition.  It remains to show that we can decorate $\rho_3$ in an
accepting way to also satisfy amalgamation.

To do so, we define the $h(i)$th pair of $w_3$ to be $(\vec u_i+\vec
c_i+\vec d_i,\vec v_i+\vec c_i+\vec d_i)$.  As $\vec
u_i+\delta(\eta(a_i))=\vec v_i$, we also have $\vec u_i+\vec c_i+\vec
d_i+\delta(\eta(a_i))=\vec v_i+\vec c_i+\vec d_i$ as desired.
Consider now the $i$th gap $\gap{i}{h}$ and assume wlog.\ that
$\gap{i}{h} = \gap{i}{f}\gap{i}{g}$.  Observe that in $w_1$, we have
$\vec u_i + \vec c_{i}+\delta(\eta(\gap{i}{f}))=\vec u_i + \vec
c_{i+1}$.  Similarly, in $w_2$ we have $\vec u_i + \vec
d_{i}+\delta(\eta(\gap{i}{g}))=\vec u_i + \vec d_{i+1}$.  Then $\vec
u_i+\vec c_i+\vec d_i+\delta(\eta(\gap{i}{f}))=\vec u_i + \vec
c_{i+1}+\vec d_i$ and by monotonicity we can decorate $w_3$ coherently
along $\gap{i}{f}$ by adding $\vec d_i$ to all the pairs from~$w_1$
along that segment, and then $\vec u_i + \vec c_{i+1}+\vec
d_i+\delta(\eta(\gap{i}{g}))=\vec u_i +\vec c_{i+1}+\vec d_{i+1}$ as
desired and again by monotonicity we can decorate $w_3$ coherently
along $\gap{i}{g}$ by adding $\vec c_{i+1}$ to all the pairs
from~$w_2$ along that segment.  The produced decorated run
$(\rho_3,w_3)$ is coherent by construction and satisfies
$(\rho_1,w_1)\runembed_{f'}^P(\rho_3,w_3)$ and
$(\rho_2,w_2)\runembed_{g'}^P(\rho_3,w_3)$ as desired.  Using the same
abuse of notation as in \cref{ssub:vass-amalg} for the border cases,
because we are dealing with accepting runs, in the case of $i=0$ we
additionally have $\vec c_0 = \vec d_0 = \vec 0$, and analogously in
the case of $i = n$ we have $\vec c_{n+1} = \vec d_{n+1} = \vec 0$,
thus $(\rho_3,w_3)$ is also accepting.
\end{proof}

\begin{claim}\label{cl:counters-wqodec}
  If $S$ supports wqo decorations, then so does $S_{\eta,\alpha}$.	
\end{claim}
\begin{proof}
  Observe that a decoration of a run $(\rho,w_1\cdots w_n)\in
  R_{\eta}$ with a sequence $x_1\cdots x_n\in X^*$ over a
  wqo~$X$ is equivalent to a decoration of the run $\rho\in R$ with
  the sequence $(w_1,x_1)\cdots(w_n,x_n)\in (P\times X)^*$ over the
  wqo alphabet $\Nat^{d}\times\Nat^d\times X$.
\end{proof}

\subsection{Algebraic Extension}
\label{sec:alg-closure}

We introduce now, as a generalisation of the case of context-free
languages presented in \cref{par:example-cfgs}, how to support
amalgamation in the algebraic closure of a class of languages.

Given a class of languages $\cC$, a \emph{$\cC$-grammar} is a tuple
$\Grammar=(N, \Sigma, S, \{L_A\}_{A \in N})$, where $S \in N$ and each
$L_A$ is a language from $\cC$ with $L_A \subseteq (N \cup \Sigma)^*$.
For $A\in N$ and $u,v,w\in (N\cup\Sigma)^*$, we write $uAv \deriv uwv$ if $w \in L_A$.  The \emph{language
generated by $\Grammar$} is 
\[ L(\Grammar)\eqdef \{ w \in \Sigma^* \mid S \derivs w \}, \]
where $\derivs$ is the reflexive transitive closure of $\deriv$.
For example, every context-free grammar can be seen as a
$\mathsf{Fin}$-grammar, where $\mathsf{Fin}$ is the class of finite
languages.  The class of context-free languages is also defined by
the so-called ``extended'' context-free grammars allowing regular
expressions in their productions, i.e., $\Reg$-grammars.

\begin{definition}[Algebraic extension]
  Given a class of languages $\cC$, we denote by $\Alg(\cC)$ the
  \emph{algebraic extension} of $\cC$, that is, the class of all
  languages recognised by $\cC$-grammars.
\end{definition} 

We are going to show that if $\cC$ is well-behaved, then so is
$\Alg(\cC)$.

\begin{restatable}{theorem}{algamalgamation}
\label{thm:alg-amalgamation}
  If $\cC$ is a class of languages that supports concatenative
  amalgamation and wqo decorations, then so does its algebraic
  extension $\Alg(\cC)$.
\end{restatable}

Let us fix a $\cC$-grammar~$\Grammar$ and write $\Model_A$ for the
amalgamation system with wqo decorations that recognises the language
$L_A$.  Just like with context-free grammars in \cref{par:example-cfgs}, the
derivations of a $\cC$-grammar can be viewed as trees, with
nodes labelled either $\epsilon$ or with pairs of non-terminals $A$ and runs $\rho$ of
$\Model_A$.  We call $\rho$ the \emph{explanation} of the expansion of
the non-terminal $A$ at this step.  
More specifically, if $\rho$ is a run in $\Model_A$ with $\yield{\rho} \in \Sigma^*$, then $\leaf{(A \to \rho)}$ is a tree. 
Otherwise, let $X_1 X_2 \cdots X_n$ be the projection of $\yield{\rho}$ to $N$ 
and $t_1, \ldots, t_n$ be $X_1$-, $\ldots, X_n$-rooted trees. 
Then $\tree{(A \to \rho)}{t_1, \ldots, t_n}$ is a tree as well.
For the remainder of this section, if we write $\yield{\rho}$ or $\can{\rho} = u_0 X_1 u_1 \cdots X_n u_n$, we assume that $X_i \in N$ and $u_j \in \Sigma^*$. We write $\memo_\rho(i)$ for the map $[1,n] \to [1,\canonlen{\rho}]$ that associates to every $X_i$ its position in the canonical decomposition.

\paragraph*{Canonical decompositions}

\begin{figure}
	\begin{subfigure}{\columnwidth}
		\begin{center}
			\begin{tikzpicture}	
				\node [] (a) at (0,0) {$A$};
				\draw [-] (a) -- (-.5,-1) -- (.5,-1) -- (a);
				\node at (0,-1.2) {$w$};
				
				\node [] (b) at (2,.5) {$B$};
				\node [] (b') at (2,-.3) {$A$};
				\draw [-,dashed] (b) -- (b');
				\draw [-] (b') -- (1.5,-1.3) -- (2.5,-1.3) -- (b');
				\draw [] (b) -- (1,-1.3) -- (1.5,-1.3);
				\draw [] (b) -- (3,-1.3) -- (2.5,-1.3);
				\draw [-{latex},color=highlight,thick] (a) -- (b');
				\node at (1.25,-1.5) {$u$};
				\node at (2,-1.5) {$w$};
				\node at (2.75,-1.5) {$v$};
			\end{tikzpicture}
		\end{center}
		\caption{Mapping to a subtree adds letters to the left and right.}
		\label{fig:subtree-mapping}
	\end{subfigure}
	\hfill
	\begin{subfigure}{\columnwidth}
		\begin{center}
			\begin{tikzpicture}[tree/.style={regular polygon,regular polygon sides=3,draw}]
				\node (a) at (0,0) {$A$};
				\node [tree,label=-90:$w_1$] (a') at (-.8,-1.5) {$t_1$};
				\node [tree,label=-90:$w_2$] (a'') at (.8,-1.5) {$t_2$};
				\draw [-] (a) -- (a'.north);
				\draw [-] (a) -- (a''.north);
				
				\node (b) at (5,0) {$A$};
				\node [tree,label=-90:$w_1$] (b'') at (3.4,-1.5) {$t_1$};
				\node [tree,label=-90:$v_1$] (b''') at (5,-1.5) {$s_1$};
				\node [tree,label=-90:$w_2$] (b'''') at (6.6,-1.5) {$t_2$};
				
				\draw [-] (b) -- (b''.north);
				\draw [-] (b) -- (b'''.north);
				\draw [-] (b) -- (b''''.north);
				
				\draw [-{latex},color=highlight,thick] (a) edge [bend left] (b);
			\end{tikzpicture}
		\end{center}
		\caption{Larger runs in the underlying system $\Model_A$ interleave letters.}
		\label{fig:underlying-run-interleaving}
	\end{subfigure}
	\caption{Gaps in the canonical decomposition of trees.}
\end{figure}
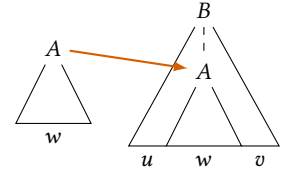
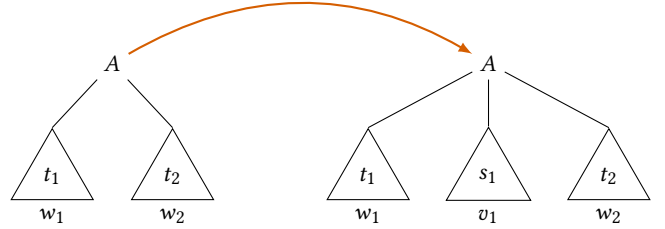

Suppose for the grammar $\Grammar$, we have derivation trees $\tau = \tree{(A \to \rho_0)}{\ldots}$ and $\pi = \tree{(B \to \rho_1)}{\ldots}$ and that $\tau \runembed \pi$. Additional letters in the output of $\pi$ can come from one of two sources: From the mapping of $\tau$ to a subtree of $\pi$ (\autoref{fig:subtree-mapping}) or from the image of $\tau$ in $\pi$ being explained by a larger run (\autoref{fig:underlying-run-interleaving}). Separating these two sources motivates the canonical decomposition for derivation trees.

\begin{restatable}{definition}{algextcanondecomp}
	Assume $\tau = \tree{(A \to \rho)}{t_1, \ldots, t_n}$ and $\can{\rho} = u_0 X_1 u_1 \cdots X_n u_n$. We define $\can{\tau} = \epsilon \cdot u_0 \cdot \epsilon \cdot \can{t_1} \cdot \epsilon \cdot u_1 \cdots u_{n-1} \cdot \epsilon \cdot \can{t_n} \cdot \epsilon \cdot u_n \cdot \epsilon$. 
	
	Intuitively, we wrap the canonical decomposition of $\tau$ itself and of each child $t_i$ in $\epsilon$ on either side to delimit gap words produced by a mapping of $t_i$ to a non-trivial descendant from those obtained by runs larger than $\rho$ in the image of $\tau$.
\end{restatable}

This also yields the expected definition of $\yield{\cdot}$: If $\tau$ consists just of a leaf node, then we define $\yield{\tau} = \yield{\rho}$; if $\tau$ is not just a leaf node, then we set $\yield{\tau} = u_0 \yield{t_1} u_1 \cdots \yield{t_n} u_n$. We write $\cT(\Grammar)$
for the set of all the trees of~$\Grammar$, and
$R_\Grammar$ for the $S$-rooted ones.  Then
$L(\Grammar)=\bigcup_{\rho\in R_\Grammar}\yield\rho$ as desired.

The embedding between trees is similar to the one we used for
context-free grammars in \cref{par:example-cfgs}, but needs to be
generalised slightly: when mapping $\tau_1 = \tree{(A \to
  \rho)}{\dots}$ to $\subtree{\tau_2}{p} = \tree{(A \to
  \sigma)}{\dots}$, we require that $\rho$ embeds into $\sigma$ such
that the corresponding subtrees also embed.  Formally, let $\tau_1$ and $\tau_2$ be trees from
$\cT(\Grammar)$.  Denote the run embeddings between runs of the
various systems $\{\Model_A\}_{A \in N}$ by $\preceq$.
We write $\tau_1 \runembed \tau_2$ if there exists a subtree
$\subtree{\tau_2}{p}$ such
that
\begin{restatable}{enumerate}{algextembed}
	\item\label{algextea} $\tau_1 =
		\tree{(A \to \rho)}{t_1, \ldots, t_n}$, $\subtree{\tau_2}{p} = \tree{(A \to \sigma)}{t'_1, \ldots, t'_k}$ and $\rho \embedwith[\preceq]{f} \sigma$, and
	\item\label{algexteb} $t_i \runembed t'_{g(i)}$ for all $i\in[1,n]$, where $g = \memo_\sigma^{-1} \circ f \circ \memo_\rho$.
\end{restatable}

For the details of the following statements that together show \cref{thm:alg-amalgamation}, we refer to \citeFullOrAppendix{appendix:alg-ext}.

\begin{restatable}{lemma}{algextwqo}\label{lem:algextwqo}
	$(\cT(\Grammar), \runembed)$ is a well-quasi-order.
\end{restatable}
\vspace{-.5em}
\begin{restatable}{lemma}{algextwqodeco}
	$\Grammar$ supports well-quasi-ordered decorations.
\end{restatable}

\vspace{-.5em}
\begin{restatable}{lemma}{algextcomposes}
	If $\tau_0 \embedwith{f} \tau_1$ and $\tau_1 \embedwith{g} \tau_2$, then $\tau_0 \embedwith{g \circ f} \tau_2$.
\end{restatable}

\vspace{-.5em}
\begin{restatable}{lemma}{algextconcalg}
	\label{lem:algext-concat-amalgamation}
	If $\tau_0, \tau_1, \tau_2$ are all $A$-rooted trees such that $\tau_0 \embedwith{f} \tau_1$ and $\tau_0 \embedwith{g} \tau_2$, then for every choice of $i \in [0, \canonlen{\tau_0}]$ there exists an $A$-rooted tree $\tau_3$ with $\tau_1 \embedwith{f'} \tau_3$ and $\tau_2 \embedwith{g'} \tau_3$ such that
	
	\begin{enumerate}
		\item $f' \circ f = g' \circ g$ (we write $h$ for this composition),
		\item $\gap{j}{h} \in \{\gap{j}{f},\gap{j}{g}, \gap{j}{g}\gap{j}{f}\}$ for every $j \in [0, \canonlen{\tau_0}]$, and in particular
		\item $\gap{i}{h} = \gap{i}{f}\gap{i}{g}$ for the chosen $i$. 
	\end{enumerate}
\end{restatable}

\subsection{Valence Automata}\label{sub:va}
In this subsection, we give a rough sketch of why, armed with the two
aforementioned mechanisms of \emph{adding counters} and
\emph{algebraic extensions}, we cover the entire class of valence
automata over graph monoids, except for those where
Turing-completeness is known.

In a \emph{valence automaton}, we have finitely many control states,
and a (potentially infinite-state) storage mechanism that is specified
by a monoid.  In the framework of valence automata
over graph monoids (see~\cite{DBLP:phd/dnb/Zetzsche16,DBLP:conf/rp/Zetzsche21} for overviews), one
considers valence automata where the monoid is defined by a finite
(undirected) graph~$\Gamma$ that may have self-loops.  The resulting
\emph{graph monoid} is denoted $\M\Gamma$.  The exact definition of
valence automata and graph monoids can be found in
\citeFullOrAppendix{appendix-valence-automata}, but will not be necessary for this
sketch.

In \cite[Thm.~3.1]{DBLP:journals/iandc/Zetzsche21}, it is shown that if $\Gamma$
contains certain induced subgraphs (essentially: a path on four nodes
or a cycle on four nodes), then valence automata over $\M\Gamma$
accept all recursively enumerable languages.  Thus, for
\cref{main-valence}, we may assume that these do not occur. Using arguments similar to~\cite[Lemma 5.9]{DBLP:journals/iandc/Zetzsche21}, one can observe that all the
\emph{remaining} monoids $\M\Gamma$ belong to the class $\PD$ (for
``potentially decidable''), where $\PD$ (i)~contains the trivial
monoid $1$ and (ii)~for all monoids $M$, $N$ in $\PD$, the class $\PD$
also contains $M*N$, $M\times\B$, and $M\times\Z$.  Here, $\B$ is the
so-called \emph{bicyclic monoid}, which corresponds to a single
so-called ``partially blind,'' $\N$-valued, counter: valence automata over $\B$ are
essentially 1-dimensional VASS.  Moreover, $M*N$ denotes the free
product of monoids.  Without giving an exact definition, they can be
thought of as \emph{stacks}: valence automata over $M*N$ are
essentially valence automata with stacks, where each entry is a
configuration of the storage mechanisms described by $M$ or $N$.
Moreover, $\Z$ corresponds to a so-called ``blind,'' $\Z$-valued counter.

It follows from known results on valence automata that going from $M$
and $N$ to $M*N$ results in languages in the {algebraic extension}
$\Alg(\VA(M)\cup\VA(N))$, where $\VA(M)$ is the class of languages
accepted by valence automata over $M$.  Moreover, going from $M$ to
$M\times\B$ can be seen as \emph{adding a counter} as in our
definition of $\cC+\N$ for language classes $\cC$.  Valence automata
over $M\times\Z$ can be simulated by valence automata over
$M\times\B\times\B$, so that $M\times\Z$ can be treated similarly.  As
this exhausts all the potentially decidable graph monoids and $\VA(M)$
is always a full trio~\citep[Thm.~4.1]{DBLP:journals/tcs/FernauS02},
\cref{main-valence} follows; see \citeFullOrAppendix{appendix-valence-automata} for
more details.

\section{Conclusion}\label{sec:conclusion}
We hope that we have demonstrated the surprisingly flexible nature of
amalgamation systems.  Their structure is at once simple enough to be a fit for several computational models, and powerful enough to be able to answer a number of open problems.

We think that this approach merits further investigation. In particular, we are interested in the following questions:

\begin{enumerate}[(a)]
	\item Which other problems are decidable for amalgamation systems?
	\item Are there amalgamation systems that are not valence systems?
	\item Is there a natural, non-trivial class that subsumes
          amalgamation systems and their algorithmic properties?
\end{enumerate}

\subsubsection*{Complexity}
\newcommand{\compP}{\mathsf{P}}
Although we show that the problems
(\labelcref{sup-decidable})--(\labelcref{emptiness-decidable}) in
\cref{main-amalgamation} are inter-reducible, their complexity can differ
widely. The reductions to emptiness in \cref{sec:properties} are all logspace
many-one, assuming we can compute the image of a rational transduction in
logspace (this is true in all the models we work with here, see e.g.~\cite[Thm.
4.1]{DBLP:journals/tcs/FernauS02}). Thus, under this assumption, the problems
(\labelcref{sup-decidable})--(\labelcref{pdc-computable}) are at least as hard
as the emptiness problem. However, the problems (\labelcref{dc-computable}),(\labelcref{unary-effectively-regular}),(\labelcref{pdc-computable}) can be harder than emptiness:
For example, for context-free languages, emptiness is $\compP$-complete, whereas NFAs for (\labelcref{dc-computable}),(\labelcref{unary-effectively-regular}),(\labelcref{pdc-computable}) may be exponential-sized~\cite{DBLP:conf/lata/BachmeierLS15,anand_et_al_CONCUR}.
Another example is the class of coverability languages of VASS, which (as a subclass of the reachability languages of VASS) also supports amalgamation. Here, emptiness is $\EXPSPACE$-complete~\cite{DBLP:journals/tcs/Rackoff78,Lipton76}, whereas NFAs for (\labelcref{dc-computable}),(\labelcref{unary-effectively-regular}),(\labelcref{pdc-computable}) require Ackermannian size~\cite{DBLP:conf/mfcs/AtigMMS17}.

For some classes of infinite-state systems, the complexity of some of the
problems (\labelcref{sup-decidable})--(\labelcref{pdc-computable}) even remains
open, whereas the complexity of emptiness is known. For example, the
complexity of separability by piecewise testable languages is not known for
context-free languages, whereas emptiness is well-known to be $\compP$-complete.

\begin{acks}
\raisebox{-9pt}[0pt][0pt]{\includegraphics[height=.8cm]{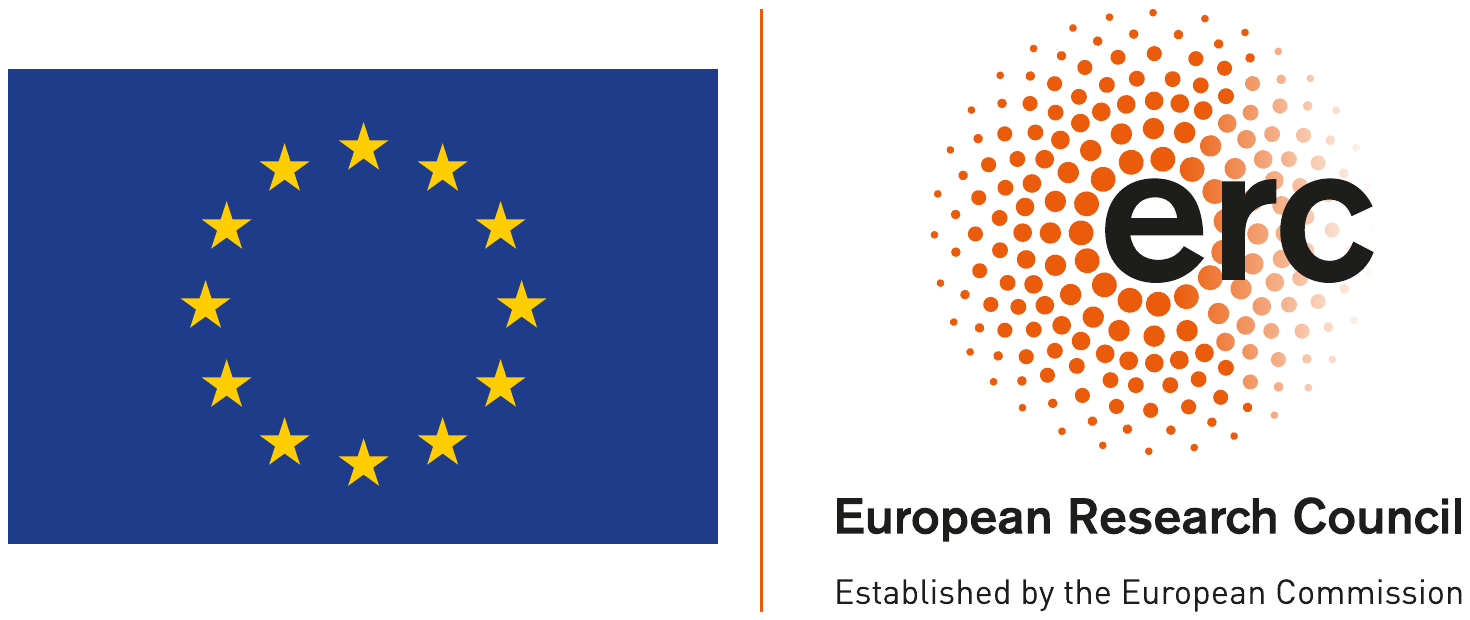}}
Funded by the European Union (\grantsponsor{501100000781}{ERC}{http://dx.doi.org/10.13039/501100000781}, FINABIS, \grantnum{501100000781}{101077902}) and \grantsponsor{DFG}{DFG}{https://www.dfg.de/} (Project~\grantnum{DFG}{389792660 TRR 248--CPEC}). %
Views and opinions expressed are however those of the authors only and do not necessarily reflect those of the European Union
or the European Research Council Executive Agency.
Neither the European Union nor the granting authority can be held responsible for them.

We are grateful to J\'{e}r\^{o}me Leroux for the discussion
about~\cite{Leroux2019}.  Moreover, we thank Pascal Baumann and Roland Meyer
for discussions about computing the set of minimal runs in VASS.
\end{acks}

\label{beforebibliography}
\newoutputstream{pages}
\openoutputfile{acm.pg}{pages}
\addtostream{pages}{\getpagerefnumber{beforebibliography}}
\closeoutputstream{pages}
\bibliographystyle{ACM-Reference-Format}
\bibliography{references}

\ifFull{%
	\clearpage
	\appendix\pagenumbering{roman}
	\setlength\textwidth{330pt}
	\setlength\oddsidemargin{61pt}
	\setlength\evensidemargin{61pt}
	\onecolumn
	\section{Minimal runs of VASS are not computable}\label{appendix:minimal-runs}
	\newcommand{\zero}{\mathsf{zero}}
In this appendix, we prove that there is no algorithm to compute a basis for
the set of runs of a (two-dimensional) VASS. Note that the run embedding for
VASS is a partial ordering, so there is always a finite set of minimal runs,
and computing a basis is equivalent to computing this set.

We use a reduction from reachability in
two-counter machines.  These are $2$-VASS with zero tests meaning, they have
two additional types of edge labels: $\zero_1$ and $\zero_2$, which test the
first, resp. second, counter for zero, with the obvious semantics.

Given a two-counter machine $(Q,q_0,\Delta,q_f)$, where $Q$
is the set of states, $q_0$ is the initial state, finite
transition set $\Delta\subseteq Q\times(\Z^2\cup
\{\zero_1,\zero_2\})\times Q$, and final state $q_f\in Q$,
it is well-known to be undecidable whether there is a run
from the configuration $(q_0,0,0)$ to $(q_f,0,0)$.

Given a two-counter machine as above, we define a $2$-VASS with state set $Q$,
initial state $q_0$, and final state $q_f$ as follows.  The input
alphabet is $\Sigma=\{z_1,z_2\}$ and it has the following transitions: 
\begin{align*}
&(p,\vec{u},\varepsilon,q)  & &\text{for every transition $(p,\vec{u},q)$ in $\cA$} \\
&(p,(0,0),z_1,q)  & &\text{for every transition $(p,\zero_1,q)$ in $\cA$} \\
&(p,(0,0),z_2,q)  & &\text{for every transition $(p,\zero_2,q)$ in $\cA$}.
\end{align*}
Thus, the $2$-VASS entirely ignores the zero tests, but it reads a letter $z_i$
whenever the two-counter machine performs a zero test on counter $i$.  We say
that a run of the $2$-VASS is \emph{faithful} if all the transitions labelled
$z_1$ or $z_2$ are actually executed in a configuration where the first, resp.\
second counter is zero.  Then clearly, the two-counter machine has a run if and
only if our $2$-VASS has a faithful run.

\begin{claim}
The $2$-VASS has a faithful run if and only if one of its minimal runs is faithful. 
\end{claim}

If this is shown, it clearly follows that the minimal runs are not computable:
Otherwise, we could compute them and check if one of them is faithful. The
``if'' direction is trivial, so suppose none of the minimal runs is faithful.
Then each of the minimal runs contains a step of one of the following forms: 
\begin{align*}
&((x_1,x_2),t,(x_1,x_2)) & &\text{where $t$ is labelled by $z_1$ and $x_1>0$, or} \\
&((x_1,x_2),t,(x_1,x_2)) & &\text{where $t$ is labelled by $z_2$ and $x_2>0$}
\end{align*}
However, this implies that \emph{every} run of our $2$-VASS contains such a
step. In particular, none of them can be faithful.

	\section{Results on priority downward closures}\label{appendix-priority-order}

In this appendix, we prove the results about priority downward closures.

\subsection{Overview}
For the implication \impl{emptiness-decidable}{pdc-computable} of \cref{main-amalgamation}, we show that if emptiness is decidable, we can compute priority downward
closures. To compute the priority downward closure of an input language $ L $, we need to show that $ L\pd = D $ for some downward closure regular language $ D $. The difficult part of this computation is to decide whether our input language $L$
satisfies $D\subseteq L\pd$.
Our algorithm uses a strategy from \cite{DBLP:conf/icalp/Zetzsche15}, namely to 
decompose $D$ into ideals. 
Somewhat more precisely, we do the above ideal decompositions not for the general priority ordering, but for the simple block order, which we define now.

\subsubsection*{The simple block order}
To strip away some technicalities, we will work with a slightly different ordering for which downward closure computation is an equivalent problem. For each $d\in\N$, we define the alphabet $\Sigma_d=[0,d]$ and the \emph{simple block ordering} over $\Sigma_d^*$. We think of $\Sigma_d$ as an alphabet with priorities $[0,d]$, except that there is only one letter of each priority. If $u,v\in\Sigma_d^*$ and $m\in\Sigma_d$, $m>0$, is the largest letter occurring in $u$ and $v$, then we define $u\so v$ if and only if
\[ u=u_0mu_1\cdots mu_k,~~v=v_0mv_1\cdots mv_{\ell} \]
with $u_0,\ldots,u_k,v_0,\ldots,v_\ell\in\Sigma_{m-1}^*$
and there is a strictly monotone map $\varphi\colon[0,k]\to[0,\ell]$ with $u_i\so v_{\varphi(i)}$ for every $i\in[0,k]$. (In particular, $k=0$ means that $m$ does not occur in $u$) Thus, $\so$ is defined recursively w.r.t.\ the occurring priorities. To cover the base case, if $u,v\in 0^*$, then we simply have $u\so v$ if and only if $|u|\le|v|$. 
For a word $ w_1mw_2m\cdots mw_nm $, we call the words $ w_i $ which are enclosed between two consecutive $ m $'s, $ m $-blocks. Thus intuitively, $ u $ is \emph{simple block smaller} than $ v $, if on splitting both words along the highest priority letter $ m $, the $ m $ blocks from $ u $ are monotonically and recursively simple block smaller than those from $ v $. 
\begin{example}\label{ex:block-order}
	For $d=1$, we have $ 0\so 00\not\so 010\so 1010 $, but $ 010\not\so 1010 $.
\end{example}
Then for $L\subseteq\Sigma_d^*$, we define as above
\[ L\sd = \{u\in\Sigma_d^* \mid \exists v\in L\colon u\so v\}. \]
The task of computing an NFA for $L\sd$ for a given language $L$ is called \emph{computing simple block downward closures}.  

\newcommand{\out}[1]{#1_{out}}
\newcommand{\jump}[1]{#1_{skip}}
In the following, we show that the simple block order is a rational relation.
\begin{lemma}
	Let $ \Sigma = [0,d] $, a rational transducer $ T $ can be constructed in polynomial time, such that for every language $ L\subseteq \Sigma^* $, $ L\sd = TL $.
\end{lemma}
\begin{proof}	
	For the simple block order consider the transducer that has two states $ \out{l} $  and $ \jump{l} $ for every letter $ l $, %
	and a sink state $ t $, and for every state $ p $ it reads a letter $ l $, and for every letter $ l $, 
	
	\begin{itemize}
		\item if the state is $ \out{l} $, it represents highest last outputted letter was $ l $. From this state, there are transitions,
		\begin{itemize}
			\item if the input letter is $ l $, it stays at $ l $, on outputting and skipping the letter.
			\item otherwise, on reading a letter $ p $, goes to $ \out{p} $ and $ \jump{p} $, respectively on outputting and skipping the letter
		\end{itemize}
		
		\item if the state is $ \jump{l} $, it represents highest last skipped letter was $ l $. From this state, there are transitions,
		\begin{itemize}
			\item if the input letter is smaller, it stays at $ \jump{l} $ on skipping, and goes to $ t $ on outputting the letter.
			\item if the input letter is $ l $, then stays at $ \jump{l} $ on skipping, and goes to $ \out{l} $ on outputting the letter.
			\item if the input letter $ p $ is greater, it goes to $ \jump{p} $ on skipping, and goes to $ \out{p} $ on outputting the letter.
		\end{itemize}
	\end{itemize}
	
	The starting state is $ \jump{0} $, and any run that does not end at $ t $ is accepting.
	
	Intuitively, the transducer makes sure that on skipping a letter $ l $, every subsequent lower letter is skipped until a letter equal or greater is output. This ensures that two $ l $-blocks are not merged, i.e., between two consecutive letters which are not dropped, no bigger priority letter is dropped.
	
	We argue that the transducer restricted to the state set $ S_k =  \{\out{l},\jump{l}|l\leq k\}\cup \{t\} $ outputs the set of words smaller than a word $ w\in \Sigma_k $.
	
	The base case is trivial, as there will be no edge to the sink state. Now suppose that the claim holds for some $ S_{k-1} $. Then for $ S_{k} $, let $ u\in \Sigma_{k} $. Then for any word $ v $ which is small block smaller than $ u $, the $ (k-1) $-blocks in $ v $ that map to that of $ u $, can be recognized by $ S_{k-1} $, and every $ (k-1) $-block that is skipped can be skipped by $ S_{k-1} $, and after skipping, outputting another $ (k-1) $-block can happen via $ \out{k} $ by  outputting a $ k $. 
	
	For any word $ v $ that is not simple block smaller than $ u $, then consider the first three consecutive letters in $ u $, $ xyz $, such that $ x,z<y $, $ x,z $ are output and $ y $ is not. Then the run will reach $ \jump{y} $ and next output letter will be smaller, leading to the sink state, hence the transducer will not output $ v $.
\end{proof}

\begin{restatable}{lemma}{blockOrderEquivalence}\label{blockOrderEquivalence}
	If $\mathcal{C}$ is a full trio, then priority downward closures are
computable for $\mathcal{C}$ if and only if simple block downward closures are
computable for $\mathcal{C}$. 
\end{restatable}
We will show this in \cref{sec:appendix-priority-order-proofs}.  In \cite{anand_et_al_CONCUR},
the authors also introduce a ``block order'' (slightly different from our
simple block order) and also show that downward closure computation of it is
equivalent to that for the priority order. 
\newcommand{\Down}{\mathsf{Down}}
\newcommand{\Atom}{\mathsf{Atom}}
\newcommand{\Ideal}{\mathsf{Ideal}}
\subsubsection*{Ideals}
An \emph{ideal} in a WQO $(X,\le)$ is a downward closed set
$I\subseteq X$ where for any $u,v\in I$, there is a $w\in I$ with $u\le w$ and $v\le w$. What makes these useful is that in a WQO, every downward closed set can be written as a finite union of ideals. Moreover, we will see that establishing $I\subseteq L\sd$ for the language $L$ of an amalgamation system can be done by enumerating runs. To this end, we rely on a syntax for specifying ideals.
\begin{restatable}{lemma}{blockIdeal}\label{lemma:blockIdeals}
	The ideals of $(\Sigma_d,\so)$ are precisely the sets in $\Ideal_d$, where
	\begin{align*}
		\Atom_d&=\{Id\cup \epsilon\mid I\in \Ideal_{d-1}\}\cup  \{(Dd)^*\mid D\in \Down_{d-1}\},\\	
		\Ideal_0&= \{0^n \mid n\in\N\}\cup \{0^*\},\\	
		\Ideal_d&= \{ X_1\cdots X_nA\mid X_i\in \Atom_d,~A\in \Ideal_{d-1}  \},
	\end{align*} 
	and $\Down_d$ is the set of all downward closed subsets of $\Sigma_d^*$ with respect to simple block order.
\end{restatable}
See \cref{sec:appendix-priority-order-proofs} for a proof. Now an algorithm for computing
the simple block downward closure of a language $L\subseteq\Sigma_d^*$ can do the
following. It enumerates finite unions $I_1\cup\cdots\cup I_n$ of ideals
$I_1,\ldots,I_n$. For each such union, it checks two inclusions:
$L\sd\subseteq I_1\cup\cdots\cup I_n$ and $I_1\cup\cdots\cup I_n\subseteq
L\sd$. The former inclusion is easy to check: Since $I_1\cup\cdots\cup I_n$ is
a regular language, we can just use decidable emptiness and closure under
regular intersection to check whether $L\sd\cap
\Sigma_d^*\setminus(I_1\cup\cdots\cup I_n)=\emptyset$. The inclusion
$I_1\cup\cdots\cup I_n\subseteq L\sd$ is significantly harder to establish,
and it will be the focus of the remainder of this subsection.

\subsubsection*{Pseudo-ideals}
First, observe that it suffices to decide $I\subseteq L\sd$ for an individual
ideal $I$.  As the second step, we will simplify ideals even further to pseudo-ideals. A \emph{pseudo-ideal of priority $0$} is an ideal of the form $0^{\le
n}$ or $0^*$. A \emph{pseudo-ideal of priority $d>0$} is an ideal of the form
$(Id)^*$ or $I_1d\cdots I_nd$, where $I$ and $I_1,\ldots,I_n$ are pseudo-ideals of
priority $d-1$. Thus, intuitively, we rule out subterms of the form $(Dd)^*$ with some
downward-closed $D\subseteq\Sigma_{d-1}^*$. Despite being less expressive, deciding inclusion of pseudo-ideals is sufficient for inclusion of arbitrary ideals:
\begin{restatable}{proposition}{reductionToSimpleIdeals}\label{reduction-to-simple-ideals}
Given an ideal $I\subseteq\Sigma_d^*$ and a class $ \mathcal{C} $ of languages closed under rational transduction, we can construct finitely many pseudo-ideals
$J_j\subseteq\Sigma^*$ and a rational transduction $T$ such that for any
language $ L\in\mathcal{C} $ and $L\subseteq\Sigma_d^* $, we have $I\subseteq L\downarrow_S$ if and only if $~\cup_jJ_j\subseteq TL\downarrow_S$.
\end{restatable}
This implies that if we have an algorithm to decide $J\subseteq L\sd$ for
pseudo-ideals $J$, then this can even be done for arbitrary ideals.
Essentially, the idea is to emulate subterms $I=(Dd)^*$ with
$D=I_1\cup\cdots\cup I_n$ by a new term $J=((I_1d\cdots I_nd)e)^*$, where $e>d$
is a fresh priority. The transduction modifies the words in $L$ so that after
every occurrence of $d$, an occurrence of $e$ is potentially inserted. Note
that then, in order for $J$ to be included, each of the ideals $I_1,\ldots,I_n$
need to occur arbitrarily often, which corresponds to inclusion of $I=(Dd)^*$.
We will show this in \cref{sec:appendix-priority-order-proofs}.

\subsubsection*{Ideal inclusion via amalgamation}
Let us now see how to establish an inclusion $I\subseteq L\sd$ for pseudo-ideals $I$ using run embeddings. We begin with an example. Suppose we want to verify that the ideal $(0^*1)^*$ in included in our language. Here, we need to check if for every $k\ge 0$, there is a word with $\ge k$ factors, each containing $\ge k$ contiguous $0$'s, and they are separated by $1$'s. Intuitively, this is more complicated that the SUP and a proof using e.g.\ grammars seems involved. However, using run amalgamation, this amounts to checking a simple kind of witness: Namely, three runs $\rho_0\runembed\rho_1\runembed \rho_2$ such that (i)~some gap of $\rho_0\runembed\rho_1$ between some positions $i<j$ of $\rho_1$ contains a $1$ and (ii)~some gap of $\rho_1\runembed\rho_2$, which is between $i$ and $j$ in $\rho_1$, belongs to $0^+$. Notice that then, by amalgamating $\rho_2$ with itself above $\rho_1$ again and again, we can create arbitrarily long blocks of contiguous $0$'s. The resulting run $\rho_2'$ still embeds $\rho_1$ and thus $\rho_0$, such that one gap of $\rho'_2$ in $\rho_0$ contains both our long block of $0$'s and also a $1$. This means, if we amalgamate $\rho'_2$ again and again above $\rho_0$, we obtain arbitrarily many blocks of our long $0$-blocks. This yields runs that cover all words in $(0^*1)^*$.

\subsubsection*{Witnesses for ideal inclusion}
We will now see how inclusions $I\subseteq L\sd$ can always be verified by such run constellations, which we call ``$I$-witnesses''.
\newcommand{\witness}{\mathcal{W}}
For a word $ w\in \Sigma\cup \{\epsilon\} $, $ w|_{\Sigma} $ denotes the restriction of $ w $ over $ \Sigma $. We call a word $ w' $ a factor of $ w = w_1\cdots w_n\in \Sigma\cup\{\epsilon\} $, if $ w|_{\Sigma}\in \Sigma^* w'\Sigma^* $, i.e., $ w' $ is an infix of restriction of $ w $ over $ \Sigma $. For a word $ w= w_0w_1\cdots w_n $, by $ w[i,j] $ we denote the word $ w_i\cdots w_j $ for $ i<j $.

A finite subset $ \witness $ of runs $ R $ is called an $ I $-witness if%

\begin{itemize}
	\item for $ I = 0^{\leq k} $ for some $ k\in \N $, there exists is a run $ \rho\in\witness $ such that $ 0^k $ is a factor of $ \can{\rho}[\overleftarrow{l}, \overrightarrow{l}] $, for some $ 0 \leq \overleftarrow{l}, \overrightarrow{l}\leq  \canonlen{\rho} $. 
	
	Then $ \rho $ is said to witness $ I $ between $ \overleftarrow{l} $ and $ \overrightarrow{l}$.
	
	\item for $ I=0^* $, there exist runs $ \rho,\psi\in \witness $ such that $ \psi\runembed_f \rho $ and  $ \gap{i}{f} = 0^l $ for some $ l>0 $, $ i\in[0,\canonlen{\psi}]$ and $ f\in E(\psi, \rho) $. 
	
	Then $ \rho $ is said to witness $ I $ between $ \overleftarrow{l} $ and $ \overrightarrow{l}$, if $ \gap{i}{f} $ is a factor of $ $ is a factor of $ \can{\rho}[\overleftarrow{l}, \overrightarrow{l}] $.
	
	\item for $ I = I_1a\cdots aI_na $, there exists a run $ \rho\in \witness $ and $ \overleftarrow{l}=l_0\leq l_1<l_2<\cdots <l_n \leq \overrightarrow{l} $ such that $ \rho $ is an $ I_i $-witness between $ l_{i-1} $ and $ l_i $, and $ \can{\rho}[\overleftarrow{l}, \overrightarrow{l}] \in\s_a^*$.
	
	Then $ \rho $ is said to witness $ I $ between $ \overleftarrow{l} $ and $ \overrightarrow{l}$.
	
	\item for $ I = (I'a)^* $, there exist runs $ \rho, \psi\in\witness $ such that 
	$ \psi\runembed_f \rho $ and $ \rho $ is a witness for $ (I'a)^l $ between $ f(i) $ and $ f(i+1) $, and $ \rho[f(i), f(i+1)]\in \Sigma_{a}^* $, for some $ l>0 $, $ i\in[0,\canonlen{\psi}-1]$ and $ f\in E(\psi, \rho) $.
	
	Then $ \rho $ is said to witness $ I $ between $ \overleftarrow{l} $ and $ \overrightarrow{l}$, if $ \gap{i}{f} $ is a factor of $ \can{\rho}[\overleftarrow{l}, \overrightarrow{l}] $.
	
\end{itemize}
With this notion of $I$-witnesses, we can prove:
\begin{restatable}{lemma}{containementOfIdeal}\label{lemma:containmentOfIdeal}
	For every pseudo-ideal $I$ and every amalgamation system for $L\sd$, we have $ I\subseteq L\sd$ if and only if the system possesses an $ I $-witness.
\end{restatable}
To illustrate the proof idea, let us see how to show that there is always a
witness for $I=(0^*1)^*$.  Suppose we have an amalgamation system $S$ for
$L\sd$ and we know $I\subseteq L\sd$. Let $M$ be the maximal number of factors
in the canonical decomposition of minimal runs of $S$. Consider the sequence
$w_1,w_2,\ldots$ with $w_i=(0^i1)^{2M}$ for $i\ge 1$. Then $w_1,w_2,\ldots\in
I\subseteq L\sd$, so there must be runs $\rho_1$ and $\rho_2$ with
$\rho_1\runembed \rho_2$ with $\can{\rho_1}=w_i$ and $\can{\rho_2}=w_j$.
Then clearly, every non-empty gap of $\rho_2$ in $\rho_1$ belongs to $0^+$.
Moreover, any embedding of a minimal run $\rho_0$ of $S$ into $\rho_1$ will
have some gap containing two $1$'s, and thus have $10^i1$ as a factor. Then
the runs $\rho_0,\rho_1,\rho_2$ clearly constitute an $I$-witness.
We will show this in \cref{sec:appendix-priority-order-proofs}.

\subsubsection*{The algorithm} 
We now have all the ingredients to show that decidable emptiness in a class of amalgamation systems that forms a full trio implies computable priority downward closures. First, by \cref{blockOrderEquivalence}, it suffices to compute $L\sd$ for a given language $L$. Second, according to \cref{reduction-to-simple-ideals} and \cref{lemma:containmentOfIdeal}, deciding whether $I\subseteq L\sd$ for an ideal $I$ is recursively enumerable. Therefore, we proceed as follows. We enumerate all finite unions $I_1\cup\cdots\cup I_n$ of ideals and try to establish the inclusion $I_1\cup\cdots\cup I_n\subseteq L\sd$. Once we find such a finite union where inclusion holds, we check whether $L\sd\subseteq I_1\cup\cdots\cup I_n$. The latter is decidable: By closure under rational transductions, we can construct an amalgamation system for $L\sd\cap (\Sigma^*\setminus(I_1\cup\cdots\cup I_n))$ and check it for emptiness. Since we know that for every downward-closed set, there exists a finite union of ideals, our algorithm will eventually discover a finite union $I_1\cup\cdots\cup I_n$ with $L\sd=I_1\cup\cdots\cup I_n$.

Finally, note that \impl{pdc-computable}{emptiness-decidable} holds as well,
because $L\ne\emptyset$ if and only if $L\pd\ne\emptyset$, meaning we decide
emptiness of $L$ by computing an NFA for $L\pd$ and check that for emptiness.

\subsection{Detailed proofs}\label{sec:appendix-priority-order-proofs}
\subsubsection{Equivalence of Simple Block Order and Priority Order}

\blockOrderEquivalence*
\renewcommand{\so}{\preceq}
\begin{proof}
	In ~\cite{anand_et_al_CONCUR}, it was shown that priority downward closure can be computed if and only if block downward closure can be computed, where we say $ u\bo v $, if
	\begin{enumerate}[i.]
		\item if $ u,v\in \equalletters{p}^* $, and $ u\so v $ ($ u $ is subword smaller than $ v $), or
		\item if 
		\begin{eqnarray*}
			u&=& u_0x_0u_1x_1\cdots x_{n-1}u_{n}\\
			\text{and, }v&=& v_0y_0v_1y_1\cdots y_{m-1}v_{m}
		\end{eqnarray*}
		where $  x_0,\ldots x_{n-1},y_0,\ldots, y_{m-1} \in \equalletters{p}$, and for all $i\in[0,n]$, we have $ u_i,v_i\in \lowerletters{p-1}^*$ (the $u_i$ and $v_i$ are called \emph{$ p $} blocks), and there exists a strictly monotonically increasing map $ \phi:[0,n]\rightarrow [0,m] $, which we call the \emph{witness block map}, such that 
		\begin{enumerate}
			\item $ u_i\bo v_{\phi(i)} $, $ \forall i $,\label{def:generalized is recursive}
			\item $ \phi(0)= 0 $, \label{def:first-to-first}
			\item $ \phi(n) =m $, and\label{def:last-to-last}
			\item $ x_i\so v_{\phi(i)}y_{\phi(i)}v_{\phi(i)+1}\cdots v_{\phi(i+1)} $, $ \forall i \in [0,n-1] $.\label{def:generalized is subword}
		\end{enumerate}
	\end{enumerate}
	
	Intuitively, we say that $ u $ is \emph{block smaller} than $ v $, if either
	\begin{itemize}
		\item both words have letters of same priority, and $ u $ is a subword of $ v $, or,
		\item the largest priority occurring in both words is $ p $. Then we split both words along the priority $ p $ letters, to obtain sequences of $ p $ blocks of words, which have words of strictly less priority. Then by item \ref{def:generalized is recursive}, we embed the $ p $ blocks of $ u $ to those of $ v $, such that they are recursively block smaller. Then with items \ref{def:first-to-first} and \ref{def:last-to-last}, we ensure that the first (and last) $ p $ block of $ u $ is embedded in the first (resp., last) $ p $ block of $ v $. We will see later that this constraint allows the order to be multiplicative. Finally, by item \ref{def:generalized is subword}, we ensure that the letters of priority $ p $ in $ u $ are preserved in $ v $, i.e. every $ x_i $ indeed occurs between the embeddings of the $ p $ block $ u_{i} $ and $ u_{i+1} $.
	\end{itemize}

	Then we now show that block downward closures can be computed if and only if simple block downward closures can be computed.
	
	Since $ u\bo v\implies u\sbo v $, it is trivial that $ L\downarrow_S $ is computable if $ L\downarrow_B $ is computable.
	
	For the other direction, suppose that the alphabet is $ \Sigma =  [0,d] $. Then we consider a new alphabet $ \Sigma' = \{0,0', 0^{\epsilon},\ldots d,d', d^{\epsilon}\} $ such that $ 0<0' < 0^{\epsilon}<1<1'<1^{\epsilon}<\dots <d<d'<d^{\epsilon} $. By $ \new $, we denote $ \Sigma'\backslash \Sigma $. Let for a word $ w = w_1aw_2a\cdots a w_n $, where $ w_i\in \Sigma_{a-1}^* $ and $ a $ is the highest priority letter in $ w $, by $ border(w) $ we define the word $ b_1\cdot border(w_1)\cdot b_1\cdot  a\cdot  border(w_2)\cdot a\cdots a\cdot  border(w_n-1)\cdot a\cdot  b_n\cdot  border(w_n)\cdot  b_n$, where $ b_1 $ (and $ b_n $) is $ (a-1)^\epsilon $ if $ w_1=\epsilon $ (resp., $ w_n = \epsilon$), else it is $ (a-1)' $. Moreover, $ border(w) = w $ if $ w\in 0^* $. Intuitively, we bound the first and the last $ a $ blocks of the word recursively, and we also distinguish whether these blocks are $\epsilon$ or not.
	
	Let $ border(L) = \{border(w)\mid w\in L\} $. Then we claim that $ L\downarrow_B = (border(L)\downarrow_S \cap~ \expression_d )|_{\Sigma} $, where $ \expression_d $ is defined recursively as follows.
	\begin{eqnarray*}
		\expression_0 =& 0^*\\
		\expression_a =& (a^\epsilon \epsilon a^\epsilon + a' \expression_{a-1} a')\cdot (a\expression_{a-1})^*a\cdot (a^\epsilon \epsilon a^\epsilon + a' \expression_{a-1} a')
	\end{eqnarray*}

	Note that $ L' =border(L)\downarrow_S \cap~\expression_d $ is the language of simple block smaller words where $ 0',0^\epsilon\cdots d',d^\epsilon $ are not dropped. Furthermore, the $ L|_\Sigma $ is restriction of words in $ L $ to letters in $ \Sigma $.
	
	Now we prove the claim. For one direction suppose $ u\in L\downarrow_B $. Then there exists a word $ v\in L $ such that $ u\bo v $. It suffices to show by induction on the maximum priority letter $ d $ in $ u $ and $ v $ (which has to be the same by definition of block order) that $ border(u)\sbo border(v)$. For the base case, i.e. $ d=0 $, since $ border(u)=u $ for any $ u\in 0^* $, this trivially holds.
	
	Then for the induction step, let the statement holds true for some $ d-1\in\N $. Then assuming $ u = u_1du_2\cdots d u_k $ and $ v = v_1 d v_2\cdots dv_l  $, there exists a $ \bo $ witness $ \phi:[1,k]\rightarrow [1,l]$, with $ \phi(1) =1 $ and $ \phi(k) =l $. Then we show that $ \phi $ is also a witness for $ border(u)\sbo border(v) $. Let  $ border(u) = u'_1du'_2\cdots d u'_k $ and $ border(v) = v'_1 d v'_2\cdots dv'_l  $. Induction hypothesis immediately implies that $ u'_i\sbo v'_{\phi(i)} $ for $ i\in [2;k-1] $. We argue that $ u'_1\sbo v'_{\phi(1)} = v'_1 $ (argument is analogous for the last $ d $ block). If $ u'_1 = d^\epsilon \epsilon d^\epsilon $, then $ v'_1 = d^\epsilon \epsilon d^\epsilon $, since $ \epsilon$ is only block smaller than $ \epsilon$. Then clearly, $ u'_1\sbo v'_1 $. Now if $ u'_1 = d' x d' $, where $ x\in\expression_{a-1} $, then $ v'_1=  d'yd'$, where $ y\in\expression_{a-1} $. Again by induction hypothesis, $ x\sbo y $, and hence $ u'_1\sbo v'_1 $. Hence, $ border(u)\sbo border(v) $.

	Now for the other direction, we suppose that $ u\in border(L)\downarrow_S \cap~\expression_d $. Then there exist $ u'\in \Sigma^*, $ and $v, v'\in \Sigma'^{*} $, such that 
	\begin{itemize}
		\item $ v'=border(u') $, $ v = border(u) $
		\item $ v\sbo v' $, and
		\item $ v'|_{\new} = v|_{\new} $, i.e. no bordering letters are dropped.
	\end{itemize}
	Then we need to show that $ u\bo u' $: then since $ u'\in L $ (by definition of $ u $) implying that $ u\in L\downarrow_B $. In the rest of the proof we will show that $ u'\bo u $. Since $ v\sbo v' $, there exists a witness $ \phi $. We show that $ \phi $ is a witness for $ u\bo u $, by induction on $ d $. The base case is trivial as $ border(x) = x $ when $ x\in 0^* $. Now, for the induction step let the highest letter in $ u $ and $ u' $ is $ d $.
	
	We show that the first (analogously, last) $ d $ block of $ v $ maps to that of $ v' $, i.e. $ \phi(1) = 1 $. Since $ v $ and $ v' $ are bordered words, the borders $ d' $ and $ d^\epsilon $ are added in the first and the last blocks of both words. Then the first block of $ v $ must be mapped to that of $ v' $ to map these borders, i.e. $ \phi(1) =1 $. Then assuming $ v= v_1d\cdots d v_k $ and $ v' = v'_1d\cdots dv'_l $, we have that $ v_1\sbo v'_1 $, then $ v_1|_\Sigma\bo v'_1|_\Sigma $. Similarly, $ v_k|_\Sigma\bo v'_l|_\Sigma $. For other block, the induction hypothesis, immediately implies $ v_i\bo v_{\phi(i)} $. Then $ \phi $ is a witness for $ u\bo u' $. Hence, $ u\in L\downarrow_B $. This completes the proof of the lemma.
\end{proof}

\subsubsection{Simple Block Order Ideals}

\paragraph{Upward closure} Let $ (X,\leq) $ be a WQO. Let $ Y\subseteq X $, then upward closure of $ Y $, denoted as $ Y\uparrow_{\leq} $, is defined as the set of all the elements in $ X $ which are bigger than an element in $ Y $, i.e.,
$ Y\uparrow_{\leq}~:= \{ x\in X | \exists y\in Y, y\leq x  \} $. For the purpose of this appendix, we will mean $ Y\uparrow_{\leq} $ with $ Y\uparrow $.

We now show that the downward closed sets are finite union of ideals which are defined as follows.
\paragraph{Ideals} Let $ (X,\leq) $ be a WQO. A subset $ I $ of $ X $ is called an ideal, if it is
\begin{itemize}
	\item downward closed, i.e. if $ u\in I $ and $ v\leq u $, then $ v\in I $, and
	\item up-directed, i.e. $ \forall\ u,v\in I, \exists\ z\in I $ such that $ u\leq z $ and $ v\leq z $.
\end{itemize}

\newcommand{\calj}{\ensuremath{\mathcal{J}}}
\newcommand{\bs}{\backslash}

\begin{lemma}\label{idealSet}
	Let $ \calj $ be a subset of ideals, such that 
	\begin{enumerate}
		\setlength\itemsep{0em}
		\item the complement of any filter can be written as a finite union of ideals from $ \calj $, and
		\item the intersection of any two ideals of $ \calj $ is a finite union of ideals from $ \calj $.
	\end{enumerate} 
	Then $ \calj $ is the set of all the ideals.	
\end{lemma}

\blockIdeal*
\begin{proof}
	Let $ Ideals_a $ be the set of all the ideals with respect to simple block order for alphabet $ \Sigma_a = [0,a] $. We first show that every set in $ Id_d $ is an ideal, i.e., $ Id_d \subseteq Ideals_d $. Observe that the elements of $ Id_0 $ are indeed ideals, because for singletons, the subword order and the simple block order coincide. So, now assume that $ Id_{a-1}=Ideals_{a-1} $.
	
	We now show that $ Id_a\subseteq Ideals_a $. Let $ X = (X_1\cdots X_nA)\in Id_a $. We need to show that $ X $ is downward closed and up-directed. 
	
	\paragraph*{Downward closed} Suppose $ u = u_1u_2\cdots u_nu_A\in X $ such that $ u_i\in X_i $ and $ u_A\in A $, and let $ v= v_1av_2a\cdots av_k $ with $ v
	\sbo u $. Then there exists a strictly monotonically increasing map $ \phi$ that maps $ a $ blocks of $ v $ to those of $ u $. Then consider the map $ \psi:[1,k]\rightarrow \{1,\ldots, n, A\} $ that maps each $ a $ block $ v_i $ to $ j $ if $ \phi(i) $-th $ a $ block of $ u $ lies in $ u_j $. Note that this map is well defined as each $ u_i $ terminates with an $ a $, so no $ a $ block of $ v $ can be mapped to a $ a $ block of $ u $ that splits between two $ u_i $'s. Then each $ v_ia\cdots a v_ja $ such that $ \psi(i)=\psi(j) = p $ is in $ X_p $, by induction hypothesis. Now suppose $ S = \{i|\psi(j) = i, \text{ for some } j\} $. Then $ v\in X_{i_1}\cdots X_{i_q}\subseteq X_1\cdots X_nA $, where $ S=\{i_1,\ldots i_q\} $.
	
	\paragraph*{Upward directed}  Let $ u, v \in X $. Suppose $ u=u_1\dots u_nu_A $ and $ v=v_1\dots v_nv_A $, such that $ u_i,v_i\in X_i $ and $ u_A,v_A\in A $. 
			Construct the word $ z=z_1\cdots z_nz_A $ as follows
			\begin{eqnarray*}
					z_i&=&\begin{cases}
							u_iv_i, \text{ if }  X_i=(Da)^* \\
							w_i, \text{ if }  X_i=Ia  \text{ and } u_i,v_i\bo w_i
						\end{cases}\\
					z_A&=& w_A, \text{ where } u_A,v_A\bo w_A
				\end{eqnarray*}
			It is again easy to notice that $ u,v\bo z $. Hence, $ X $ is up-directed.

	We have shown above that $ Id_d\subseteq Ideals_d $. To show that $ Id_d= Ideals_d $, we use the \cref{lemma:blockIdeals}, and show that $ Id_d $ satisfies the two preconditions of the lemma.
	
	\begin{itemize}
		\item (Complement of a filter is a finite union of $ Id_a $ ideals) Let $ u\in \Sigma_a^* $, and we need to show that $ \Sigma_a^*\bs u\uparrow $ is a finite union of ideals from $ Id_a $. The proof is by an induction on the number of $ a $ blocks in $ u $.
		
		When $ u $ has only 1 $ a $ block, $ u\in\Sigma_{a-1}^* $. Then $ \Sigma_a^*\bs u\uparrow= \Sigma_{a-1}^*\bs u\uparrow = \cup_{i=1}^k I_i $, where $ I_i\in Id_{a-1}\subseteq Id_a $ and $ k\in\N $. Hence, the base case holds. For the induction hypothesis, assume that the required result holds when $ u $ has $ n-1 $ $ a $ blocks. 
		
		Let $ u $ has $ n $ many $ a $ blocks. Then $ u $ can be written as $ vaw $ where $ v\in \Sigma_{a-1}^* $ and $ w $ has $ n-1 $ many $ a $ blocks.
		\begin{claim} \label{claim:equalityOfFilters}
			\begin{eqnarray*}
				\Sigma_a^*\bs\ vaw\uparrow &=& \left((\Sigma_{a-1}^*\bs\ v\uparrow)a\right)^*(\Sigma_{a-1}^*\bs\ v\uparrow)\\
				&& \cup \left(\left((\Sigma_{a-1}^*\bs\ v\uparrow)a\right)^* (\Sigma_{a-1}^*a)(\Sigma_{a}^*\bs\ w\uparrow)\right)\downarrow_S
			\end{eqnarray*}
		\end{claim}
		\begin{proof}[Proof of \cref{claim:equalityOfFilters}]
			$ (\subseteq) $ Let $ z\in \Sigma_a^*\bs\ vaw\uparrow $. Then $ u=vaw\not\sbo z $, and we have the following two cases,%
			\begin{enumerate}
				\item Case 1: If $ v $ can not be mapped to a $ a $ block of $ z $, then all the $ a $ blocks of $ z $ come from $ (\Sigma_{a-1}^*\bs\ v\uparrow) $. Hence,  \[ z\in \left((\Sigma_{a-1}^*\bs\ v\uparrow)a\right)^*(\Sigma_{a-1}^*\bs\ v\uparrow) \subseteq RHS . \]
				
				\item Case 2: If $ v $ can be mapped to a $ a $ block of $ z $ then let this be the $ i^{th} $ block of $ z=z_1az_2a\cdots az_k $. Hence, the first $ i-1 $ blocks come from $ (\Sigma_{a-1}^*\bs\ v\uparrow) $, and the $ i^{th} $ block comes from $ \Sigma_{a-1}^* $. This implies that $ z_1az_2a\cdots az_{i}a\in \left((\Sigma_{a-1}^*\bs\ v\uparrow)a\right)^* (\Sigma_{a-1}^*a) $. 
				
				Since $ u\not\sbo z $ and $ va\sbo z_1az_2a\cdots z_ia $, $ w\not\sbo z_{i+1}a\cdots a z_n $, i.e. $ w\in (\Sigma_{a}^*\bs\ w\uparrow) $.
				Hence,\[ z\in \left((\Sigma_{a-1}^*\bs\ v\uparrow)a\right)^* (\Sigma_{a-1}^*a)(\Sigma_{a}^*\bs\ w\uparrow) \subseteq RHS . \]
			\end{enumerate}
			Hence, $ LHS\subseteq RHS $.
			
			$ (\supseteq) $ This containment can be seen by going backwards in the arguments for the other containment.
		\end{proof}
		Since  $ (\Sigma_{a-1}^*\bs\ v\uparrow)$ is a finite union of ideals from $ Id_{a-1} $, 
		\begin{eqnarray*}
			\left((\Sigma_{a-1}^*\bs\ v\uparrow)a\right)^*(\Sigma_{a-1}^*\bs\ v\uparrow) &=& (\cup_i I_i)a(\cup_j I_j)\\
			&=& \bigcup_{i,j} I_iaI_j
		\end{eqnarray*}
		where $ I_i, I_j\in Id_{a-1} $. Since $ I_iaI_j $ is an element in $ Id_a $, $ \Sigma_{a-1}^*\bs\ v\uparrow$ is a finite union of ideals from $ Id_{a-1} $.
		
		Using similar arguments, it can be shown that 
		
	$ \left((\Sigma_{a-1}^*\bs\ v\uparrow)a\right)^* (\Sigma_{a-1}^*a)(\Sigma_{a}^*\bs\ w\uparrow) $ is also a finite union of ideals from $ Id_a $.
		
		Hence, the complement of a filter is a finite union of $ Id_a $ ideals.

	\item (Intersection of $ Id_a $ ideals is a finite union of $ Id_a $ ideals) We prove a stronger property, where ideals are defined over $ Atom'_a = Atom_a \cup {Ia|I\cap Id_{a-1}} $. When $ \Sigma = \{0\} $, the ideals are of the form $ 0^* $ or $ \{0^i|0\leq i\leq n \}$ for some $ n $, and the intersection of two ideals can only be another ideal.
	
	Suppose that the statement holds for some $ \Sigma_{a-1} $. Then let $ I_1=X_1X_2\cdots X_{k}A $ and $ I_2=Y_1Y_2\cdots Y_{l}B $ be two ideals from $ Id_a $. We then show by induction on the sum of the number of atoms in each ideal, i.e. $ s=k+l $. The base cases are:
	\begin{enumerate}
		\item when $ s=0 $: then $ I_1=A\in Id_{a-1} $ and $ I_2=B\in Id_{a-1} $, then by induction hypothesis, $ I_1\cap I_2 $ a finite union of ideals.
		\item when $ s=1 $: then $ I_1=A\in Id_{a-1} $ and $ I_2=X_1B$ where $ X_1\in Atom'_a $ and $B \in Id_{a-1} $. Then $ I_1\cap I_2 = (A\cap B )\cup (A\cap B_1 ) $, where $ X_1= B_1a $ or $ X_1=(B_1a)^* $. But both intersections are finite union of ideals.
	\end{enumerate}

	Then suppose the statement holds for all ideals $ I_1=X_1X_2\cdots X_{k}A  $ and $ I_2=Y_1Y_2\cdots Y_{l}B $, i.e. for some $ s=k+l $. Now suppose the sum of numbers of atoms in ideals be $ s+1 $. Then without loss of generality let $ I_1=X_1X_2\cdots X_{k}X_{k+1}A  $ and $ I_2=Y_1Y_2\cdots Y_{l}B $. To reduce notational clutter, we write $ I_1= X_1XA $ and $ I_2=Y_1YB $, with canonical $ X $ and $ Y $.
	
	Then we show that $ I_1\cap I_2   $ is a finite union of ideals. We consider the following cases, depending on the types of atoms $ X_1 $ and $ Y_1 $:
	\begin{enumerate}
		\item if $ X_1 = A_1a $ and $ Y_1=B_1a $, \\then $ I_1\cap I_2= (A_1\cap B_1)a (XA\cap YB) $
		\item if $ X_1 = A_1a $ and $ Y_1=(B_1a\cup \{\epsilon\}) $, \\
		then $ I_1\cap I_2= (A_1aXA\cap B_1aYA) \cup (A_1aXA\cap YB)  =  ((A_1\cap B_1)a (XA\cap YB)) \cup (A_1aXA\cap YB)$,
		\item if $ X_1 = A_1a $ and $ Y_1=(B_1a)^* $, \\ 
		then $ I_1\cap I_2 = I_1 \cap (YB\cup B_1aI_2) = (I_1\cap YB)\cup (I_1\cap B_1aI_2) = (I_1\cap YB)\cup ((A_1\cap B_1)a(XA\cap I_2))  $,
		\item if $ X_1 = (A_1a\cup \{\epsilon\}) $ and $ Y_1=(B_1a\cup \{\epsilon\})$, \\
		then $ I_1\cap I_2 = [(A_1\cap B_1)a(XA\cap YB)\cup (A_1aXA\cap YB)\cup (XA\cap B_1aYB)\cup (XA\cap YB)] $,
		\item if $ X_1 = (A_1a\cup \{\epsilon\}) $ and $ Y_1=(B_1a)^* $, \\
		then $ I_1\cap I_2 = (I_1\cap YB)\cup ((A_1\cap B_1)a(XA\cap I_2))\cup (XA\cap I_2) $,
		\item if $ X_1 = (A_1a)^* $ and $ Y_1=(B_1a)^* $, \\
		then $ I_1\cap I_2 = (XA\cup A_1aI_1) \cap (YB\cup B_1aI_2)  = (XA\cap YB) \cup (XA\cap B_1aI_2) \cup (A_1aI_1\cap YB)\cup (A_1aI_1\cap B_1aI_2) $.
	\end{enumerate}
	
	The equalities above follow from basic distributivity of unions and intersections. Since each intersection is between ideals with sum of atoms at most $ s $, then using the induction hypothesis, we have that $ I_1\cap I_2 $ are finite union of ideals.
	
	\end{itemize}

Hence, by \cref{idealSet}, we have that $ Id_a= Ideals_a $.
\end{proof}

\newcommand{\be}[1]{\underline{#1}}
\newcommand{\af}[1]{\overline{#1}}

\subsubsection{From ideals to pseudo-ideals}
\reductionToSimpleIdeals*
\begin{proof}
	Since pseudo-ideals are special cases of general ideals, one direction is trivial. For the other direction, suppose that the containment is decidable for pseudo-ideals. %

	\paragraph{From ideals to flat ideals}
	We call an ideal $ I= X_1X_2\cdots X_n A $ flat, if $ X_i $ is of the form $ Id $ or $ (I_1d\cdots dI_kd)^* $. We first show that for a general ideal $ I $, there exist flat ideals $ J_j $, such that $ I\subseteq L\downarrow_S $ if and only if $ \cup_j J_j\subseteq L'\downarrow_S $, for some language $ L' $.
	
	Let $ I =  X_1\cdots X_nA $ be an ideal, where $ X_i\in Atom_d, A\in Id_{d-1} $. First, since some atoms are of the form $ (I'_d\cup \epsilon) $, by distributivity of concatenation over union, we have that $ I $ is finite union of sets $ J_j $ of the form $ Y_1\cdot Y_kA $, where $ k\leq n $ and $ Y_i $'s are of form $ I'd $ or $ (Dd)* $. Then $ I\subseteq L\downarrow_S $ if and only if $ J_j\subseteq L\downarrow_S $ for all $ j $. For each $ J_j $, we give construct a flat ideal. So for simplicity of notation, we assume that $ X_i $'s are of the form $ I'd $ or $ (Dd)^* $. Moreover, since $ I\subseteq L\downarrow_S $ if and only if $ Id\subseteq Ld\downarrow_S$, we may also assume that $ I= X_1\cdots X_n $.

	Since every downward closed set is a finite union of ideals, then if $ X_i = (Dd)^* $, then we may replace $ X_i $ with $ (I_1dI_2d\ldots I_kd)^* $, where $ D=\cup_{j\in[1,k]} I_j $ is the ideal decomposition of $ D $. Suppose we obtain $ I' $ by replacing such $ X_i $s, then it is easy to see that $ I\subseteq L\downarrow_S $ if and only if $ I'\subseteq L\downarrow_S $. So we may assume that $ X_i $ is of the form $ Id $ or $ (I_1d\ldots dI_kd)^* $.
	
	Now consider the alphabet $ \s'_d =\{  i,\af{i} | i\in\s \} $, such that $ 0<\af{0}<1<\cdots < d<\af{d} $. Consider a transducer $ T_i $ that arbitrarily adds a $ \af{i} $ after an occurrence of $ i $. Then consider the ideal $ I' = X_1\cdot \af{d} \ldots \af{d}\cdot X_n\cdot \af{d} $. And consider the language $ L' = T_1\cdots T_aL $. Since $ \mathcal{C} $ is closed under rational transduction, $ L'\in \mathcal{C} $.
	
	By induction on highest letter $ a $ in $ I $, we will show that $ I\subseteq L\downarrow_S $ if and only if $ I'\subseteq L'\downarrow_S $.
	
	For the base case, suppose $ a=1 $. Then let $ I\subseteq L\downarrow_S $, and $ u = u_1\af{1}\cdots u_k\af{1}\in I' $, where $ u_i\in \S_1 $. Then $ u'= u_1\cdots u_k\in I $, and there exists $ v'\in L $ such that $ u' $ embeds in a $ 2 $-block of $ v' $ with witness map $ \phi $. Then on adding a $ \af{1} $ after every $ 1 $-block where the last $ 1 $-block of each $ u_i $ embeds to, to obtain $ v $, we get that $ u\sbo v $. 
	
	For the reverse direction, let $ I'\subseteq L'\downarrow_S $ and  $ u = u_1\cdots u_k\in I $, where $ u_i\in X_i $. Then $ v = u_1\af{1}\cdots \af{1}u_k\in I' $, and hence there exists $ v' = v_1\af{1}\cdots \af{1}v_l\in L' $ such that $ v\so v' $. Then since each $ \af{1} $ block in $ v $ recursively embeds in $ \af{1} $ blocks in $ v' $, the $ 1 $-blocks recursively embed too. Then $ u' = v_1v_2\cdots v_l\in L $ and $ u\so u' $.
	
	Then we assume that for some $ a-1 $, $ I\subseteq L\downarrow_S $ if and only if $ I'\subseteq L'\downarrow_S $, where $ L'= T_1\cdots T_{a-1}L $. Now suppose highest letter in $ I $ is $ a $, and $ L' = T_a\cdots T_1L $. Also, let $  I' = X_1\cdot \af{a} \ldots \af{a}\cdot X_n\cdot \af{a} $. We first show that $ I\subseteq L\downarrow_S $ iff $ I'\subseteq T_aL\downarrow_S $.
	
	First, let $ I\subseteq L\downarrow_S $, and $ u = u_1\af{d}\cdots \af{d}u_k\in I' $, where $ u_i\in X_i  $. Then $ v= u_1u_2\cdots u_k\in I $. Then there exists $ v' \in L $ such that and $ v\so v' $, with a witness map $ \phi $. Consider $ u' = v_1\af{d}\cdots \af{d}v_k $ such that $ v_i\in \Sigma_d\cdot w_i \cdot d  $ where $ w_i $ is the $ \phi(i)  $-th $ d $-block of $ u $. Note that $ v'\in T_aL $ and $ u\so u' $, since $ u_i\so v_i $.
	
	Now, suppose that $ I'\subseteq T_aL\downarrow_S $. Then let $ u = u_1\cdots u_k\in I $, where $ u_i\in X_i $. Then $ v = u_1\af{d}\cdots \af{d}u_k\in I' $, and hence there exists $ v' = v_1\af{d}\cdots \af{d}v_l\in T_aL $ such that $ v\so v' $. Then since each $ \af{d} $ block in $ v $ recursively embeds in $ \af{d} $ blocks in $ v' $, the $ d $-blocks recursively embed too. Then $ u' = v_1v_2\cdots v_l\in L $ and $ u\so u' $. 
	
	Since $ X_i $'s are ideals enclosed between $ \af{d} $'s in $ I'$, then by induction hypothesis $ X_i\subseteq T_aL\downarrow_S $ iff $ X_i' \subseteq T_1\cdots T_{a-1}T_aL\downarrow_S $, where $ X_i' $ is a flat ideal obtained by eliminating the downward closed sets in the Kleene stars. Hence, $ I\subseteq L\downarrow_S $ if and only if $ I'' = X'_1\af{d}\cdots X'_n \af{d} \subseteq T_1\cdots T_{a-1}T_aL\downarrow_S $.
	
	\paragraph{From flat ideals to pseudo-ideals}
	
	Since flat ideals are almost in form of pseudo-ideals, except only when $ X_i $ is of the form $ (I_1dI_2d\cdots I_kd)^* $, for the simplicity of the proof, we show how we reduce from $ (I_1dI_2d\cdots I_kd)^* $ to $ (Id)^* $. The generalization is simple extension. 
	
	Consider the alphabet $ \s'_d =\{ \be{i}, i,\af{i} | i\in\s \} $, such that $ \be{0}<0<\af{0}<\be{1}<1<\cdots < \be{d}<d<\af{d} $. Consider a transducer $ T_i $ that arbitrarily adds a $ \af{i} $ after an occurrence of $ i $, and a transducer $ T_i' $ that in arbitrary $ \af{i} $-blocks, replaces every $ i $ with an $ \be{i} $, and adds a $ i $ after the final replacement in the $ \af{i} $-block. For example, on a word $ 0101010 $, one of the outputs of $ T_1 $ is $ 01\af{1}0101\af{1}0 $, and on this word, $ T_1' $ outputs $ 01\af{1}0\be{1}0\be{1}1\af{1}0 $.
	
	Let $ X = (I_1dI_2d\cdots I_kd)^* $, where $ I_i $'s are flat ideals over $ \s_{d-1} $, then consider $ X'=(I_1\be{d}I_2\cdots \be{d}I_k\be{d}d )^* $. Note that $ X' $ is a pseudo-ideal. Then we show by induction over the highest letter $ a $ in flat ideals, that $ X\subseteq L\downarrow_S $ if and only if $ X'\subseteq T'_aT'_{a-1}\cdot T'_1L\downarrow_S $.
	
	For the base case, when $ a=1 $, then $ X= (I_11\cdots 1I_k1)^* $, where $ I_i $ is either $ 0^* $ or $ 0^{\leq k} $, and $ L' =  (I_1\be{1}\cdots \be{1}I_k\be{1}1)^* $. If $ X\subseteq L\downarrow $, and $ u = u_11\cdots u_n1\in X' $, where $ u_i\in \s_{\be{1}}^*\be{1} $. Then $ u'  = u'_11\cdots u'_n1 $, such that every $ \be{1} $ is replaced with $ 1 $, and the last $ \be{1} $ before every $ 1 $ is dropped. Then $ u'\in I $. Then there is a word $ v'= v'_1v'_2v'_3 $, where $ v'_2 $ is a $ 2 $-block such that $ u'\sbo v'_2 $, with a witness map $ \phi $. Let's say $ u_i $ has $ s_i $ many $ \be{1} $-blocks. Then on replacing all the $ 1 $'s with $ \be{1} $'s in $ v'_2 $, except the ones that appear at $ s_1+s_2+\ldots +s_i $-th $ 1 $'s for every $ i $, and adding $ \be{1} $, before them, we get a word $ v $. Then it is easy to observe that $ u\sbo v $, since $ u $ embeds within $ v_2 $: map $ i $-th $ 1 $-block of $ u $ to $ i $-th $ 1 $-block of $ v_2 $, and recursively map $ \be{1} $-blocks respecting $ \phi $.
	
	If $ I'\subseteq T'_1L\downarrow_S $ then going backward in the argument above, we get that $ I\subseteq T'_1L\downarrow_S $. 
	
	Then for some $ a $, the argument is similar with induction hypothesis over flat ideals of smaller highest letter, with the observation that adding, removing, and replacing $ \be{i} $ as per $ T'_i $, preserves the blocks of $ (i-1) $'s, and never splits them, which can continue to embed respecting their original embedding. Also, observe now that $ X' $ is a pseudo-ideal.

	Now to see this generalizes to any flat ideals, we observe that flat ideals are of the form $ X_1X_2\cdots X_n $ where $ X_i $ are of the form $ Id\af{d} $ or $ (I_1dI_2d\cdots I_xd)^*\af{d} $, i.e. every $ X_i $ is enclosed with highest priority letter in the ideal. Then within each $ \af{d} $-block the embedding is respected in the transformation from flat ideals to pseudo-ideals. So, we just replace $ X_i $ with $ X_i' $ as defined above, and apply $ T'_a\cdots T'_1 $ to $ T_a\cdots T_1L $.

	The two transformation reduce the containment problem of general ideals in a downward closed language to that for psuedo-ideals (via flat ideals).

\end{proof}

\subsubsection{Proof that pseudo-ideal is contained in downward closure if and only if $ I $-witness exists.}

A finite subset $ \witness $ of runs $ R $ is called an $ I $-witness if%

\begin{itemize}
	\item for $ I = 0^{\leq k} $ for some $ k\in \N $, there exists is a run $ \rho\in\witness $ such that $ 0^k $ is a factor of $ \can{\rho}[\overleftarrow{l}, \overrightarrow{l}] $, for some $ 0 \leq \overleftarrow{l}, \overrightarrow{l}\leq  \canonlen{\rho} $. 
	
	Then $ \rho $ is said to witness $ I $ between $ \overleftarrow{l} $ and $ \overrightarrow{l}$.
	
	\item for $ I=0^* $, there exist runs $ \rho,\psi\in \witness $ such that $ \psi\runembed_f \rho $ and  $ \gap{i}{f} = 0^l $ for some $ l>0 $, $ i\in[0,\canonlen{\psi}]$ and $ f\in E(\psi, \rho) $. 
	
	Then $ \rho $ is said to witness $ I $ between $ \overleftarrow{l} $ and $ \overrightarrow{l}$, if $ \gap{i}{f} $ is a factor of $ $ is a factor of $ \can{\rho}[\overleftarrow{l}, \overrightarrow{l}] $.
	
	\item for $ I = I_1a\cdots aI_na $, there exists a run $ \rho\in \witness $ and $ \overleftarrow{l}=l_0\leq l_1<l_2<\cdots <l_n \leq \overrightarrow{l} $ such that $ \rho $ is an $ I_i $-witness between $ l_{i-1} $ and $ l_i $, and $ \can{\rho}[\overleftarrow{l}, \overrightarrow{l}] \in\s_a^*$.
	
	Then $ \rho $ is said to witness $ I $ between $ \overleftarrow{l} $ and $ \overrightarrow{l}$.
	
	\item for $ I = (I'a)^* $, there exist runs $ \rho, \psi\in\witness $ such that 
	$ \psi\runembed_f \rho $ and $ \rho $ is a witness for $ (I'a)^l $ between $ f(i) $ and $ f(i+1) $, and $ \rho[f(i), f(i+1)]\in \Sigma_{a}^* $, for some $ l>0 $, $ i\in[0,\canonlen{\psi}-1]$ and $ f\in E(\psi, \rho) $.
	
	Then $ \rho $ is said to witness $ I $ between $ \overleftarrow{l} $ and $ \overrightarrow{l}$, if $ \gap{i}{f} $ is a factor of $ \can{\rho}[\overleftarrow{l}, \overrightarrow{l}] $.
	 
\end{itemize}

For a word $ w\in \Sigma\cup \{\epsilon\} $, $ w|_{\Sigma} $ denotes the restriction of $ w $ over $ \Sigma $. We call a word $ w' $ a factor of $ w = w_1\cdots w_n\in \Sigma\cup\{\epsilon\} $, if $ w|_{\Sigma}\in \Sigma^* w'\Sigma^* $, i.e., $ w' $ is an infix of restriction of $ w $ over $ \Sigma $. For a word $ w= w_0w_1\cdots w_n $, by $ w[i,j] $ we denote the word $ w_i\cdots w_j $ for $ i<j $. Given a set of runs $ S $, by the amalgamation closure of $ S $ we mean the set of runs that can be produced by amalgamating runs in $ S $.

\containementOfIdeal*
\begin{proof}
Let $ I $ be a pseudo-ideal over $ \Sigma= \Sigma_a $. Suppose $ I\subseteq L\downarrow_S $. 
\begin{itemize}
	\item If $ I=0^{\leq k} $, then for $ u=0^k $ there exists a run $ \rho $ in the system recognizing $ L $, such that $ u\sbo \yield{\rho}|_{\Sigma} = v$. Then $ 0^k $ is a factor of $ \can{\rho}[\overleftarrow{l}, \overrightarrow{l}] $ for some $ \overleftarrow{l} $ and $ \overrightarrow{l} $. Hence, $ \{\rho\} $ is an $ I $-witness between $ 0 $ and $ \canonlen{\rho} $.
	
	\item If $ I = 0^* $, then suppose there is no $ I $-witness, i.e. for every pair of runs $ \psi\runembed_f \rho $, every gap word is either $ \epsilon^* $ or it contains a letter $ p>0 $. If every gap word is $ \epsilon^* $, clearly $ I\not\subseteq L\downarrow_S $. Then suppose $ r\in \N  $ ($ r>0 $) be the maximum number such that $ 0^r $ is a factor of a gap word. Then since $ I\subseteq L\downarrow_S $, $ 0^{3r+2} $ must be a factor of a run $ \rho $, then for any run that embeds in to $ \rho $, the factor $ 0^{3r+2} $ splits over at least $ 3 $ gap words, due to the maximality of $ r $. But then there would be a gap word which is $ 0^l $ for some $ l>0 $, which is a contradiction to non-existence of an $ I $-witness. 
	
	\item If $ I=I_1a\cdots aI_na $, then consider the set of runs $ R'\subseteq R $ that yield simple block bigger words than any word in $ I $ (we say $ R' $ covers $ I $). If $ R' $ is finite, then there is no Kleene star in the pseudo-ideal. Hence there is a run among $ R' $ which yields $ maximal(I) $ and this run witnesses $ I $. Otherwise, if $ R' $ is an infinite set, then since $ (R, \runembed) $ is a WQO, $ R' $ has finitely many minimal runs. Among these minimal runs, consider a minimal set of these minimal runs whose amalgamation closure $ R'' $ covers $ I $. Then due to the up-directedness of pseudo-ideals, we can choose a sequence of runs $ \rho_1 \runembed \rho_2\runembed\cdots $ from $ R'' $ such that $ \{\rho_1,\rho_2,\ldots \} $ covers $ I $: for $ \rho_1 $ take the smallest run that embeds each run from the minimal set of minimal run (which corresponds to the join of yields of minimal runs). 
	
	Then observe that each run in $ R'' $ yields $ n $ many $ a $'s. So we can construct an amalgamation system $ A_i $ that produces only the $ i $-th $ a $ block of the yields of the runs in $ R'' $ for every $ i\in[1,n] $: this can be done since amalgamation systems are closed under rational transduction. Then since $ R'' $ covers $ I $, it also covers $ I_i $ for $ i\in [1,n] $. So, by induction hypothesis, there is a run that witnesses $ I_i $ in $ A_i $, for every $ i\in[1,n] $. Then there is a run $ \rho'_i $ in $ R'' $ which witnesses $ I_i $. But every run in $ R'' $ embeds $ \rho_1 $, hence $ \rho_1 $ is a witness for $ I $.
	
	\item  If $ I=(I'a)^* $, then consider the set of runs $ R'\subseteq R $ that yield $ I\cap L\downarrow_S=I $. Since $ I\subseteq L\downarrow_S $, hence $ (I'a)^k\subseteq L\downarrow_S $ for $ k\in\N $. Let $ R'_i $ be the set of runs that witness $ (I'a)^i $ which is an pseudo-ideal of the type above.
	
	Then again consider the sequence of runs $ \rho_1, \rho_1\cdots $ such that $ \rho_i\in R'_i $. Since the set of runs is a WQO over run embeddings, there is a subsequence $ \rho'_1,\rho'_2\cdots $ such that every run embeds into the next run. Now consider a run $ \rho'_t $ in this sequence which belongs to $ R'_t $, where $ t > \canonlen{\rho'_1} $. Let $ \rho'_1\runembed_f \rho'_t $. Then there must exist $ i\in [1,\canonlen{\rho'_1}] $, such that $ \rho'_t $ witnesses $ (I'a)^l $ for some $ l $ in an interval of a gap. Moreover, by the definition of $ R' $, the gap interval only contains the letters from $ \Sigma_{a} $. And hence, $ \rho'_t $ witnesses $ I $ between the interval of the gap word.
\end{itemize} 

Now for the other direction, suppose that $ I $-witness exists in the system recognizing $ L\downarrow_S $.
\begin{itemize}
	\item If $ I=0^{\leq k} $ for some $ k\in \N $, then the $ I $-witness yields a word $ w $ that contains $ 0^k $ as a factor. Then for any word $ u\in I $, $ u\sbo w $, implying that $ u\in L\downarrow_S $.
	
	\item If $ I=0^* $, then the $ I $-witness contains two runs $ \rho,\psi $ such that $ \psi\runembed_f\rho $ and $ \gap{i}{f} = 0^l$ for some $ i $. Suppose $ u= 0^k \in I $, then can construct a sequence of runs $ \rho=\rho_1, \rho_2,\ldots$, $ \rho_i $ is obtained by amalgamating $ \rho_{i-1} $ with $ \psi $. Then the yield of $ \rho_i $ contains $ 0^{il} $ as a factor. Then there exists $ i $ such that $ i\times l\geq k $, and then yield of $ \rho_i $ is simple block bigger than $ u $.
	
	\item If $ I = I_1aI_2a\cdots aI_na $, then there is a run $ \rho $ that witnesses $ I_i $ between some $ l_i $ and $ l_{i+1} $. Let $ u = w_1aw_2a\cdots aw_na\in I $ such that $ w_i\in I_i $. Then since $ \rho $ witnesses $ I_1 $ between $ l_1 $ and $ l_2 $, we can obtain a run $ \rho_1 $ such that $ \rho\runembed_f \rho_1 $ that yields a simple block bigger word than $ w_1 $ between some $ f(i) $ and $ f(i+1) $, and witnesses $ I_2a\cdots I_na $ in an interval after $ f(i+1) $. Continuing the same way, we obtain runs $ \rho_2, \ldots \rho_n $ such that $ \rho_i $ contains a simple block bigger word than $ w_1aw_2a\cdots w_ia $ before position $ j $ and witnesses $ I_{i+1}a\cdots I_na $ after position $ j $. Then $ u\sbo \yield{\rho_n} $, implying $ I\subseteq L\downarrow_S $.
	
	\item if $ I = (I'a)^* $, then there exist two runs $ \rho, \psi $ in the witness set such that $ \psi\runembed_f \rho $ and $ \rho $ is a witness for $ (I'a)^l $ between $ f(i) $ and $ f(i+1) $, and $ \rho[f(i), f(i+1)]\in \Sigma_{a}^* $, for some $ l>0 $, $ i\in[0,\canonlen{\psi}-1]$ and $ f\in E(\psi, \rho) $. Let $ u= w_1aw_2a\cdots w_na\in I $. Then we can amalgamate $ \rho $ $ k $ many times with $ \psi $ to obtain a run that witnesses $ I'a $ in $ k $ disjoint intervals. Then with the arguments as above, we can get a run $ \rho' $ such that $ u\sbo \yield{\rho'} $.
\end{itemize}

\end{proof}

	\section{Details on counter extensions}\label{appendix:counter-ext}
	\counterextlangsysequiv*

\begin{proof}
  By definition, $L(S_{\eta,\alpha})=\bigcup_{(\rho,w)\in
  R_{\eta}}\alpha(\yield\rho)$ and
  $L_{\eta,\alpha}=\alpha(\eta^{-1}(N_d)\cap L)$, thus it suffices to
  show that $\bigcup_{(\rho,w)\in
  R_{\eta}}\yield\rho=\eta^{-1}(N_d)\cap L$.  Consider a word
  $w=a_1\cdots a_n$ from $L=L(S)$.  There is a run $\rho\in R$ such
  that $w=\yield\rho$ is accepted by~$\rho$.  Let us show that
  $\eta(w)\in N_d$ (by showing that it has an accepting run in the
  VASS for the language~$N_d$) if and only if there exists an
  accepting decoration $u$ of~$\rho$.

  If $(\rho,u)$ is an accepting decorated run for some $u=(\vec
  u_1,\vec v_1)\cdots(\vec u_n,\vec v_n)$, then $q(\vec
  u_i)\xrightarrow{\eta(a_i)}q(\vec v_i)$ for all $i\in[1,n]$ in our
  VASS because $\vec v_i = \vec u_i + \delta(\eta(a_i))$ and $\eta$
  was assumed to be tame, $q(\vec v_i)=q(\vec u_{i+1})$ for all
  $i\in[1,n-1]$ because $\vec v_i=\vec u_{i+1}$, $q(\vec u_1)=q(\vec
  0)$ because $\vec u_1=\vec 0$, and $q(\vec v_n)=q(\vec 0)$ because
  $\vec v_n=\vec 0$.  Thus there is a run of the VASS
  since \begin{equation}\label{eq-langsysequiv} q(\vec 0)=q(\vec
  u_0)\xrightarrow{\eta(a_1)}q(\vec u_1)\xrightarrow{\eta(a_2)}q(\vec
  u_2)\cdots q(\vec u_{n-1})\xrightarrow{\eta(a_n)}q(\vec u_n)=q(\vec
  0)\;.  \end{equation} This shows that
  $\eta(w)=\eta(a_1)\cdots\eta(a_n)\in N_d$.

  Conversely, if $\eta(w)\in N_d$, then there is a run of the VASS
  for~$N_d$ such that \eqref{eq-langsysequiv} holds, and we can
  decorate $\rho$ with the sequence of pairs $u\eqdef (\vec u_0,\vec
  u_1)(\vec u_1,\vec u_2)\cdots(\vec u_{n-1},\vec u_n)$; then
  $(\rho,u)\in R_{\eta}$ is an accepting decorated run.
\end{proof}

	\section{Details on algebraic extensions}\label{appendix:alg-ext}
	Let $\cC$ be a class of languages with concatenative amalgamation and
well-quasi-ordered decorations and let $\Grammar = (N, T,
S, \{L_A\}_{A \in N})$ be a $\cC$ grammar. For each $A \in N$, let
$\Model_A$ be the amalgamation system with wqo decorations recognising
$L_A$. Let $\preceq$ be the embedding in $\Model_A$.

\subsection{Well-Quasi-Orderedness and Decorations}
Let $\tau_1$ and $\tau_2$ be trees of $\Grammar$. We recall the
definition of the embedding $\runembed$ in $\Grammar$, being
$\tau_1 \runembed \tau_2$ if there exists a subtree
$\subtree{\tau_2}{p}$ such that

\algextembed*
\algextwqo*

\begin{proof}
We rely on \citeauthor{nash-williams63}'s \emph{minimal bad sequence
argument}~\cite{nash-williams63}.  Assume for the sake of contradiction that
$(\cT(\Grammar), \runembed)$ is not a wqo.  Then we can construct a
\emph{minimal} infinite bad sequence of trees $t_0,t_1,\dots$, where
minimality means that for all~$i$, any sequence
$t_0,t_1,\dots,t_{i-1},t,\dots$ where $t$ is a (strict) subtree of
$t_{i}$, i.e., $t=\subtree{t_i}{p}$ for some $p\neq\varepsilon$, is
good.  To construct such a sequence, we start by selecting a
tree~$t_0$ minimal for the subtree ordering among all those that may
start an infinite bad sequence; this $t_0$ exists because the subtree
ordering is well-founded.  We continue adding to this sequence by
selecting a minimal~$t_i$ for the subtree ordering among all the
trees~$t\in\cT(\Grammar)$ such that there exists an infinite bad
sequence starting with $t_0,t_1,\dots,t_{i-1},t,\dots$.  At every
step, the constructed sequence $t_0,\dots,t_i$ is bad.  The infinite
sequence remains bad: for every $i<j$, $t_0,\dots,t_j$ is a bad
sequence, hence $t_i\not\runembed t_j$.

Let $S_i\eqdef\{t\in\cT(\Grammar)\mid\exists
p\neq\varepsilon.t=\subtree{t_i}{p}\}$ be the set of subtrees of~$t_i$
and $S\eqdef\bigcup_{i \geq 0} S_i$.

\begin{claim}
	\label{lem:tree-embed-wqo-restricted}
	$(S, \runembed)$ is a wqo.
\end{claim}
\begin{proof}
  Assume $(S, \runembed)$ is not a wqo.  Then there is an infinite bad
  sequence $s_0, s_1,\ldots$.  Let $i$~be minimal such that $s_0\in
  S_i$.  Since $\bigcup_{j\leq i}S_j$ is finite, without loss of
  generality we may assume that each $s_k$ originates from a set
  $S_\ell$ with $\ell\geq i$.
	
  Consider the sequence $t_0, \ldots, t_{i-1}, s_0, s_1, \ldots$.  As
  $s_0$ is a strict subtree of $t_i$, by the minimality assumption of
  $t_i$, this sequence is good.  Since the sequences $t_1, \ldots,
  t_{i-1}$ and $s_0, s_2, \ldots$ are both bad, there must exist $j,
  k$ with $j < i$ such that $t_j \runembed s_k$.  However, this means
  that there exists a subtree $\subtree{s_k}{p}$ satisfying the
  conditions of~$\runembed$.  As $s_k\in S_\ell$ for some
  $\ell\geq i$, there exists $p'\neq\varepsilon$ such that
  $\subtree{s_k}{p}=\subtree{t_\ell}{p'p}$.  Thus $t_j \runembed
  t_\ell$ with $j<i\leq\ell$, a contradiction with the fact that
  $t_0,t_1,\ldots,t_\ell$ is bad.
\end{proof}

We return to the proof that $(\cT(\Grammar), \runembed)$ is a
well-quasi-order.

As there are only
finitely many symbols in $N$, there is $A\in N$ and an infinite bad
subsequence $t_{i_0}, t_{i_1}, \ldots$ of $(t_i)_i$ where all the
$t_{i_j}$'s are $A$-rooted; let us write $\rho_j$ for the run of
$\Model_A$ labelling the root
of~$t_{i_j}=\tree{(A \to \rho_{j})}{\dots}$.%

If any $\rho_{j}$ has $\can{\rho_{j}} \in \Siep^*$, then
$t_{i_j}=\tree{(A \to \rho_{j})}{}$ is a leaf.  Because $\Model_A$ is
an amalgamation system and therefore~$\preceq$ is a wqo, there exists
$\ell>j$ such that $\rho_j\preceq\rho_\ell$.  Then $t_{i_j}\runembed
t_{i_\ell}$ because condition~\eqref{algextea} holds by assumption
and condition~\eqref{algexteb} is vacuous in this case, a contradiction.

We therefore assume that each $\rho_{j}$ has
$\can{\rho_{j}}= u_{j,0} B_{j,1} u_{j,1} \cdots B_{j,k_j} u_{j,k_j}$ for some $k_j>0$ and
non-terminals $B_{j,1}\cdots B_{j,k_j}\in N$; then
$t_{i_j}=\tree{(A\to\rho_{j})}{\subtree{t_{i_j}}{1},\dots,\subtree{t_{i_j}}{k_j}}$.
We decorate each $\rho_{j}$ with its sequence of children
$w_j\eqdef\subtree{t_{i_j}}{1}\cdots\subtree{t_{i_j}}{k_j}$, which all belong
to~$S$.
This gives rise to an infinite sequence $(\rho_j,w_j)_{j}$ of
decorated runs in $\Deco^S(R)$.  Because $(S, \runembed)$ is
a wqo by \cref{lem:tree-embed-wqo-restricted} and
$\Model_A$ supports wqo decorations by assumption, there is a pair
$j<\ell$ with \eqref{algextea} $\rho_{i_j}\embedwith[\preceq]{f}\rho_{i_\ell}$ and
\eqref{algexteb} $\subtree{t_{i_j}}{k} \runembed \subtree{t_{i_k}}{f(k)}$ for all
$k\in[1,k_j]$.  This however implies that $t_{i_j} \runembed
t_{i_\ell}$, again a contradiction.
\end{proof}

We can at this point show that $\Grammar$ also supports wqo decorations. 

\algextwqodeco*

\begin{proof}
Assume we decorate every terminal symbol of a tree with symbols from a
wqo $X$.  This is equivalent to decorating the runs $\rho$ with the
same symbols, introducing a new symbol $1$ incomparable
from all the elements of~$X$ for the output letters from $N$.  Then
$X\cup\{1\}$ is also a wqo.

Then the embedding $\runembed^X$ between decorated trees of $\Grammar$
can be defined by replacing the order $\preceq$ on the inner runs with
$\preceq^{X \cup\{1\}}$. Since $\cC$ supports wqo decorations,
$\preceq^{X \cup\{1\}}$ is a well-quasi-ordering.  The proof
of \cref{lem:algextwqo} shows that $\runembed^X$ is a well-quasi-order
as well.
\end{proof}

\subsection{Admissible Embeddings}
Recall the definition of the canonical decomposition of trees of $\cC$-grammars:

\algextcanondecomp*

The nested structure of the decomposition requires us to define some additional notation to address the letters contributed by specific subtrees to the canonical decomposition of some tree. We define $\canonidx_{p}(\tau)$ as the offset of the canonical decomposition of $\subtree{\tau}{p}$ in the decomposition of $\tau$. That is, if the canonical decomposition of $\tau$ is $a_1 \cdots a_n$ and the canonical decomposition of $\subtree{\tau}{p}$ is $b_1 \cdots b_m$ then $b_i$ corresponds to $a_{\canonidx_{p}(\tau) + i}$.

Recall that we write $\memo_\rho$ for the map associating with every occurence of a non-terminal $X_i$ in $\yield{\rho}$ its position in the canonical decomposition of $\rho$. Then if $p = \varepsilon$, we have

\begin{align*}
	\canonidx_\epsilon(\tau) &= 0.
\end{align*}

If $p = i \cdot p'$, we have 
\begin{itemize}
	\item the length of the canonical decomposition of $\rho$ up to $X_i$,
	\item the combined length of the canonical decomposition of every $t_j$ with $j < i$,
	\item less the individual letters $X_1, \ldots X_{i-1}$,
	\item two $\epsilon$ markers for every $u_j$ with $j \leq i$,
	\item and finally the position of $p'$ in the canonical decomposition of $t_i$.
\end{itemize}

Put together, we have

\begin{align*}
	\canonidx_{i \cdot p'}(\tau) &= \memo_\rho(i) + \left(\sum_{j = 1}^{i-1} \canonlen{t_j}\right) + i + \canonidx_{p'}(t_i)
\end{align*}

More generally, we have $\canonidx_{p \cdot q}(\tau) = \canonidx_p(\tau) + \canonidx_q(\subtree{\tau}{p})$. Observe also that if $\subtree{\tau}{(p \cdot i)}$ and $\subtree{\tau}{(p \cdot [i + 1])}$ are defined, then $\canonidx_{p \cdot (i+1)}(\tau) = \canonidx_{p \cdot i}(\tau) + \canonlen{\subtree{\tau}{(p \cdot i)}} + (\memo_\rho(i+1) - \memo_\rho(i)) + 1$. 

If $\subtree{\tau}{p} \runembed \subtree{\tau'}{p'}$ and $f \in E(\subtree{\tau}{p}, \subtree{\tau'}{p'})$, we write $\liftmap[\subtree{\tau}{p}][\subtree{\tau'}{p'}]{f}$ for the lifting of $f$ to $\tau$ and $\tau'$, a function from $[\canonidx_p(\tau) + 1, \canonidx_p(\tau) + \canonlen{\subtree{\tau}{p}}]$ to $[\canonidx_{p'}(\tau') + 1, \canonidx_{p'}(\tau') + \canonlen{\subtree{\tau'}{p'}}]$ given by $\hat f^{\subtree{\tau}{p}}_{\subtree{\tau'}{p'}}(i + \canonidx_p(\tau)) = f(i) + \canonidx_{p'}(\tau')$.

If $\tau$ and $\tau'$ are trees from $\Grammar$, than the set of admissible embeddings $E(\tau, \tau')$ is isomorphic to all the ways to embed $\tau$ into $\tau'$.

Let $\tau = \tree{(A \mapsto \rho)}{t_1, \ldots, t_l}$. 
Each $p \in \Nat^*$ such that $\subtree{\tau'}{p} \allowbreak=\allowbreak \tree{(A \to \rho')}{t'_1, \ldots, t'_k}$, $\rho \embedwith[\preceq]{\phi} \rho'$, and $t_i \runembed t'_{\phi(i)}$  corresponds to a set of admissible embeddings. 
Let $f_i \in E(t_i, t'_{\phi(i)})$. We write $f'_i$ for $\liftmap[\subtree{\tau}{i}][\subtree{\tau'}{p \cdot \phi(i)}]{f_i}$. Note that the domains and co-domains of each $f'_i$ are necessarily disjoint. Therefore we may take their union $g = \bigdisunion f_i$. This $g$ induces a unique (partial) admissible embedding in $E(\tau, \tau')$. We are missing the mapping for the terminal letters in the canonical decomposition of $\rho$, as well as the $\epsilon$-components, which we assign as follows:

Let $\nu \eqdef \memo^{-1}_{\rho'} \circ \phi \circ \memo_\rho : [1,l] \to [1,k]$ be the subtree index map between $t_1, \ldots, t_l$ and $t'_1, \ldots, t'_k$. Let $\beta_\rho(i) \eqdef \min (\{ j-1 \mid i < \memo_\rho(j) \} \cup \{l\})$ be the \emph{block index} of a non-terminal letter at position $i$ of the canonical decomposition of $\rho$. For convenience, we assume $\canonidx_{0}(\tau) = 0$ and $\memo_\rho(l+1) = \canonlen{\rho} + 1$

We have 

\begin{align*}
	1 \mapsto &~\canonidx_p(\tau') + 1 \\
	\canonlen{\tau} &~\canonidx_p(\tau') + \canonlen{\subtree{\tau'}{p}} \\
	\canonidx_i(\rho) \mapsto &~\canonidx_p(\tau') + \canonidx_{\nu(i)}(\subtree{\tau'}{p}) \tag{$1 \leq i \leq l$} \\
	\canonidx_i(\rho) + \canonlen{t_i} + 1 \mapsto &~\canonidx_p(\tau') + \canonidx_{\nu(i)}(\subtree{\tau'}{p}) + \canonlen{t'_{\nu(i)}} + 1 \tag{$1 \leq i \leq l$}\\
	\canonidx_i(\rho) + \canonlen{t_i} + j \mapsto &~ \canonidx_{p}(\tau') \cdot \canonidx_x(\subtree{\tau'}{p}) + (\phi(d) - \memo_{\rho'}(x)) + 1 
		\tag{$1 \leq i \leq l$, $2 \leq j \leq (\memo_\rho(i+1) - \memo_\rho(i))$}\\
		\tag{where $d=\memo_\rho(i) + j-1, x = \beta_{\rho'}(\phi(d)$}
\end{align*}

Intuitively, we may assume that the $\epsilon$-component directly to the left of the $i$-th gets mapped to the one directly to the left of the image of $i$, and the one directly to the right gets mapped to the one directly to the right of the image. Non-terminals get mapped to the corresponding non-terminal in $\rho'$ by $\phi$.

Note that different choices of $p \in \Nat^*$ induce different assignments for these values. If we assume that we have two different embeddings of $\tau$ but the same path $p$ for both, then either the underlying run embedding $\phi$ must be different, which leads to a distinct lifting for the subtrees, or at least one subtree $t_i$ has a different embedding into the same subtree $t'_{\phi(i)}$ and by induction we may assume that this corresponds to a distinct embedding in $E(t_i, t'_{\phi(i)})$. In brief, we may conclude that there is a one-to-one correspondence between tree embeddings and admissible embeddings.

\subsection{Composition}
\algextcomposes*

Let $\tau_0 = \tree{(A \to \rho_0)}{\dots}, \tau_1, \tau_2$ be trees from $\Grammar$ and $\tau_0 \embedwith{f} \tau_1 \embedwith{g} \tau_2$. The maps $f$ and $g$ correspond to specific embeddings between $\tau$, $\tau'$ and $\tau''$. In particular, let $p$ be the path corresponding to the mapping of $\tau_0$ into $\tau_1$. We have $\subtree{\tau_1}{p} = \tree{(A \to \rho_1)}{\dots}$ and $\rho_0 \embedwith[\preceq]{\phi} \rho_1$. Let $q$ be the path corresponding to the mapping of $\subtree{\tau_1}{p}$ into $\tau_2$. We have $\subtree{\tau_2}{q} = \tree{(A \to \rho_2)}{\dots}$ and $\rho_1 \embedwith[\preceq]{\psi} \rho_2$. Due to the structure of $\Model_A$, we know that $\phi$ and $\psi$ can be composed such that $\rho_0 \embedwith[\preceq]{\psi \circ \phi} \rho_2$. Then the path $q$ and the embedding of the children of $\tau$ along $\psi \circ \phi$ is also a valid embedding of $\tau_0$ into $\tau_2$.

If we expand the composition of $f$ and $g$, we get

\begin{align*}
	g(f(i)) &= g(\widehat{f'}^{\tau_0}_{\subtree{\tau_1}{p}}(i)) \tag{for $f' \in E(\tau_0, \subtree{\tau_1}{p})$} \\
	&= g(f'(i) + \canonidx_p{(\tau_1)}) \\
	&= \widehat{g'}^{\subtree{\tau_1}{p}}_{\subtree{\tau_2}{q}}(f'(i) + \canonidx_{p}(\tau_1)) \tag{for $g' \in E(\subtree{\tau_1}{p}, \subtree{\tau_2}{q})$} \\
	&= g'(f'(i)) + \canonidx_q(\tau_2) \\
	&= \widehat{\left(g' \circ f'\right)}^{\tau}_{\subtree{\tau_2}{q}}(i)
\end{align*}

which is what we would get from the direct mapping of $\tau$ into $\subtree{\tau_2}{q}$. 

An analogous line of reasoning holds for the case of $\tau_0 = \leaf{(\varepsilon)}$

\subsection{Concatenative Amalgamation}
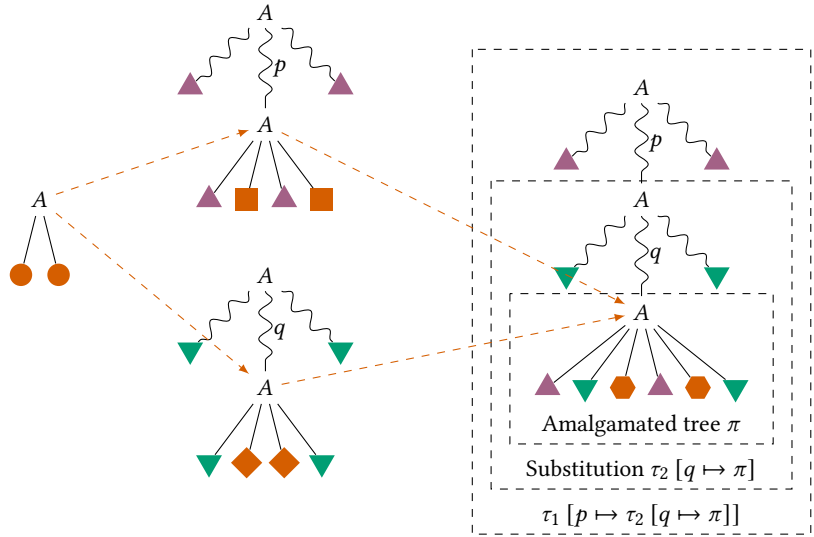
\begin{figure}
	\begin{center}
		\begin{tikzpicture}[
			dot/.style={fill,circle,minimum size=3pt},
			dota/.style={fill,diamond,rotate=45,minimum size=5pt,color=highlight},
			dotb/.style={fill,diamond,minimum size=2.5pt,color=highlight},
			dotc/.style={fill,regular polygon,regular polygon sides=6,minimum size=3pt,color=highlight},
			tri/.style={fill,regular polygon,regular polygon sides=3,minimum size=2pt,inner sep=2pt,fill=highlight-2},
			tri2/.style={tri,rotate=180,fill=highlight-3}
			]
			\node (a) at (0,0) {$A$};
			\node [dot,color=highlight] at (-.25,-1) (a1) {};
			\node [dot,color=highlight] at (.25,-1) (a2) {};
			\draw [-] (a) -- (a1);
			\draw [-] (a) -- (a2);
			
			\node (b1-head) at (3,2.5) {$A$};
			\node [tri] (b1-head-l) at (2,1.5) {};
			\node [tri] (b1-head-r) at (4,1.5) {};
			\node (b1) at (3,1) {$A$};
			\node [tri] at (2.25,0) (b1-1) {};
			\node [dota] at (2.75,0) (b1-2) {};
			\node [tri] at (3.25,0) (b1-3) {};
			\node [dota] at (3.75,0) (b1-4) {};
			\draw [-,decorate,decoration={snake}] (b1-head) -- (b1) node [midway] {$p$};
			\draw [-,decorate,decoration={snake}] (b1-head) -- (b1-head-l);
			\draw [-,decorate,decoration={snake}] (b1-head) -- (b1-head-r);
			\draw [-] (b1) -- (b1-1);
			\draw [-] (b1) -- (b1-2);
			\draw [-] (b1) -- (b1-3);
			\draw [-] (b1) -- (b1-4);
			
			\node (b2-head) at (3,-1) {$A$};
			\node [tri2] (b2-head-l) at (2,-2) {};
			\node [tri2] (b2-head-r) at (4,-2) {};
			\node (b2) at (3,-2.5) {$A$};
			\node [tri2] at (2.25,-3.5) (b2-1) {};
			\node [dotb] at (2.75,-3.5) (b2-2) {};
			\node [dotb] at (3.25,-3.5) (b2-3) {};
			\node [tri2] at (3.75,-3.5) (b2-4) {};
			\draw [-,decorate,decoration={snake}] (b2-head) --  (b2) node [midway] {$q$};
			\draw [-,decorate,decoration={snake}] (b2-head) -- (b2-head-l);
			\draw [-,decorate,decoration={snake}] (b2-head) -- (b2-head-r);
			\draw [-] (b2) -- (b2-1);
			\draw [-] (b2) -- (b2-2);
			\draw [-] (b2) -- (b2-3);
			\draw [-] (b2) -- (b2-4); 
			
			\node (c-head-1) at (8,1.5) {$A$};
			\node [tri] (c-head-1-l) at (7,.5) {};
			\node [tri] (c-head-1-r) at (9,.5) {};
			\node (c-head-2) at (8,0) {$A$};
			\node [tri2] (c-head-2-l) at (7,-1) {};
			\node [tri2] (c-head-2-r) at (9,-1) {};
			\node (c) at (8,-1.5) {$A$};
			\node [tri] (c1) at (6.75,-2.5) {};
			\node [tri2] (c2) at (7.25,-2.5) {};
			\node [dotc] (c3) at (7.75,-2.5) {};
			\node [tri] (c4) at (8.25,-2.5) {};
			\node [dotc] (c5) at (8.75,-2.5) {};
			\node [tri2] (c6) at (9.25,-2.5) {};
			\draw [-,decorate,decoration={snake}] (c-head-1) --  (c-head-2) node [midway] {$p$};
			\draw [-,decorate,decoration={snake}] (c-head-1) -- (c-head-1-l);
			\draw [-,decorate,decoration={snake}] (c-head-1) -- (c-head-1-r);
			\draw [-,decorate,decoration={snake}] (c-head-2) --  (c) node [midway] {$q$};
			\draw [-,decorate,decoration={snake}] (c-head-2) -- (c-head-2-l);
			\draw [-,decorate,decoration={snake}] (c-head-2) -- (c-head-2-r);
			\draw [-] (c) -- (c1);
			\draw [-] (c) -- (c2);
			\draw [-] (c) -- (c3);
			\draw [-] (c) -- (c4);
			\draw [-] (c) -- (c5);
			\draw [-] (c) -- (c6);
			
			\draw [dashed] (6.25,-1.25) rectangle (9.75,-3.25);
			\node at (8,-3) {Amalgamated tree $\pi$};
			
			\draw [dashed] (6,.25) rectangle (10,-3.85);
			\node at (8,-3.6) {Substitution $\treesubst{\tau_2}{q}{\pi}$};
			
			\draw [dashed] (5.75,2) rectangle (10.25,-4.45);
			\node at (8,-4.2) {$\treesubst{\tau_1}{p}{\treesubst{\tau_2}{q}{\pi}}$};
			
			\draw [-{latex},dashed,color=highlight] (a) edge [] (b1);
			\draw [-{latex},dashed,color=highlight] (a) edge [] (b2);
			
			\draw [-{latex},dashed,color=highlight] (b1) edge [] (c);
			\draw [-{latex},dashed,color=highlight] (b2) edge [] (c);
		\end{tikzpicture}
	\end{center}
	\caption{Amalgamation of trees $\tau_1$ and $\tau_2$ (middle) over base tree $\tau_0$ (left).}
	\label{fig:tree-amalg}
\end{figure}

\algextconcalg*

If $\tau_0 = \leaf{(\varepsilon)}$ and therefore also $\tau_1$ and $\tau_2$, the statement trivially holds. We therefore consider the interesting case.

As an intuition for the amalgamation of trees, see \autoref{fig:tree-amalg}. To make the construction of the large tree easier, we introduce notation for the substitution of subtrees, as in \cite[Sec.~3]{Leroux:2019:pvass-embedding}. If $\tau = \tree{(A \to \rho)}{\dots}$, $\pi$ are trees and $\subtree{\pi}{p} = \tree{(A \to \rho')}{\dots}$, we inductively define $\treesubst{\pi}{p}{\tau}$ as

\begin{align*}
	\treesubst{\pi}{\epsilon}{\tau} &= \tau \\
	\treesubst{\tree{(B \to \rho)}{t_1, \ldots, t_n}}{(i \cdot p')}{\tau} &= \tree{(B \to \rho)}{t_1, \ldots, t_{i-1}, \treesubst{t_i}{p'}{\tau}, t_{i+1}, \ldots, t_n}
\end{align*}

Note that this operation maintains all labels along the path $p$ and both $\tau$ and $\subtree{\pi}{p}$ are $A$-rooted. The result is therefore a valid tree of $\Grammar$ again. 

Substituting a subtree by a larger tree makes the entire tree larger:

\begin{lemma}
	\label{lem:treesubst-is-possible}
	If $\subtree{\pi}{p} \runembed \tau$ and $\tau$ and $\subtree{\pi}{p}$ are both $A$-rooted, then $\pi \runembed \treesubst{\pi}{p}{\tau}$ (and $\pi$ and $\treesubst{\pi}{p}{\tau}$ are both $B$-rooted).
\end{lemma}

\begin{proof}
	If $p = \varepsilon$, then this is trivially true. Otherwise let $p = i \cdot p'$. Let $t_i$ be the $i$-th child of $p$. By induction, we have $t_i \runembed \treesubst{t_i}{p'}{\tau}$. Then the definition of the tree embedding means we have $\pi \runembed \treesubst{\pi}{i}{\treesubst{t_i}{p'}{\tau}} = \treesubst{\pi}{p}{\tau}$.
\end{proof}

We can now proceed with the proof of \cref{lem:algext-concat-amalgamation}. Refer also to \cref{fig:tree-amalg} for a visual example.

\begin{proof}
	Let $\tau_1, \tau_2$ be trees such that $\tau_0 \embedwith{f} \tau_1$ and $\tau_0 \embedwith{g} \tau_2$. Recall that this means there are $p, p'$ such that $\subtree{\tau_1}{p} = \tree{(A \to \rho_1)}{t'_1, \ldots, t'_{k'}}$, $\rho_0 \embedwith[\preceq]{\phi} \rho_1$ and $t_i \runembed t'_{\phi(i)}$; similarly $\subtree{\tau_2}{p'} = \tree{(A \to \rho_2)}{t''_1, \ldots, t''_{k''}}$, $\rho_0 \embedwith[\preceq]{\psi} \rho_2$ and $t_i \runembed t''_{\psi(i)}$.
	
	As $\cC$ has concatenative amalgamation, we can construct a run $\rho_3$ in the system associated with $A$ such that $\rho_1 \embedwith[\preceq]{\phi'} \rho_3$ and $\rho_2 \embedwith[\preceq]{\psi'} \rho_3$ such that $\phi' \circ \phi = \psi' \circ \psi$. 
	
	We now construct a tree $\pi = \tree{(A \to \rho_3)}{t'''_1, \ldots, t'''_{k'''}}$. For each $j$ in $[1, k''']$, we can distinguish three cases:
	\begin{itemize}
		\item $j$ is in the image of $\phi'$ and not in the image of $\psi'$. Then we set $t'''_j = t''_{\phi'^{-1}(j)}$ (corresponding to the upward triangular nodes in \cref{fig:tree-amalg}).
		\item $j$ is not in the image of $\phi'$, but is in the image of $\psi'$. Then we set $t'''_j = t''_{\psi'^{-1}(j)}$ (corresponding to the lower triangular nodes in \cref{fig:tree-amalg}).
		\item $j$ is in the image of both $\phi'$ and $\psi'$. Then $j$ is in the image of $\phi' \circ \phi$. We have $t_{(\phi' \circ \phi)^{-1}(j)} \runembed t'_{\phi'^{-1}(j)}$ and $t_{(\phi' \circ \phi)^{-1}(j)} \runembed t''_{\psi'^{-1}(j)}$ and all three are $X$-rooted. We can construct, by induction, an $X$-rooted tree $t'''_j$ such that $t'_{\phi'^{-1}(j)} \runembed t'''_j$ and $t''_{\psi'^{-1}(j)} \runembed t'''_j$ (corresponding to the hexagonal nodes in \cref{fig:tree-amalg}).
	\end{itemize}
	
	We have $\subtree{\tau_1}{p} \runembed \pi$ and both are $A$-rooted. By \autoref{lem:treesubst-is-possible}, we can substitute $\pi$ for $\subtree{\tau_1}{p}$. Then we have $\pi \runembed \treesubst{\tau_1}{p}{\pi}$. Analogously we have $\subtree{\tau_2}{p'} \runembed \treesubst{\tau_1}{p}{\pi}$ and both sub-trees are $A$-rooted. We can therefore again substitute $\treesubst{\tau_1}{p}{\pi}$ for $\subtree{\tau_2}{p'}$. We get that $\tau_1$ and $\tau_2$ both embed into $\treesubst{\tau_2}{p'}{\treesubst{\tau_1}{p}{\pi}}$. Equivalently, we may swap the order of substitutions to obtain
	$\treesubst{\tau_1}{p}{\treesubst{\tau_2}{p'}{\pi}}$.
	
	Let $i$ be a gap of $\tau_0$. We distinguish the following cases:
	\begin{itemize}
		\item $i$ is the first or the last gap. By the definition of the admissible embeddings, the first and last $\epsilon$ in the canonical decomposition of $\tau_0$ gets mapped to the first or last $\epsilon$-component of $\pi$. It follows that the content of the gap comes only from the context of the final substitutions and depending on the order of these substitutions we have $\gap{h}{0} = \gap{f}{0}\gap{g}{0}$ or vice versa and equivalently for $\gap{h}{\canonlen{\tau_0}}$.
		\item $i$ is the gap between two symbols that occur on the level of $\tau_0$ and not one of its children. Then the content of the gap is induced by the gap language $\gap{\phi' \circ \phi}{k}$, which satisfies the concatenative amalgamation property.
		\item $i$ is the gap between an $\epsilon$-component and a letter of a child $t_j$, or occurs entirely within $t_j$. Then a simple inductive argument shows that the concatenative amalgamation property holds.
	\end{itemize}
\end{proof}

	\section{Details on valence automata}\label{appendix-valence-automata}
	Recall that a \emph{monoid} is a set with a binary associative
operation and a neutral element. Intuitively, in a valence automaton
over a monoid $M$, each edge is labelled by an input word and an
element of the monoid. Then, an execution from an initial state to a final
state is valid if the product of the monoid elements is the identity.
Unless stated otherwise, we denote the operation by juxtaposition and
the neutral element by $1$.

\subsection{Valence Automata}
Formally, a \emph{valence automaton} over
a monoid $M$ is an automaton $\cA=(Q,\Sigma,M,\Delta,q_0,F)$, where $Q$ is a
finite set of \emph{states}, $\Sigma$ is a finite alphabet, $\Delta\subseteq
Q\times\Sigma^*\times M\times Q$ is a finite set of \emph{edges}, $q_0\in Q$ is
its \emph{initial state}, and $F\subseteq Q$ is its set of \emph{final states}.
Towards defining the language of $\cA$, we consider the following relation.
For $(q,w,m),(q',w',m')$, we write $(q,w,m)\to(q',w',m')$ if there is an edge $(q,u,x,q')\in\Delta$ such that $w'=wu$ and $m'=mx$. Then, language of $\cA$ is defined as
\[ L(\cA)\eqdef\{w\in\Sigma^* \mid \exists q\in F\colon (q_0,\varepsilon,1)\xrightarrow{*}(q,w,1) \}, \]
where $\xrightarrow{*}$ is the reflexive, transitive closure of $\to$.

\subsubsection{Graphs Monoids}
Here, we are interested in the case where the monoid $M$ is defined by a finite
graph.  In the following, by a \emph{graph} we mean a finite undirected graph
$\Gamma=(V,E)$ where self-loops are allowed. Hence, $V$ is a finite set of
vertices, and $E\subseteq\{e\subseteq V \mid |e|\le 2\}$ is its set of edges.
To each graph $\Gamma$, we associate a monoid as follows. Consider the alphabet
$X_\Gamma\eqdef\{a_v,\bar{a}_v\mid v\in V\}$, i.e., we create two letters $a_v$,
$\bar{a}_v$ for each vertex $v\in V$. 

Intuitively, we think of the letters $a_v$ as \emph{increment} or \emph{push}
instructions and each $\bar{a}_v$ as the corresponding \emph{decrement} or
\emph{pop} instructions.  Let us make this formal. On the set $X_\Gamma^*$ of
words, we define an equivalence relation. Consider the relation
\begin{align*}
 R_\Gamma&\eqdef\{(a_v\bar{a}_v,\varepsilon) \mid v\in V\} \\
&\cup\{(xy,yx) \mid x\in\{a_u,\bar{a}_u\},~y\in\{a_v,\bar{a}_v\},~\{u,v\}\in E\}. \end{align*}
We now write $w\equiv_\Gamma w'$ if $w'$ can be obtained from $w$ by repeatedly
replacing factors $x$ by $x'$ such that $(x,x')\in R_\Gamma$. First, this means
we can always delete $a_v\bar{a}_v$ for any $v\in V$: this reflects the fact
that $\bar{a}_v$ is the inverse operation of $a_v$. Moreover, if $u$ and $v$
are adjacent in $\Gamma$, then the letters of $u$ (i.e., $a_u$ and $\bar{a}_u$)
commute with the letters of $v$ (i.e., $a_v$ and $\bar{a}_v$). As another
example, if the edge $v$ has a self-loop in $\Gamma$, then we may commute $a_v$
and $\bar{a}_v$, because $(a_v\bar{a}_v,\bar{a}_va_v)\in R_\Gamma$.  Finally, we
define the monoid $\M\Gamma$ as the quotient $X_\Gamma^*/\equiv_\Gamma$. Thus,
$\M\Gamma$ is the set of equivalence classes of $X_\Gamma^*$ modulo
$\equiv_\Gamma$ and multiplication is via $[x][y]\eqdef[xy]$ (this is well-defined
since $\equiv_\Gamma$ is a congruence by definition).

For example, if $\Gamma$ consists of a single vertex, then $\M\Gamma$
is called the \emph{bicyclic monoid} and is denoted by~$\B$. Since
then $X_\Gamma=\{a_v,\bar{a}_v\}$ and the only pair in  $R_\Gamma$ is
$(a_v\bar{a}_v,\varepsilon)$, it is not difficult to see that for
$w\in X_\Gamma^*$, we have $w\equiv_\Gamma\varepsilon$ if and only if
$w$ is a well-bracketed word where $a_v$ and $\bar{a}_v$ are the
opening and closing brackets. Hence, valence automata over $\B$ are
automata with one $\N$-counter.

\subsubsection{Valence Automata over Graphs}
This allows us to define valence automata over graphs: For a graph $\Gamma$, a
\emph{valence automaton over $\Gamma$} is a valence automaton over the monoid
$\M\Gamma$. By $\VA(\Gamma)$, we denote the class of languages accepted by valence automata over $\Gamma$.

\subsection{Valence Automata as Amalgamation Systems}
In order to deduce \cref{main-valence} from \cref{main-amalgamation}, we need
to show that if the emptiness problem is decidable for $\VA(\Gamma)$, then the
language class $\VA(\Gamma)$ is a class of amalgamation systems. To this end, we show that 
$\VA(\Gamma)$ belongs to a language class obtained from the regular languages by repeatedly applying the operators $\cdot+\mathbb{N}$
and $\Alg(\cdot)$. This will follow from results in \cite{DBLP:journals/iandc/Zetzsche21}.

For a graph $\Gamma$, let $\Gamma^-$ denote the graph where all self-loops are
removed. In \cite{DBLP:journals/iandc/Zetzsche21}, it was shown that if
$\Gamma$ has one of the two graphs 
\[ \unloopedpathFour{0.7}\quad\quad\quad\quad \unloopedcycle{0.7} \]
(which are denoted $\Pfour$ and $\Cfour$, respectively) as an induced subgraph, then $\VA(\Gamma)$ is the class of all recursively enumerable languages, and in particular, the emptiness problem is undecidable for $\VA(\Gamma)$. Thus, for \cref{main-valence}, we only need to consider those graphs $\Gamma$ for which $\Gamma^-$ does not contain $\Pfour$ and $\Cfour$ as induced subgraphs. These graphs have been described in \cite{DBLP:journals/iandc/Zetzsche21}.%

\subsubsection{Graphs without $\Pfour$ and $\Cfour$}
Let $\PD$ be the smallest (isomorphism-closed) class of monoids such
that
\begin{enumerate}[(i)]
\item\label{PD1} the trivial monoid $1$ belongs to $\PD$,
\item\label{PD+c} for every
monoid $M$ in $\PD$, we also have $M\times\B$, $M\times\Z$ in $\PD$,
and
\item\label{PDAlg} for any monoids $M$ and $N$ in $\PD$, we also have $M*N$ in
$\PD$.  Here, $M*N$ denotes the \emph{free product} of monoids.  The
precise definition is not needed here---we will only need the
following: in \cite[Lem.~2]{DBLP:conf/icalp/Zetzsche13}, it is shown
that for any monoids $M,N$, we have
$\VA(M*N)\subseteq\Alg(\VA(M)\cup\VA(N))$.
\end{enumerate}
Using essentially the same proof as \cite[Lemma
5.9]{DBLP:journals/iandc/Zetzsche21} one can observe that if $\Gamma^-$ does
not contain $\Pfour$ or $\Cfour$ as an induced subgraph, then $\M\Gamma$
belongs to $\PD$. To be slightly more specific, a graph theoretic result of
Wolk~\cite{Wolk1965} implies that if $\Gamma^-$ does not contain $\Pfour$ or
$\Cfour$, then $\Gamma^-$ is a transitive forest. Here, a transitive forest is
graph obtained from a forest by adding transitive edges between any two nodes
that lie on some path from a root to a leaf. Moreover, transitive forests can
be decomposed into free products (if they contain more than connected
component) or have a vertex that is adjacent to all other nodes (indeed, in a
connected forest, such a vertex is the root of the tree). In either case, we
either obtain the form $\M\Gamma_1*\M\Gamma_2$ for smaller transitive forests
$\Gamma_1,\Gamma_2$, or $\M\Gamma_1\times\B$ or $\M\Gamma\times\Z$ (depending
on whether the root node is looped or not).

\begin{proof}[Proof of \cref{main-valence}]
By the previous discussion, it remains to show that for every language
in $\VA(M)$ with $M$ in $\PD$, we can construct an amalgamation
system.  For \eqref{PD1}~the trivial monoid $1$, $\VA(1)$ is just the
class of regular languages, so this follows
from \cref{thm:regular-amalgamation}.

Moreover, for the subcase of~\eqref{PD+c} of monoids $M\times\Z$, a
classic construction for counter systems yields the inclusion
$\VA(M\times\Z)\subseteq\VA(M\times\B\times\B)$.  Indeed, a single
$\Z$-counter can be simulated by two $\N$-counters.

Thus, it suffices to show that if $\VA(M)$ and $\VA(N)$ have
concatenative amalgamation, then so do~\eqref{PD+c} $\VA(M\times\B)$
and~\eqref{PDAlg} $\VA(M*N)$.  In the latter case, we know that
$\VA(M*N)\subseteq\Alg(\VA(M)\cup\VA(N))$~\cite[Lem.~2]{DBLP:conf/icalp/Zetzsche13}
and thus we may apply \cref{thm:alg-amalgamation}. Finally, for
$M\times\B$, it is entirely straightforward to prove that
$\VA(M\times\B)$ is included in the language class $\VA(M)+\N$, where
$\VA(M)+\N$ is defined as in \cref{sec:counter-extension}.  Indeed, a
valence automaton over $M\times\B$ can be viewed has having three
inscriptions on each edge, namely an element of $M$, an element of
$\B$ (which is a counter update for an $\N$-counter), and an input
word.  This can be directly encoded into a language in $\VA(M)+\N$.

This shows that for every $M$ in $\PD$, all the languages in $\VA(M)$
have concatenative amalgamation systems.  Finally, since each class
$\VA(M)$ is a full trio~\cite[Thm.~4.1]{DBLP:journals/tcs/FernauS02}
this shows that
\cref{main-valence} follows from \cref{main-amalgamation}.
\end{proof}

}

\label{endofdocument}
\newoutputstream{pagestotal}
\openoutputfile{acm.pgt}{pagestotal}
\addtostream{pagestotal}{\getpagerefnumber{endofdocument}}
\closeoutputstream{pagestotal}

\end{document}